\documentclass[fleqn,usenatbib]{mnras}

\usepackage[T1]{fontenc}
\usepackage{ae,aecompl}

\usepackage{graphicx}	
\usepackage{amsmath}	
\usepackage{amssymb}	
\usepackage{pdflscape}
\usepackage{epsfig}
\usepackage{epstopdf}
\usepackage{ctable}
\usepackage{romannum}
\usepackage{array}
\usepackage{multirow}
\usepackage{array}
\usepackage{appendix}
\usepackage{booktabs}

\newcolumntype{+}{>{\global\let\currentrowstyle\relax}}
\newcolumntype{^}{>{\currentrowstyle}}

\newcommand{\intl}{\textit{INTEGRAL}}

\newcommand{\chandra}{\textit{Chandra}}

\newcommand{\sw}{\textit{Swift}}
\newcommand{\ha}{H{\small $\alpha$}}

\newcommand{\hb}{H{\small $\beta$}}

\def\cygx1{Cygnus~X$-$1}
\def\gx{GX~339$-$4}
\def\grs{GRS~1915+105}
\def\v404{V404~Cygni}
\def\swj{Swift~J1753.5-0127}
\def\chis{$\chi^2$}

\def\msun{$M_{\odot}$}

\def\ergcms{erg~cm$^{-2}$~s$^{-1}$}
\def\ergcmsa{erg~cm$^{-2}$~s$^{-1}$~\AA$^{-1}$}
\def\kms{km~s$^{-1}$}
\def\wm2{W~m$^{-2}$}

\def\cm2{cm$^{-2}$}
\def\se1{s$^{-1}$}
\def\nh{$N_{\rm H}$}
\def\nhe{N_{\rm H}}
\def\Av{$A_{\rm V}$}

\def\Ave{A_{\rm V}}

\title[Optical spectroscopy of \v404]{The nova-like nebular optical spectrum of \v404\ at the beginning of the 2015 outburst decay}

\author[F. Rahoui et al.]{
Farid Rahoui$^{1,2}$\thanks{E-mail: farid@rahoui.eu},
J. A. Tomsick$^{3}$,
P. Gandhi$^{4}$,
P. Casella$^{5}$,
F. F{\"u}rst$^{6}$,
L. Natalucci$^{7}$,
\newauthor
A. Rossi$^{8}$,
A. W. Shaw$^{4}$, 
V. Testa$^{5}$,
and D. J. Walton$^{6}$
\\
$^{1}$European Southern Observatory, K. Schwarzschild-Str. 2, 85748
Garching bei M\"unchen, Germany\\
$^{2}$Department of Astronomy, Harvard University, 60 Garden street,
Cambridge, MA 02138, USA\\
$^{3}$Space Sciences Laboratory, 7 Gauss Way, University of
  California, Berkeley, CA 94720-7450, USA\\
$^{4}$Department of Physics and Astronomy, University of Southampton, Highfield, Southampton SO17~1BJ, UK\\
$^{5}$INAF-OA Roma, Via Frascati 33, I-00078 Monteporzio Catone, Italy\\
$^{6}$California Institute of Technology, 1200 East California Boulevard, Pasadena, CA 91125, USA\\
$^{7}$Istituto di Astrofisica e Planetologia Spaziali, INAF, Via Fosso del Cavaliere 100, I-00133 Roma, Italy\\
$^{8}$INAF-IASF Bologna, Area della Ricerca CNR, via Gobetti 101, I--40129 Bologna, Italy\\
}

\date{Accepted XXX. Received YYY; in original form ZZZ}

\pubyear{2016}

\begin{document}
\label{firstpage}
\pagerange{\pageref{firstpage}--\pageref{lastpage}}
\maketitle

\begin{abstract}
We report on FORS2 optical spectroscopy of the black hole X-ray binary \v404, performed at the very beginning of its 2015 outburst decay, complemented by quasi-simultaneous \sw\ X-ray and ultra-violet as well as REM near-infrared observations. Its peculiar spectrum is dominated by a wealth of emission signatures of \ion{H}{i}, \ion{He}{i}, and higher ionisation species, in particular \ion{Fe}{ii}. The spectral features are divided between broad red-shifted and narrow stationary varieties, the latter being emitted in the outer regions. Continuum and line variability at short time scale is high and we find Baldwin effect-like anti-correlations between the full-widths at half-maximum and equivalent widths of the broad lines with their local continua. The Balmer decrement \ha/\hb\ is also abnormally large at $4.61\pm0.62$. We argue that these properties hint at the broad lines being optically thick and arising within a circumbinary component in which shocks between faster optically thick and slower optically thin regions may occur. We associate it to a nova-like nebula formed by the cooling remnant of strong accretion disc winds that turned off when the mass-accretion rate dropped following the last major flare. The \ion{Fe}{ii} lines likely arise from the overlap region between this nebula and the companion star winds, whereas we favour the shocks within the nebula as responsible for the optical continuum via self-absorbed optically thin bremsstrahlung. The presence of a near-infrared excess also points towards the contribution of a strongly variable compact jet or a dusty component.

\end{abstract}




\section{Introduction}

\begin{table*}
\caption{Summary of \v404\ observations and archival data we made use of in this study. The orbital phase $\phi$ of the BH was estimate using the ephemeris given in \citet{1994Casares}.\label{logobs}}
\begin{center}
\begin{tabular}{cccccc}
\hline
Facilities&ID&Instrumental configuration&Start date (MJD) &Exp. Time (second)&$\phi$\\
\hline
\hline
ESO VLT&095.D-0459&FORS2/GC435+600V, OG590+600I&57200.2661&600, 570&0.05--0.06\\
\intl & 155700300010&JEMX-1, JEMX-2, ISGRI&57200.1179&2620, 2597, 2075&0.03--0.04\\
''& 155700320010& ''&57200.1904&2570, 2551, 2046&0.04--0.05\\
''& 155700340010& ''&57200.2629&2587, 2584, 2064&0.05--0.06\\
\sw & 00031403061&XRT, UVOT/W1&57200.1388&58, 73&0.03\\
''&00033861002&XRT, UVOT/U&57200.3925&1475, 1494&0.07--0.08\\
ESO REM&29023 (AOT29)&REMIR/$J$, $H$,$K_{\rm s}$&57200.2899&50, 50, 50&0.06\\
AAVSO&--&$V$, $I$&57200.2658&--&0.05--0.06\\
IRAM&--&NOEMA/97.5~GHz, 140.5~GHz&57199.3000&5400, 9000&0.86--0.97\\
RATAN-600&--&2.3~GHz, 4.6~GHz, 8.2~GHz, 11.2~GHz, 21.7~GHz&57199.9000&--&0.97--0.09\\

\hline
\end{tabular}
\end{center}
\end{table*}

On 2015 June 15, the Burst Alert Telescope instrument \citep[BAT,][]{2005Barthelmy} mounted on the \sw\ satellite \citep{2004Gehrels} detected the Galactic black hole (BH) X-ray binary (XRB) \v404\ after 26 years of quiescence \citep{2015Barthelmy}. The rarity of such an event triggered a collaborative multiwavelength observational campaign of unprecedented scale and despite its brevity, the extreme behaviour exhibited by the source has puzzled the community. Indeed, unlike most of other microquasars whose outbursts can be well-described by hysteresis-like (or q-shape) hardness intensity diagrams tracing spectral transitions between corona-dominated hard states and disc-dominated soft states \citep{2004Fenderb}, \v404\ underwent several hard X-ray flares with up to 10-fold flux variations within a few hours. Whether it transitioned between different spectral states is not clear, several authors having claimed that it was always caught in the hard state \citep[see e.g.,][]{2015Rodriguez, 2015Natalucci, 2016Jenke}, while \citet{2016Radhika} argue that it went through several transitions within each flare. The origin of these flaring events has not been clarified yet, but several studies point towards hard X-ray spectra stemming from Compton scattering of relatively hot photons, maybe from the jets \citep{2015Natalucci, 2016Jenke}, while \chandra\ and GTC/OSIRIS  spectroscopic observations hint at the likely presence of a strong quasi-spherical accretion disc wind (ADW) detected via narrow soft X-ray emission lines \citep{2015King} as well as optical P-Cygni profiles \citep{2016Munoz}. Finally, this extreme variability pattern was not restricted to the X-ray domain and was observed in all bands, including fast radio flares from transient jets \citep{2015Mooley,2015Tetarenko}, as well as sub-second optical photometric flickering associated with optically-thin synchrotron from compact jets \citep{2016Gandhi}. 

As one of the brightest microquasars at optical wavebands, both in outburst and quiescence, \v404\ has been extensively studied in this spectral domain. During the 1989 outburst, several authors reported very rich optical spectra dominated by strong emission signatures from \ion{H}{i} and \ion{He}{i} as well as higher ionisation elements \citep[see e.g.][]{1991Casares, 1992Gotthelf}. The continuum and lines were found to be strongly variable, and various morphologies were observed, including single to triple-peaked features as well as transient P-Cygni profiles; some of these characteristics were again observed during the 2015 outburst \citep[see e.g.][]{2016Munoz}. The system's properties are also relatively well-constrained, and \v404\ consists of a K0-3IV star orbiting a BH with a $6.08\pm0.06$~\msun\ mass function in $6.4714\pm0.0001$ days, the secondary-to-primary mass ratio being $q=0.060_{-0.005}^{+0.004}$ \citep{1992Casares, 1994Casares, 2010Khargharia}. Its distance was also assessed via accurate radio parallax measurements at $2.39\pm0.14$~kpc \citep{2009Miller-Jones}, but the inclination of the system remains relatively unknown, with values from $50^\circ$ to $70^\circ$, which leads to a poorly constrained BH mass in the range 8--15~\msun \citep{1994Shahbaz, 2010Khargharia}. 

In this paper, we report on medium-resolution optical spectroscopy of \v404\  -- complemented by X-ray, ultra-violet, and near-infrared data -- performed right after the last major flare, at the very beginning of its decay to quiescence (see \autoref{aavso}). Section~2 details the data reduction procedure, whereas Section~3 is dedicated to the spectral analysis. We discuss the outcomes and their implications in Section~4 and conclude in Section~5.

\section{Observations and data reduction}

The data set consists of quasi-simultaneous observations obtained on 2015
June 27, with (1) the FOcal Reducer/Low disperser Spectrograph 2
\citep[Obs. ID 095.D-0459, PI Rahoui; FORS2,][]{1998Appenzeller}  mounted on the UT1 Cassegrain focus at the European Southern Observatory (ESO) Very Large Telescope (VLT) at Cerro Paranal; (2) the X-Ray Telescope \citep[XRT,][]{2005Burrows} and Ultra-Violet/Optical Telescope \citep[UVOT,][]{2005Roming} on board the \sw \citep{2004Gehrels} satellite (Obs.~ID 00031403061, Public; Obs.~ID 00033861002, PI Altamirano); (3) the Joint European X-ray Monitor \citep[JEM-X,][]{2003Lund} and Imager On board {\it INTEGRAL} Satellite \citep[IBIS,][]{2003Ubertini} mounted on the INTERnational Gamma-Ray Astrophysics Laboratory \citep[\intl,][]{2003Winkler} satellite (Science Window~ID 155700300010, 155700320010, and 155700340010); and (4) the Rapid Eye Mount IR \citep[Obs. ID 29023, PI Casella; REMIR,][]{2005Calzoletti} installed on the Rapid Eye Mountain (REM) telescope at Cerro La Silla. For the sake of building a radio to X-ray spectral energy distribution (SED), we also use quasi-simultaneous radio data obtained with the RATAN-600 radio telescope at 2.3, 4.6, 8.2, 11.2, and 21.7 GHz \citep{2015Trushkin} between MJD~57199.9 and MJD~57200.7, as well as 97.5 GHz and 140.5 GHz flux densities from the NOrthern Extended Millimeter Array (NOEMA) and integrated between MJD~57199.3 and MJD~57200.1 \citep{2015Tetarenkob}.

\subsection{FORS2 observations}
\begin{figure*}
\begin{center}
\includegraphics[width=15cm]{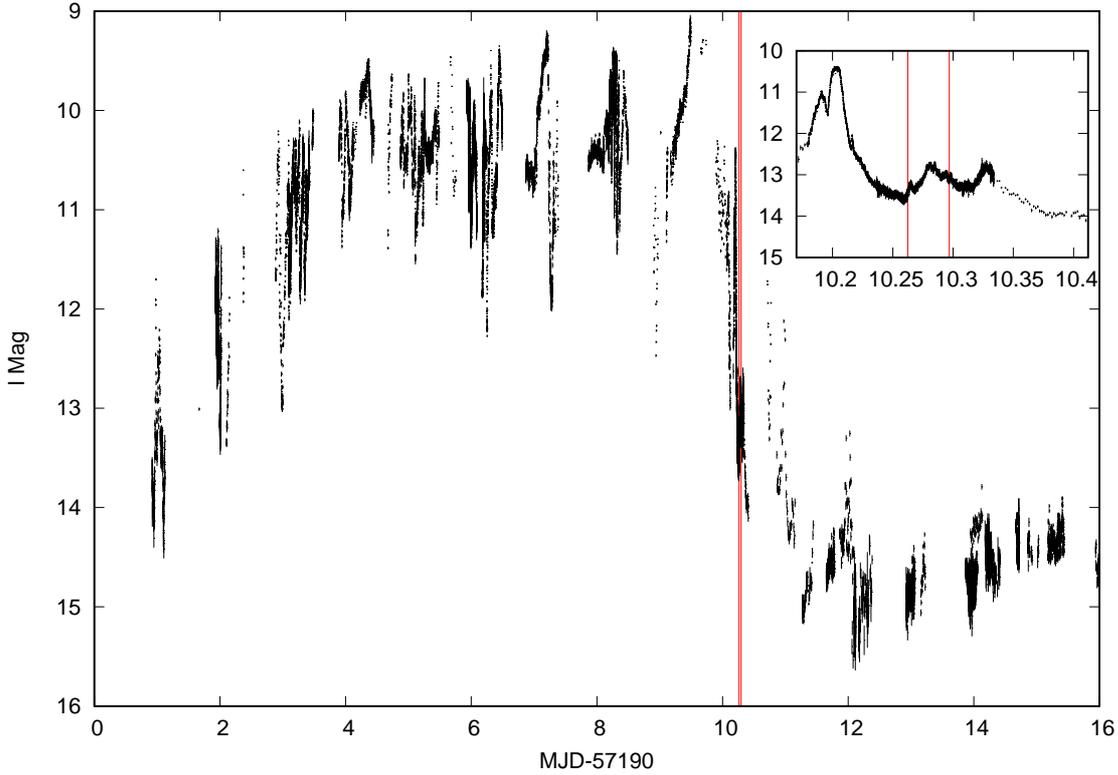}
\caption{V404 outburst as seen through I band photometry. The red vertical bars mark the time range of our FORS2 observations, and the inset displays a zoom-in centred on the FORS2 epoch.}
\label{aavso}
\end{center}
\end{figure*}

On 2015 June 27, we performed FORS2 medium-resolution spectroscopy of \v404\ (see \autoref{aavso}) with the 600V (hereafter $V$) and 600I (hereafter $I$) grisms combined with the GC435 and OG590 filters, respectively, for a total $4500-9300$~\AA\ spectral coverage. In both cases, we used the standard resolution (SR) collimator and the slit-width was set to 1\arcsec\ with a rotation angle always close to the parallactic angle. Atmospheric conditions were medium-to-good, with a thin sky transparency and a 1.91 airmass. The exposure time of each individual frame was set to 30~s and a total of 20 and 19 exposures were taken in both $V$ and $I$, respectively, which, accounting for overheads, gives a 67~s effective time resolution for variability study. The A0V spectro-photometric standard star LTT~7379 was observed in similar conditions for flux-calibration.

\begin{figure*}
\begin{center}
\includegraphics[width=15cm]{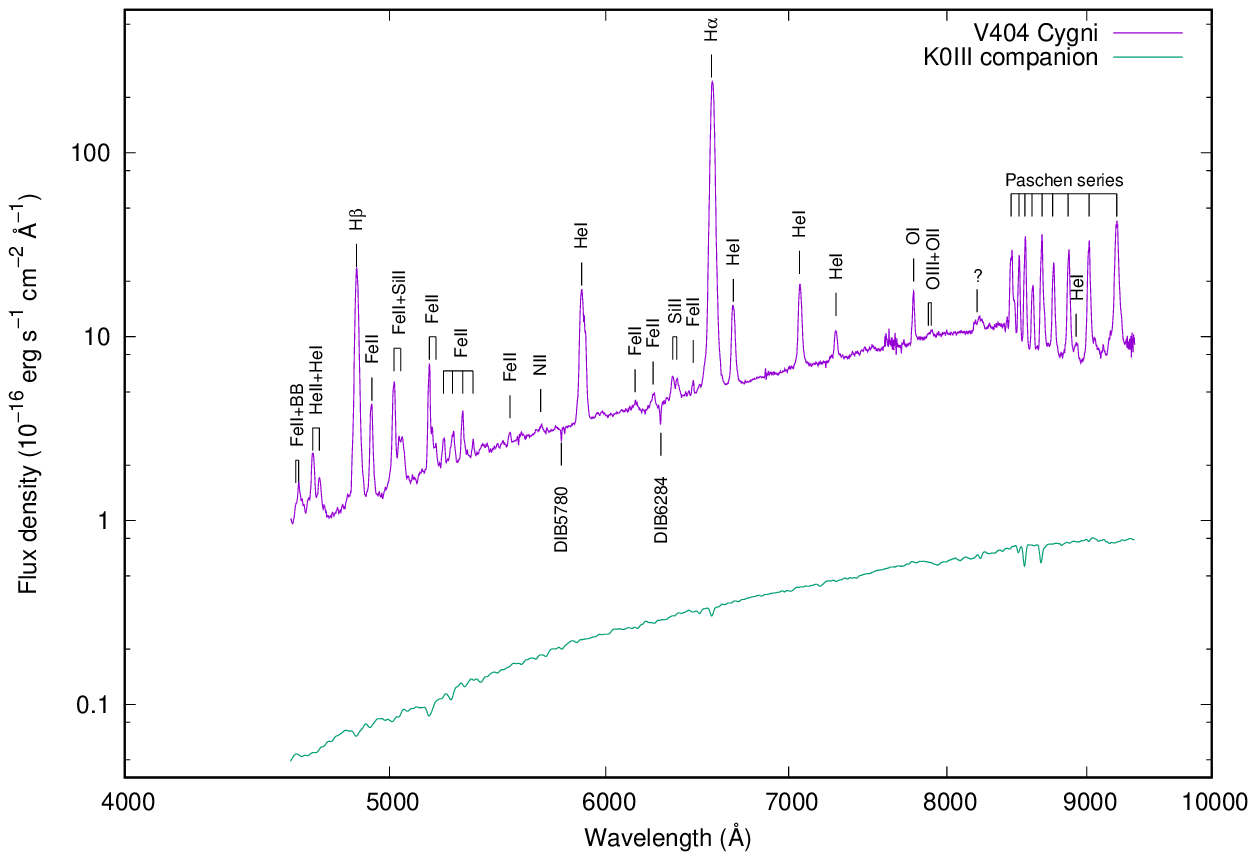}
\caption{Flux-calibrated time-averaged FORS2 spectrum of \v404\ (magenta)
  compared to that of a K0III giant star suffering from the same
  ISM extinction and scaled to \v404\ companion
  star's optical magnitudes (green). All the detected emission lines and DIBs are marked.}
\label{contspec}
\end{center}
\end{figure*}

We reduced the data using the dedicated pipeline (v.~5.3.5) implemented in the ESO data reduction environment {\tt Reflex}~v.~2.6 \citep{2013Freudling}, which follows the standard steps for optical spectroscopy reduction and produces cleaned and background subtracted 2D spectroscopic images. The wavelength-calibration was achieved by deriving a dispersion solution using Helium, Argon, and Mercury/Cadmium arc observations. We then used the routines {\tt apall}, {\tt standard}, {\tt sensfunc}, and {\tt calibrate} implemented in {\tt IRAF}~v.~2.16\footnote{ {\tt  IRAF} is distributed by the National Optical Astronomy Observatories, which are operated by the Association of Universities for Research in Astronomy, Inc., under cooperative agreement with the National Science Foundation.} to optimally extract the source, sky, and spectro-phometric standard star 1D spectra, compute the sensitivity function, and apply it to \v404\ $V$ and $I$ spectra for flux calibration. The accuracy of the dispersion solution was finally assessed by measuring the centroids of the brightest OH emission lines present in the sky spectra. We found systematic +0.6 to +0.8~\AA\ offsets, likely due to the five-hour time gap between the science and arc observations. A $-0.8$~\AA\ correction was thus applied to all line centroid measurements listed in this study. When applicable, we also corrected all listed radial velocity shifts for a 13~\kms\ barycentric-heliocentric velocity as estimated at the epoch of the observations.
\newline

The flux-calibrated spectra were  corrected for slit-losses, which we estimated assuming the slit transmission factor used by ESO for its exposure time calculators \footnote{ https://www.eso.org/observing/etc/doc/formulabook/node18.html}. It consists in an error function $erf$ that depends on the slit-width and the image quality at the focal plane of the telescope. We slightly modified it to include potential slit-centring offsets implied by the wavelength-dependent atmospheric differential refraction for a given airmass and rotation angle, as formulated in \citet{1982Filippenko}. The wavelength-dependence of the image quality along the slit was modelled using the expression given in \citet{2010Martinez}, which takes into account the combined effects of the airmass and the outer scale of the turbulences. The reference image quality was directly measured on each individual 2D spectroscopic frame at 5015~\AA\ and 7065~\AA\ in $V$ and $I$, respectively, through Gaussian fitting of the spatial profiles, and we find values in the range  0\farcs6-0\farcs8. We stress here that our observations were seeing-limited and the spectral resolution was thus better that the one expected when a source fully fills the slit. Through the fit to the full-widths at half-maximum (FWHMs) of the arc lines, we measure the slit-limited resolution between $R\sim1100-1500$ and $R\sim1400-1900$ in $V$ and $I$, respectively. The seeing-limited resolution being roughly equal to the slit-limited one divided by the image quality, we consequently reached $R\sim1200-1900$ and $R\sim2200-2900$ in $V$ and $I$, respectively.

Finally, based on the continuum level gap between $V$ and $I$ in the overlap region, we estimate that the flux calibration is accurate at 3\%, although the statistical noise is a lot lower. Moreover, we derive $V=15.72\pm0.03$ and $I=12.90\pm0.03$ through the convolution of the average flux-calibrated spectrum with Johnson and Cousins filters, whereas the average simultaneous AAVSO magnitudes during our observations are $V=15.67\pm0.05$ and $I=12.95\pm0.01$.

\subsection{\sw\ observations}

We reduced the XRT data with {\tt HEASOFT}~v.~6.17 and the 2016 January 20 calibration data base (CALDB) version. We used \texttt{xrtpipeline}~v.~0.13.1 to collect events in Windowed Timing (WT) mode to avoid pile-up. For Obs.~ID 00031403061, the source and background spectra were extracted with \texttt{xselect}~v.~2.4c using 40-pixel square boxes in the 0.4--10~keV energy range. We generated the ancillary response file (ARF) with \texttt{xrtmkarf} and used the latest version (v.~015) of the response matrices provided by the \sw\ team. We rebinned the spectrum to obtain a minimum of 50 counts per channel, and restricted the effective energy range to 0.4--5~keV as the source is not detected beyond.

In contrast, the XRT data for Obs.~ID 00033861002 appeared clearly extended towards the South-West direction without the clear presence of a point source. This is likely due to the formation of the X-ray dust scattered rings reported in \citet{2015Beardmore, 2016Vasilopoulos}. For this reason, no spectrum was extracted and we decided not to use these data for this study.
\newline

The UVOT photometry was obtained in two filters, W1 for Obs.~ID 00031403061 and U for Obs.~ID 00033861002. We produced an image in each of them with {\tt uvotimsum}. We then used
{\tt uvotsource} to extract the source in a 5\arcsec\ region and the background counts in a 15\arcsec\ source-free circular aperture, respectively. The derived \v404\ flux densities are
$(7.89\pm0.89) \times 10^{-16}$~\ergcmsa\ and $(1.79\pm0.09) \times 10^{-16}$~\ergcmsa\ in W1 and U, respectively.

\subsection{\intl\ observations}

The data were reduced with the {\tt jemx\_science\_analysis} and {\tt ibis\_science\_analysis} routines implemented in the Off-Line Scientific Analysis (OSA) v10.2 suite using the most recent calibration files. Images were first generated in 16 bands between 3.04 and 34.88~keV for the two JEM-X instruments whereas 4 bands were used for the IBIS Soft Gamma-Ray Imager \citep[ISGRI,][]{2003Lebrun}, covering the range 20 to 200 keV. A careful analysis of all produced images showed that \v404\ was detected in all JEM-X and ISGRI bands during the first two science windows but only with JEM-X below 16~keV during the third one, which is simultaneous to our FORS2 observations. We then extracted JEM-X and ISGRI spectra at \v404\ \chandra\ position and generated redistribution matrix files (RMF) rebinned to the spectral ranges of interest. 

\subsection{REM observations}

The REM data were taken on June 27 2015 with REMIR in $J$,$H$ and $K_{\rm S}$ and consist in a sequence of three images per filter, each image being constituted of five frames that were dithered using a prism wedge. The reduction process, done automatically by the robotic Automatic Quick Analysis pipeline \citep[AQuA,][]{2004Testa}, consisted in first obtaining an empty sky image from a median stack of the five dithered frames and subtracting it from each of them. Following the application of a flat-field image, the images were re-aligned and stacked to obtain a final image with a S/N equivalent to the total exposure time of the five-image sequence. Photometry was then performed using the PSF-fitting software DAOPHOT \citep{1987Stetson} and the fluxes were calibrated using some secondary standard stars in the field against the 2MASS catalogue \citep{2006Skrutskie}. We measure the following magnitudes: $11.40\pm0.06$, $11.35\pm0.05$, and $11.37\pm0.05$ in $J$, $10.72\pm0.08$, $10.80\pm0.08$, $10.85\pm0.08$ in $H$, as well as $10.25\pm0.09$, $10.24\pm0.07$, and $10.33\pm0.08$ in $K_{\rm S}$.

\section{Results}  

\autoref{contspec} displays the 4500--9300~\AA\ slit-loss corrected average spectrum of \v404. The expected contribution from its companion star, modelled by a K0III spectrum scaled to the disc contribution-corrected magnitudes derived in \citet{1993Casares} when the source was in quiescence and reddened to the ISM extinction along the line-of-sight of the source (see below), is also displayed. The detected lines, the measurements of which are listed in \autoref{speclines}, are also marked. We give their measured wavelength $\lambda_{\rm c}$ and laboratory wavelength $\lambda_{\rm l}$ in \AA, radial velocity shifts $\Delta V$ in \kms\, if any, equivalent widths $\mathring{W}$ in \AA, FWHMs in \kms, quadratically corrected from the instrumental FWHMs, as well as their intrinsic fluxes in \ergcms. Note that their underlying continuum was locally assessed with a first-order polynomial. The latter being the primary source of inaccuracy, each measurement was repeated several times with different continuum placements within the same wavelength range to obtain a set of values that eventually averaged out. The uncertainties listed in Table~\ref{speclines} are therefore the scatter to the mean rather than the statistical errors for any one fit, which are much smaller. 

\begin{table*}
\begin{center}
\caption{Optical lines in \v404\ FORS2 spectrum. ``$\dag$'' marks the features have a complex continuum making the measurements difficult; ``$\ast$'' means that the lines are detected but measurements would be unreliable.\label{speclines}}
\begin{tabular}{l|cccccc}
\hline\hline
Element&${\lambda_{\rm \small c}}^a$&${\lambda_{\rm \small l}}^b$&${\Delta V}^c$&${\mathring{W}}^d$&FWHM$^e$&${F_{\rm line}}^f$\\
\hline
&\multicolumn{6}{c}{$V$ band}\\
\hline
\ion{Fe}{ii}	&4583.8$\pm$1.8&4583.8&--&$\ast$&$\ast$&$\ast$\\
\ion{Fe}{ii}+BB	&4632.0$\pm$3.4&4629--4650&--&$\ast$&$\ast$&$\ast$\\
\ion{He}{ii}	&4686.4$\pm$1.3&4685.9&--&$\ast$&$\ast$&$\ast$\\
\ion{He}{i}		&4714.0$\pm$2.3&4713.1&--&$\ast$&$\ast$&$\ast$\\
H{$\beta$}		&4863.5$\pm$0.6&4861.3&123$\pm$37&-264.2$\pm6.0$&$934\pm31$&$277.8\pm5.4$\\
\ion{Fe}{ii}	&4923.8$\pm$0.8&4923.9&--&$-23.7\pm3.1$&$631\pm64$&$27.5\pm2.7$\\
\ion{Fe}{ii}	&5018.2$\pm$0.6&5018.4&--&$-36.1\pm5.2$&$701\pm61$&$42.2\pm4.8$\\
$\dag$\ion{Si}{ii}	&5040.0$\pm$2.4&5041.0&--&$-10.2\pm4.1$&$744\pm154$&$12.4\pm3.3$\\
$\dag$\ion{Si}{ii}	&5055.5$\pm$2.7&5056.0&--&$-13.6\pm4.5$&$940\pm189$&$18.2\pm5.1$\\
\ion{Fe}{ii}	&5169.7$\pm$0.7&5169.0&--&$-21.0\pm2.3$&$618\pm62$&$33.7\pm4.1$\\
\ion{Fe}{ii}	&5234.4$\pm$2.2&5234.6&--&$-4.6\pm1.6$&$593\pm125$&$7.8\pm2.3$\\
\ion{Fe}{ii}	&5276.2$\pm$1.8&5276.0&--&$-6.6\pm3.2$&$563\pm116$&$11.3\pm1.6$\\
\ion{Fe}{ii}	&5316.7$\pm$1.0&5316.7&--&$-9.5\pm1.7$&$543\pm112$&$16.2\pm2.0$\\
\ion{Fe}{ii}	&5363.2$\pm$3.5&5362.8&--&$-1.5\pm0.5$&$504\pm94$&$3.1\pm1.7$\\
\ion{Fe}{ii}	&5533.6$\pm$4.3&5532.1&--&$\ast$&$\ast$&$\ast$\\
\ion{N}{ii}		&5678.2$\pm$5.1&5679.6&--&$\ast$&$\ast$&$\ast$\\
\ion{He}{i} 	&5878.1$\pm$0.6&5875.6&115$\pm$31&$-80.3\pm4.1$&$982\pm59$&$259.9\pm10.6$\\
\ion{Fe}{ii}	&6150.2$\pm$7.7&6148.0&--&$\ast$&$\ast$&$\ast$\\
\ion{Fe}{ii}	&6245.4$\pm$5.5&6247.6&--&$\ast$&$\ast$&$\ast$\\
\ion{Si}{ii}	&6347.2$\pm$1.7&6347.1&--&$-5.8\pm1.9$&$668\pm135$&$24.0\pm5.6$\\
\ion{Si}{ii} 	&6370.6$\pm$2.1&6371.4&--&$-3.8\pm1.5$&$580\pm241$&$15.5\pm4.4$\\
\ion{Fe}{ii} 	&6382.9$\pm$4.2&6383.4&--&$\ast$&$\ast$&$\ast$\\
\ion{Fe}{ii}	&6457.4$\pm$3.1&6456.4&--&$\ast$&$\ast$&$\ast$\\
H{$\alpha$}		&6564.6$\pm$0.2&6562.7&76$\pm$10&$-1129.0\pm4.0$&$1056\pm13$&$5315.1\pm24.1$\\
\ion{He}{i}		&6680.9$\pm$0.5&6678.2&108$\pm$22&$-39.0\pm3.4$&$959\pm31$&$189.1\pm6.4$\\
\ion{He}{i}		&7067.5$\pm$0.5&7065.2&85$\pm$21&$-40.7\pm1.5$&$920\pm41$&$247.0\pm7.4$\\
\ion{He}{i}		&7283.4$\pm$1.0&7281.3&74$\pm$41&$-8.2\pm1.1$&$809\pm111$&$56.9\pm6.2$\\
\hline
&\multicolumn{6}{c}{$I$ band}\\
\hline
\ion{He}{i}		&7068.0$\pm$0.3&7065.2&106$\pm$13&$-41.8\pm1.2$&$906\pm26$&$256.8\pm8.0$\\
\ion{He}{i}		&7283.7$\pm$0.9&7281.3&86$\pm$37&$-8.7\pm0.8$&$8033\pm88$&$60.7\pm5.6$\\
$\dag$\ion{O}{i}		&7776.6$\pm$0.6&7773.4&110$\pm$23&$-10.6\pm1.0$&$546\pm48$&$106.1\pm6.2$\\
\ion{O}{iii} 	&7875.8$\pm$6.1&7873.5&--&$\ast$&$\ast$&$\ast$\\
\ion{O}{ii}		&7897.2$\pm$8.1&7894.6&--&$\ast$&$\ast$&$\ast$\\
?&8223.2$\pm$3.8&--&--&$\ast$&$\ast$&$\ast$\\
$\dagger$\ion{O}{i}	&8446.2$\pm$1.7&8446.4&--&$-32.8\pm3.1$&$740\pm59$&$356.0\pm18.8$\\
$\dagger$Pa (17-3)	&8468.7$\pm$2.8&8467.3&--&$-3.5\pm1.1$&$462\pm60$&$61.0\pm7.1$\\
Pa (16-3)		&8501.1$\pm$0.9&8502.5&--&$-24.8\pm2.5$&$465\pm25$&$218.1\pm10.8$\\
Pa (15-3)		&8544.6$\pm$0.8&8545.4&--&$-45.2\pm4.4$&$574\pm32$&$338.7\pm12.8$\\
Pa (14-3)		&8598.8$\pm$0.5&8598.4&--&$-20.7\pm2.2$&$585\pm30$&$184.4\pm7.4$\\
Pa (13-3)		&8665.6$\pm$0.6&8665.0&--&$-48.5\pm4.0$&$621\pm34$&$463.9\pm8.9$\\
Pa (12-3)		&8751.2$\pm$0.5&8750.5&--&$-30.1\pm2.1$&$632\pm35$&$291.1\pm9.7$\\
Pa (11-3)		&8863.5$\pm$0.5&8862.8&--&$-44.6\pm3.1$&$676\pm24$&$384.3\pm8.6$\\
Pa (10-3)		&9015.9$\pm$0.5&9014.9&20$\pm$17&$-60.2\pm3.0$&$702\pm30$&$464.9\pm22.3$\\
$\dagger$Pa (9-3)	&9230.4$\pm$0.8&9229.0&33$\pm$26&$-85.8\pm4.6$&$927\pm101$&$860.6\pm20.0$\\
\hline

\end{tabular}
\end{center}
\raggedright{\hspace{2.4cm}{{$^a$}{\scriptsize Measured wavelength in \AA}\\}
\hspace{2.4cm}{{$^b$}{\scriptsize Air laboratory wavelength in \AA}\\}
\hspace{2.4cm}{{$^c$}{\scriptsize Radial velocity shift in \kms, corrected for a 13~\kms\ heliocentric-barycentric velocity}\\}
\hspace{2.4cm}{{$^d$}{\scriptsize Equivalent widths in \AA}\\}
\hspace{2.4cm}{{$^e$}{\scriptsize Full-width at half-maximum in \kms, quadratically corrected for instrumental broadening}\\}
\hspace{2.4cm}{{$^f$}{\scriptsize Intrinsic line flux in units of $10^{-15}\,\,{\rm erg\,\,cm}^{-2}\,\,{\rm s}^{-1}$}\\}}
\end{table*}

\subsection{The optical spectroscopic content}

\v404\ optical spectrum is very rich, with a wealth of \ion{H}{i}
(\ha, \hb, the Paschen series), \ion{He}{i} and \ion{He}{ii} emission
lines. We report a weak \ion{C}{iii}+\ion{N}{iii} fluorescence complex (also called the Bowen Blend, BB) around 4640~\AA\ that likely originates from the irradiated accretion disc as our observations took place at $\phi\approx0.05$ (see~\autoref{logobs}), i.e. when the emission from the irradiated hemisphere of the companion star is not visible to the observer. We moreover detect several less common emission features, most of which we associate with \ion{Fe}{ii} along with some signatures of \ion{Si}{ii}, \ion{O}{i}, \ion{O}{ii}, \ion{O}{iii}, and \ion{N}{ii}, which were not present at the beginning of the outburst \citep{2016Bernardini}. A unidentified complex, which might as well be an instrumental feature as a blend of \ion{He}{i} and \ion{He}{ii} lines, is also present around 8223~\AA. 

This spectroscopic content is consistent with that reported in some previous studies of the 1989 outburst \citep{1991Casares,1991Wagner,1992Gotthelf}. Yet, despite having a similar continuum flux level, \ha\ and \hb\ are 50 and 20 times brighter, respectively, than during 1989 July 1-4 observations presented in \citet{1992Gotthelf}. This explains why their equivalent widths are so much larger than those previously reported, with $\mathring{W}_{\textrm{\ha}}\approx-1130$~\AA\ and $\mathring{W}_{\textrm{\hb}}\approx-$241~\AA\ in the FORS2 spectrum compared
with about -110~\AA\ and -12~\AA, respectively, in 1989 July 1-4. \ha\ also has very broad wings of $\pm2000$~\kms. Complex and variable line profiles were previously reported, including double and triple-peaked emissions as well as P-Cygni profiles \citep{2016Munoz}. In our case, although the 1100 to 1900 average resolution of the FORS2 spectra prevents a detailed analysis, it is very likely that all the features are single-peaked. \ion{H}{i}~$\lambda8865$, $\lambda9015$, and $\lambda9229$ profiles do show two local maxima, but we do not believe those are physical. 

Based on their radial velocities, the features are separated between stationary and red-shifted, pointing towards at least two components responsible for \v404\ optical emission. Most of the higher ionisation species, in particular \ion{Fe}{ii} and \ion{Si}{ii}, as well as Pa (17-3) to Pa (11-3), do not exhibit any velocity shift, although we stress that the uncertainties on the centroid measurements are sometimes quite large and we cannot rule out that some of them are actually displaced. Pa~(10-3) and Pa~(9-3) centroids do show +20 to +30~\kms\ shifts but they are likely not physical. In contrast, \hb, \ion{He}{i}~{ $\lambda5876$}, \ha, \ion{He}{i}~{$\lambda6678$}, \ion{He}{i}~{$\lambda7065$}, \ion{He}{i}~{$\lambda7281$}, and \ion{O}{i}~{$\lambda7774$ are unambiguously red-shifted, with radial displacements ranging between about +70 to +120~\kms}. 
\newline

Finally, besides features intrinsic to \v404, we searched for diffuse
interstellar bands (DIBs) and we report two centred at 5780~\AA\ and
6284~\AA. \citet{1975Herbig} showed that DIBs were
strongly correlated to the ISM extinction along the line-of-sight of
the sources in which they are detected and derived several
relationships between their equivalent widths and $E(B-V)$. Here, we
nonetheless rely on DIB5780 only as DIB6284 is likely
contaminated with some atmospheric absorption
troughs. \citet{1994Jenniskens} obtained $\mathring{W}_{\rm
    5779}=(0.647\pm0.053) \times {E(B-V)}$. We measure
  $\mathring{W}_{\rm 5779}\,=\,0.798\pm0.027$~\AA\ which leads to $E(B
-V)\,=\,1.233\pm0.116$. For an average total-to-selective extinction ratio
$R_{\rm V}\,=\,3.1$, we thus derive $\Ave\,=\,3.82\pm0.36$, in line
with previous estimates \citep[see e.g.][]{2009Hynes}. We
note that this \Av\ value is consistent with a distance in the range
2.4-2.8 kpc as derived using the 3D extinction map given in
\citet{2006Marshall}, which is similar to the
$2.39\pm0.14$~kpc distance measured through radio parallax and reported in 
\citet{2009Miller-Jones}.

\subsection{The optical spectroscopic variability}

\begin{figure*}
\begin{center}
\begin{tabular}{cc}
\includegraphics[width=8cm]{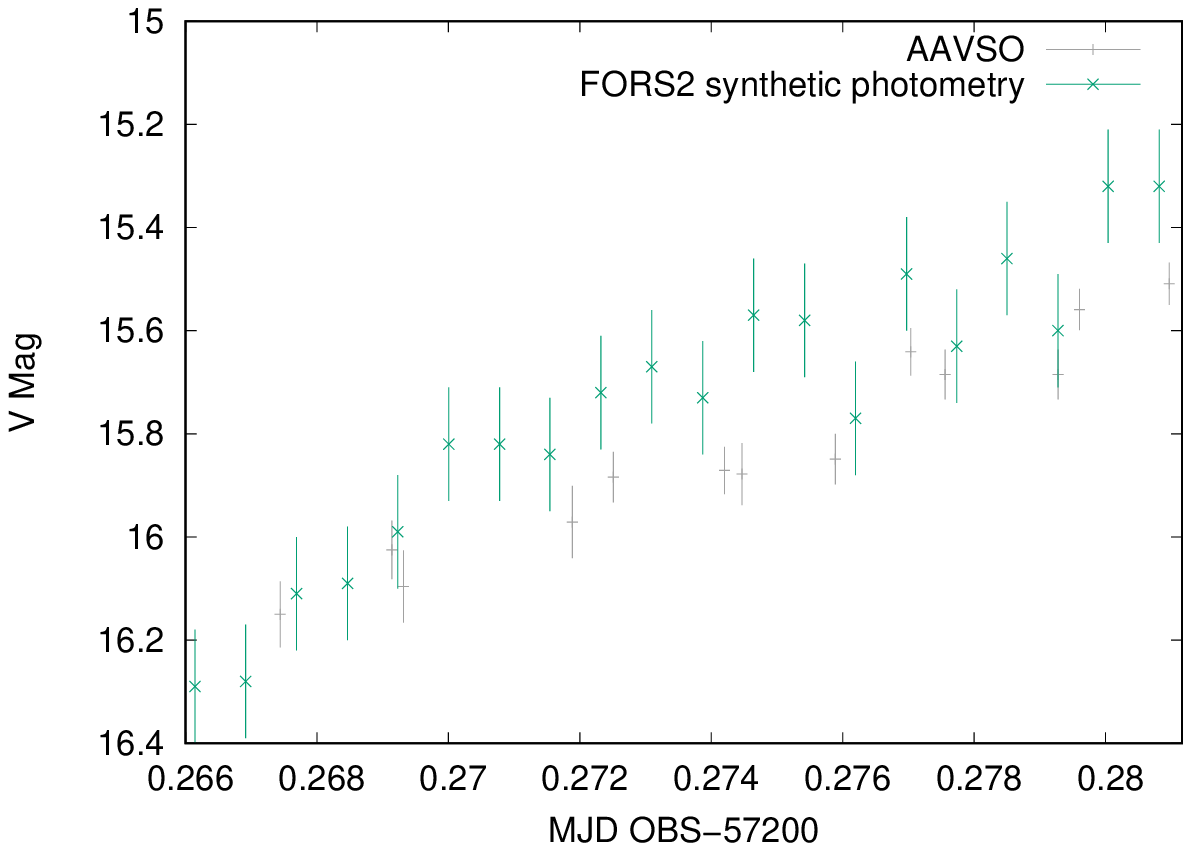}&\includegraphics[width=8cm]{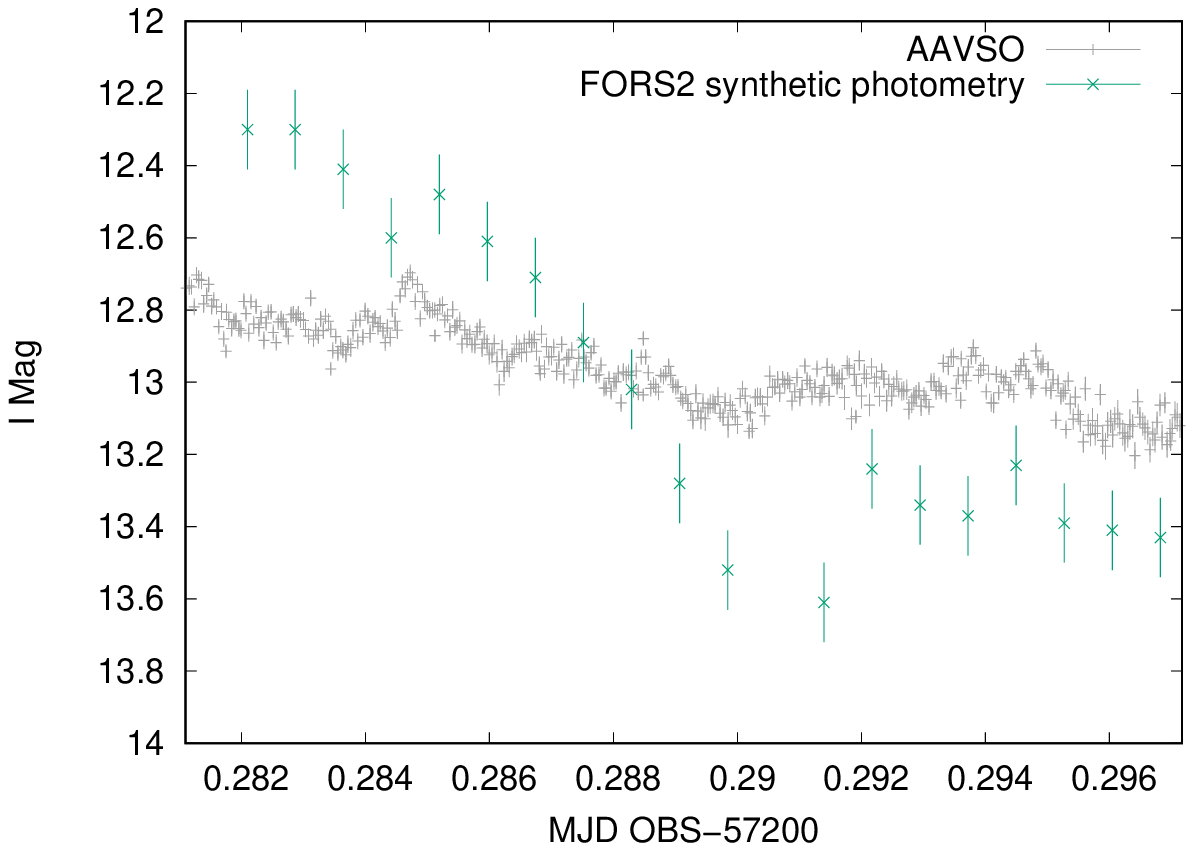}\\
\end{tabular}
\caption{Variations at 67~s time scale of the $V$ (left) and $I$ (right) synthetic magnitudes derived from the convolution of \v404\ sub-spectra with Johnson and Cousins photometric filters (green). We compare them to the simultaneous pre-validated AAVSO $V$ and $I$ magnitudes (grey).}
\label{varmag}
\end{center}
\end{figure*}

During this outburst, \v404\ emission has been highly erratic and extreme multi-wavelength variations were reported by several authors \citep{2015Tetarenko, 2015Rodriguez, 2015King, 2016Kimura, 2016Marti, 2016Rana}. A strong variability pattern with FORS2 is therefore expected, and the fact that both $V$ and $I$ observations consist of 20 and 19 sub-exposures of 30~s each, respectively, allows us to investigate changes in the spectral continuum as well as the centroids, intrinsic fluxes, equivalent widths, and FWHMs of the spectral lines on a 67~s time scale, taking into account observing overheads. \autoref{varmag} displays the evolution of the simultaneous AAVSO $V$ and $I$ pre-validated magnitudes compared with that of the synthetic magnitudes derived through the convolution of Johnson and Cousins filters with the sub-spectra. It is clear that despite some small discrepancies, the match in $V$ is quite good, which confirms the reality of the spectral variability and points towards the flux calibration of each $V$ sub-spectrum to be accurate at about 10\%. However, differences are larger in $I$, and while the evolutions of both the AAVSO pre-validated and FORS2 synthetic magnitudes are similar, flux discrepancies are wider. We did a similar comparison with the simultaneous $I$ magnitudes presented in \citet{2016Kimura} and find the same discrepancies. Large losses are likely mostly due to sky transparency variations, in particular thin clouds passing. However, it is not clear why some of our synthetic magnitudes are significantly smaller, i.e. we find large flux excesses. No problem was reported during the observations and we are not aware of any extra light entering the slit and/or the detector. The spectro-photometric standard star was moreover observed in similar conditions in both $V$ and $I$ and its flux-calibration is very good. Even if the flux level of the average $I$ spectrum is accurate, it is clear that the flux-calibration of each individual $I$ sub-spectrum was likely compromised, with mild to severe differences with photometric measurements. Although we believe that this does not impact the relevance of the variability pattern, we decided not to investigate the flux variability of the lines present in the $I$ sub-spectra. Nonetheless, their equivalent widths and FWHMs being independent on the accuracy of their local continuum, we still trace the changes in these two quantities, although we only consider the $I$ sub-spectra for which the difference between the AAVSO and synthetic magnitudes is lower than 0.2 in absolute value.

\begin{figure*}
\begin{center}
\begin{tabular}{ccc}
\includegraphics[height=3.83cm,width=5.7cm]{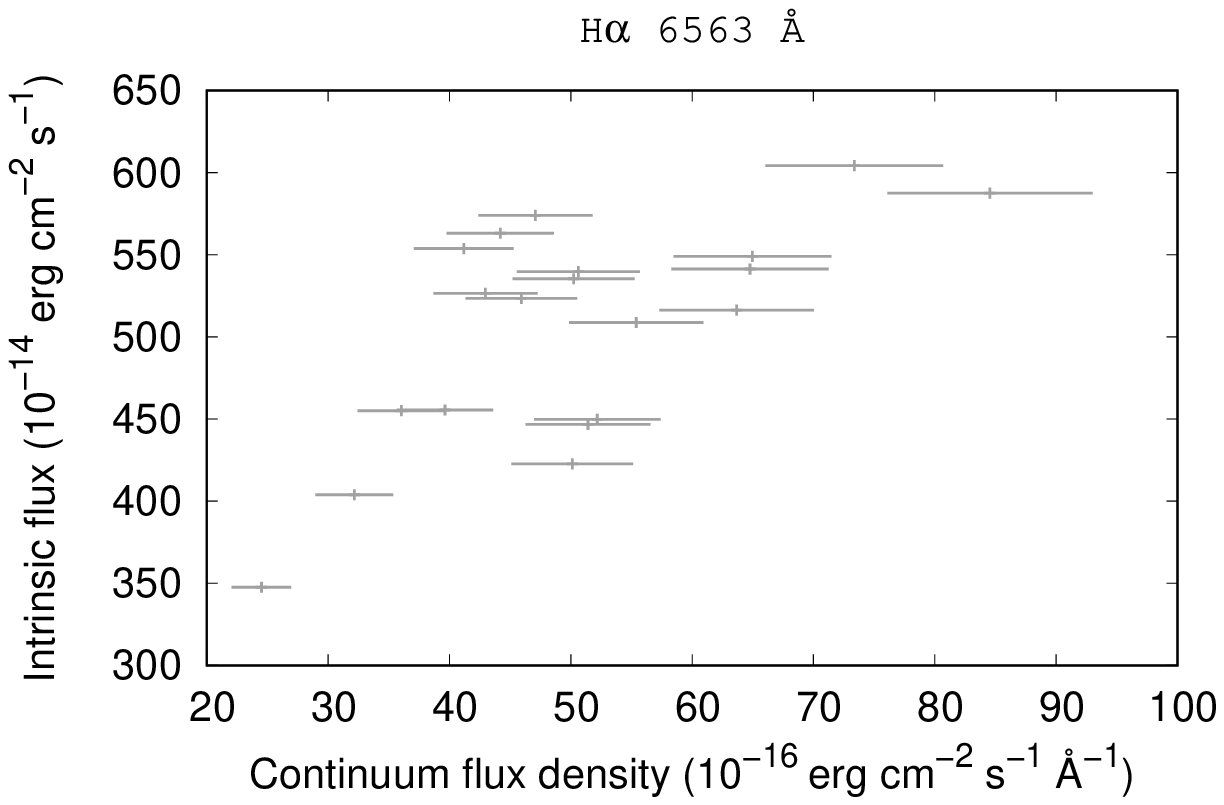}&\includegraphics[height=3.83cm,width=5.7cm]{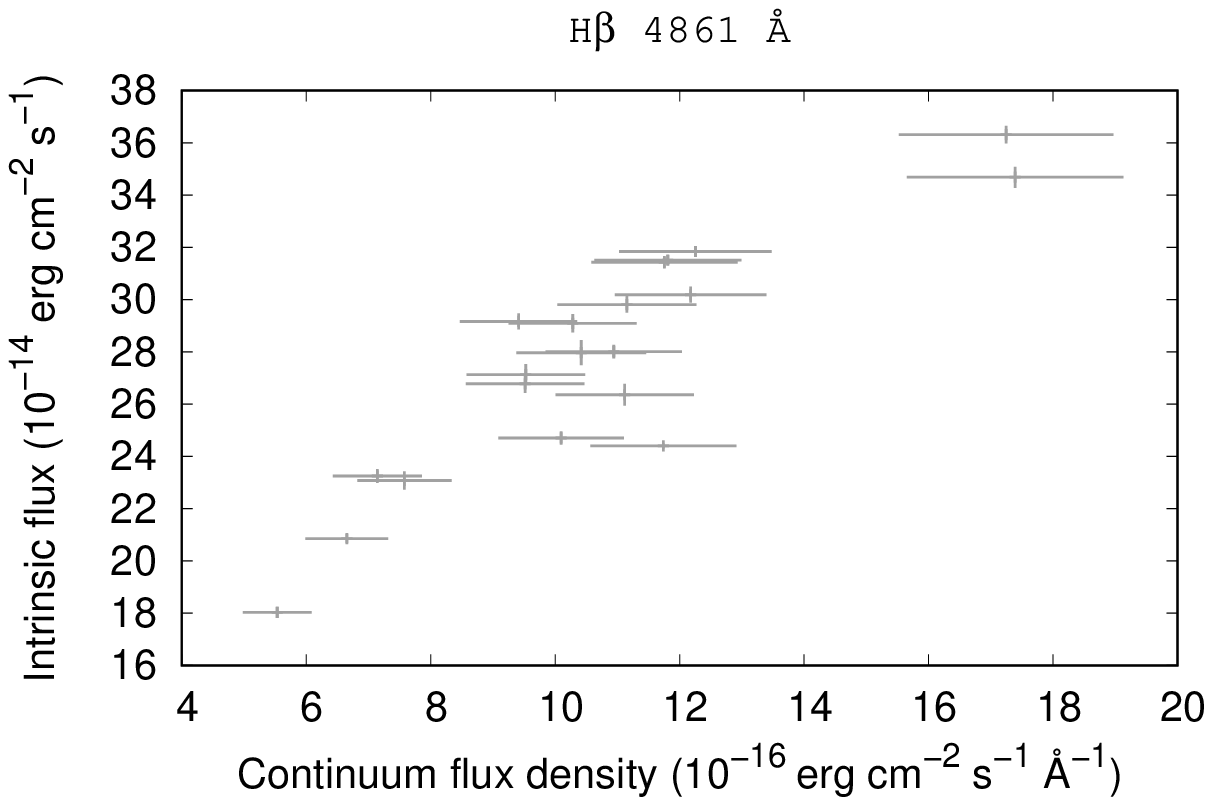}&\includegraphics[height=3.83cm,width=5.7cm]{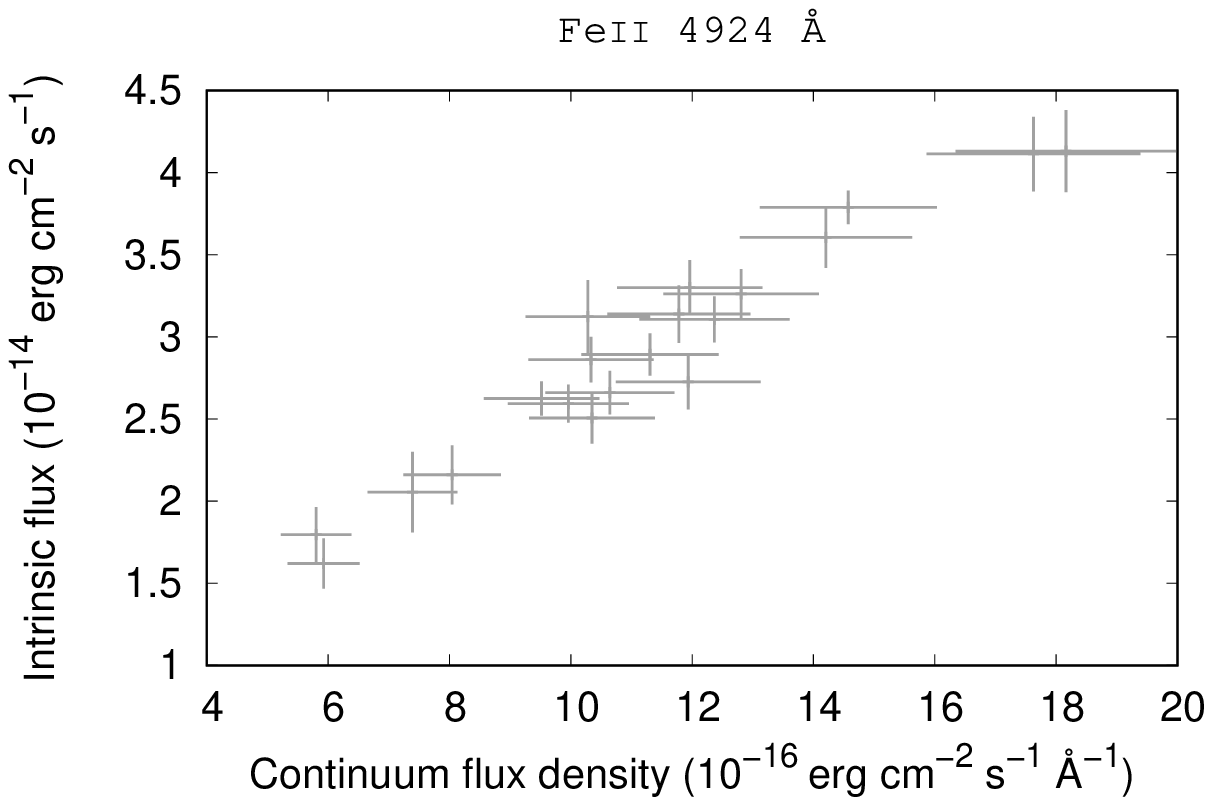}\\
\includegraphics[height=3.83cm,width=5.7cm]{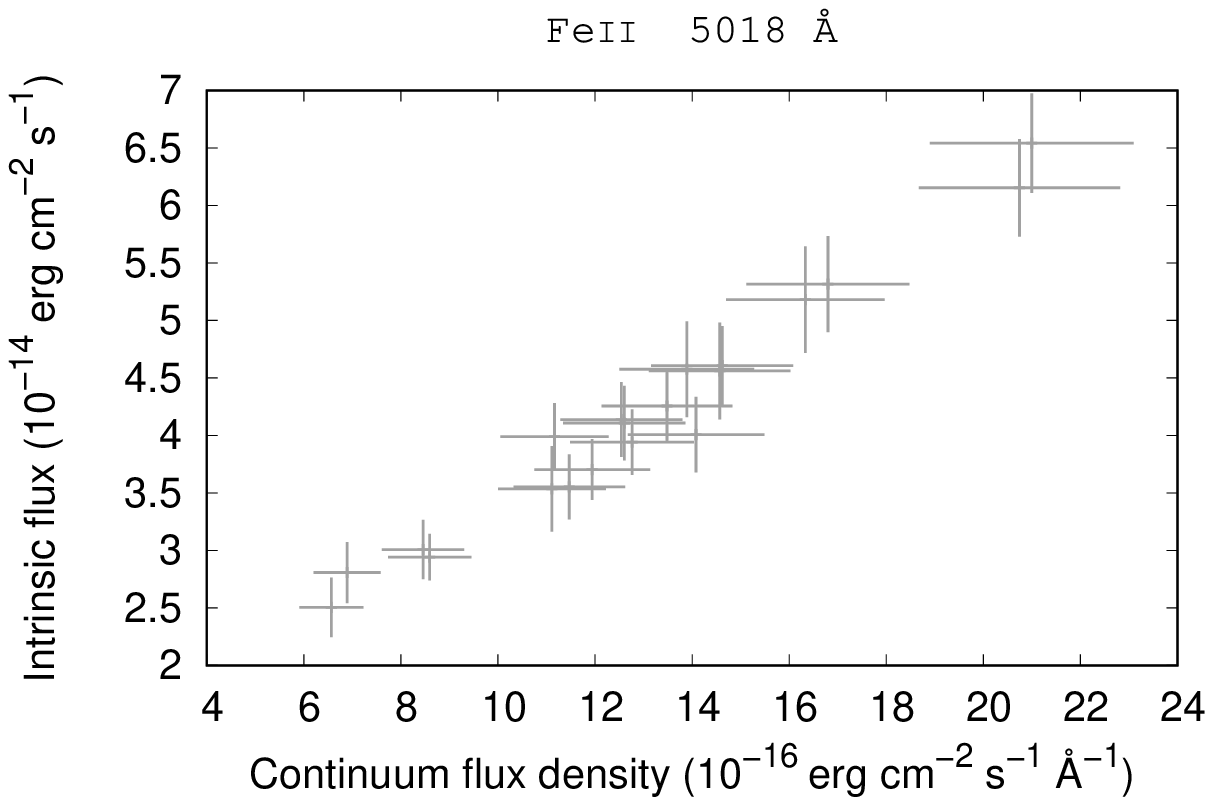}&\includegraphics[height=3.83cm,width=5.7cm]{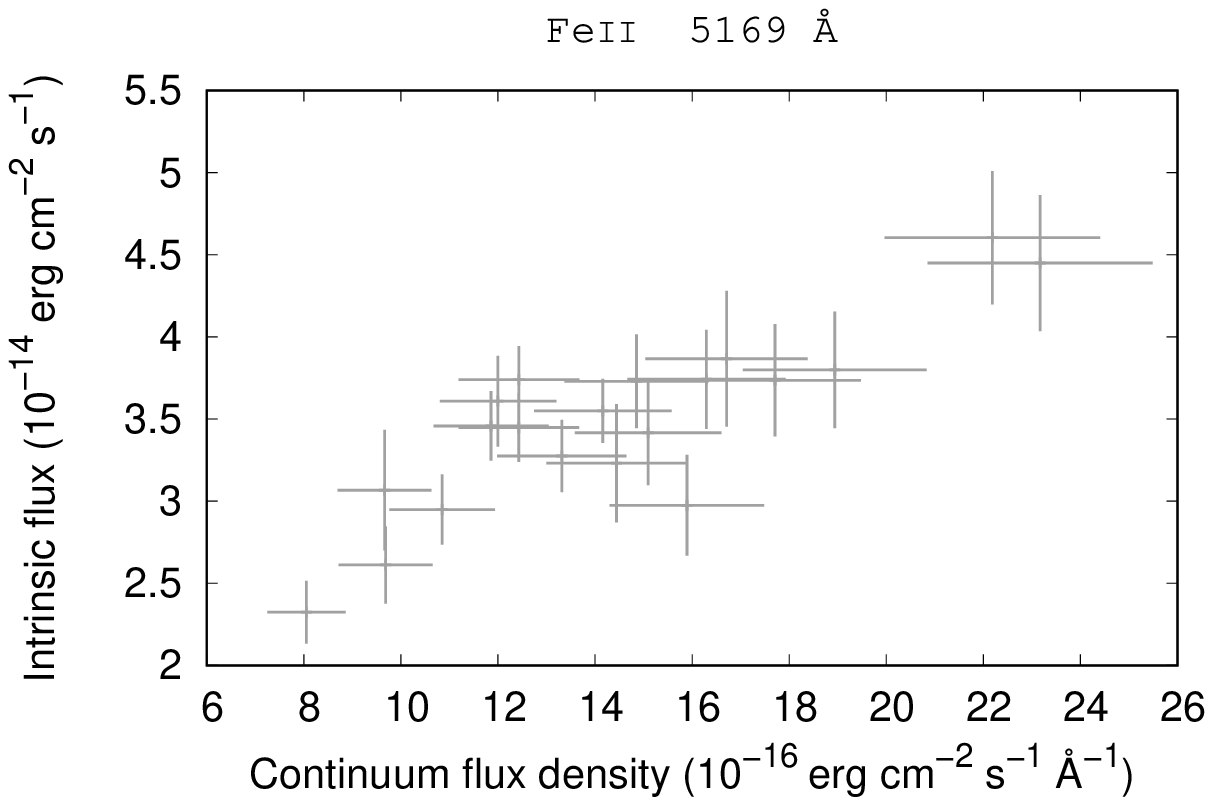}&\includegraphics[height=3.83cm,width=5.7cm]{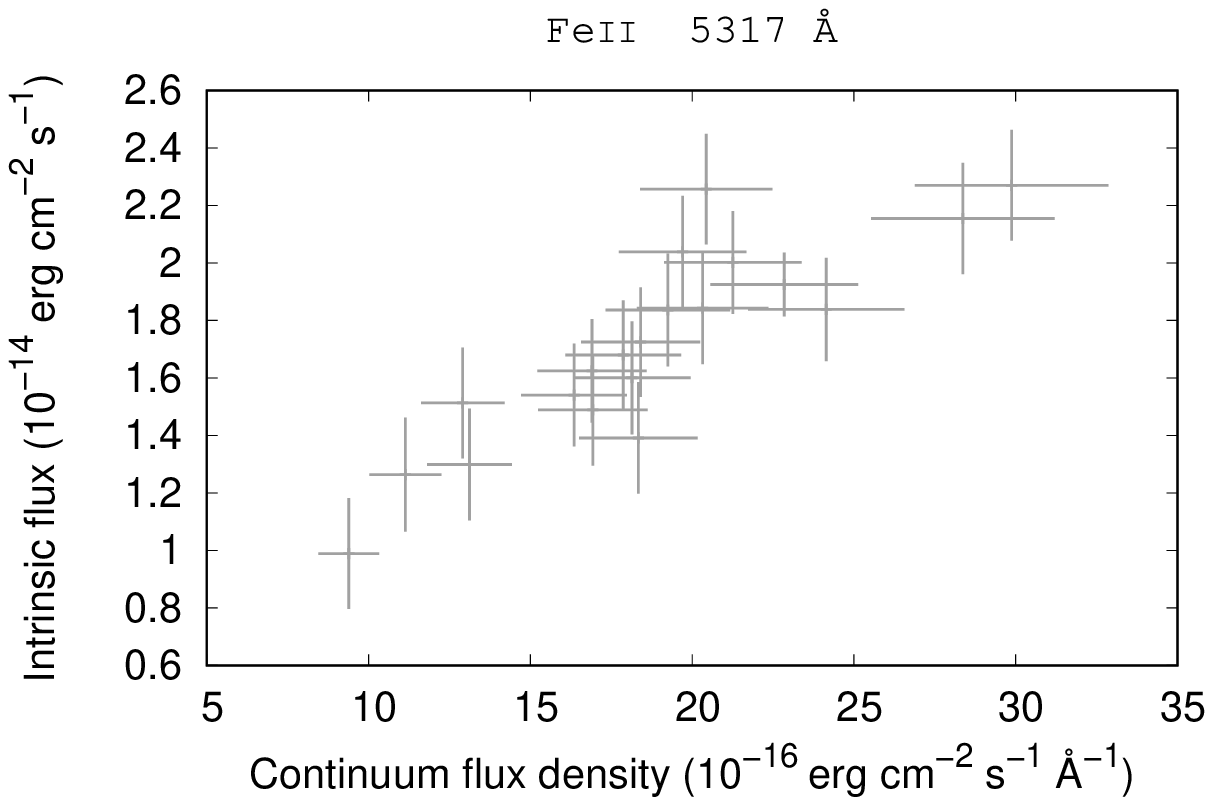}\\
\includegraphics[height=3.83cm,width=5.7cm]{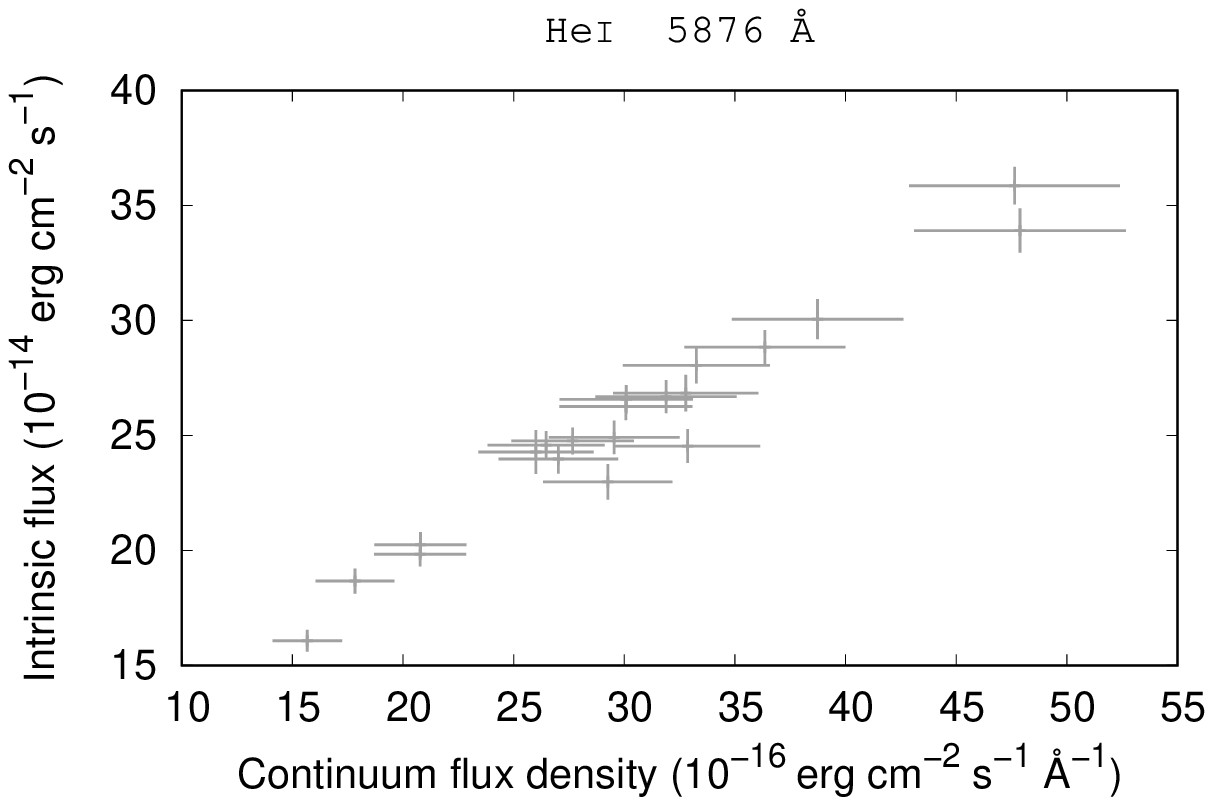}&\includegraphics[height=3.83cm,width=5.7cm]{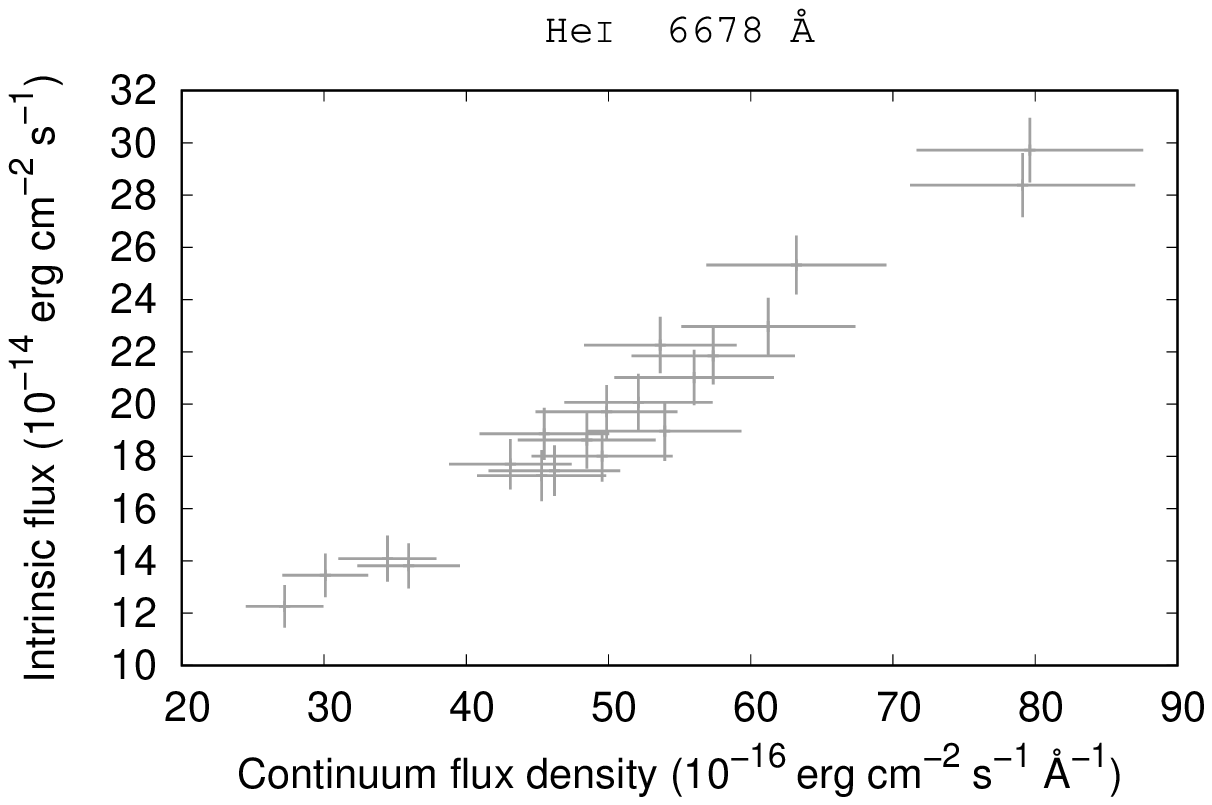}&\includegraphics[height=3.83cm,width=5.7cm]{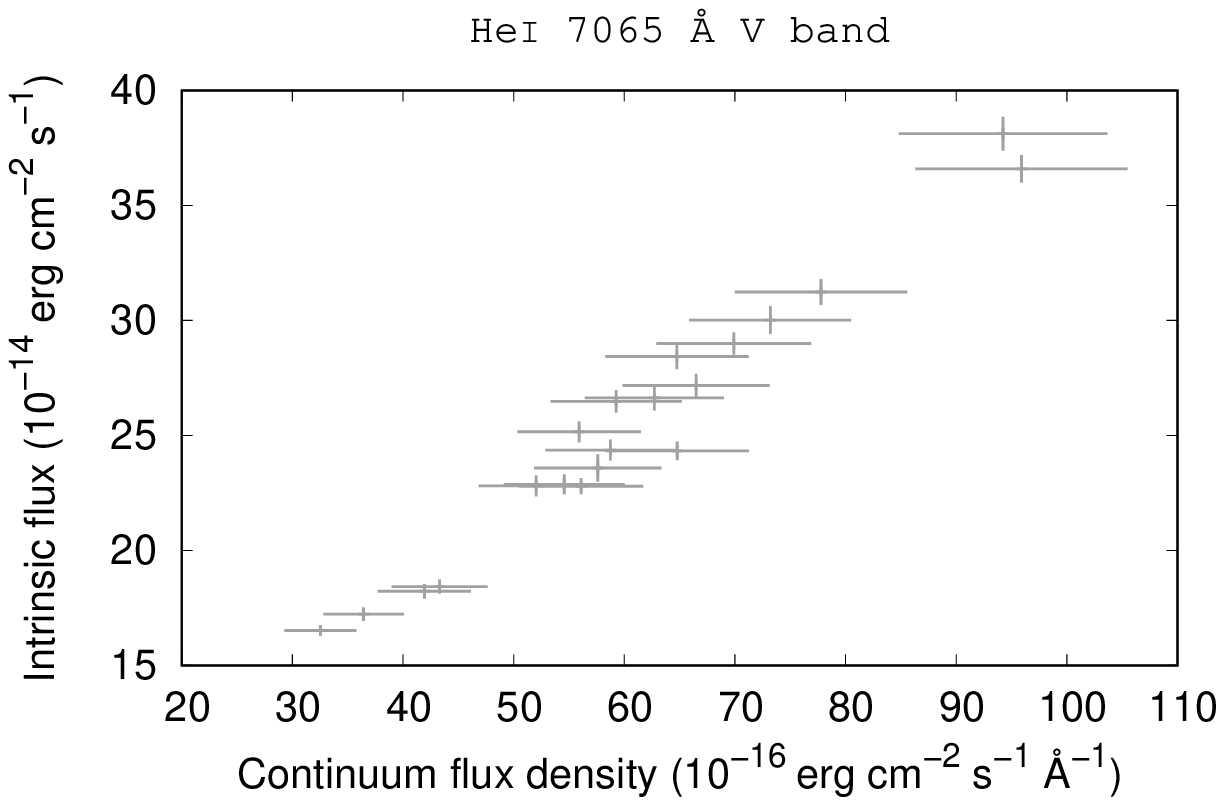}\\
\end{tabular}
\caption{Variations of the intrinsic flux of the emission lines
in our sample in function of their underlying continuum.} 
\label{var_line_flux}
\end{center}
\end{figure*}

\begin{figure*}
\begin{center}
\begin{tabular}{ccc}
\includegraphics[height=3.83cm,width=5.7cm]{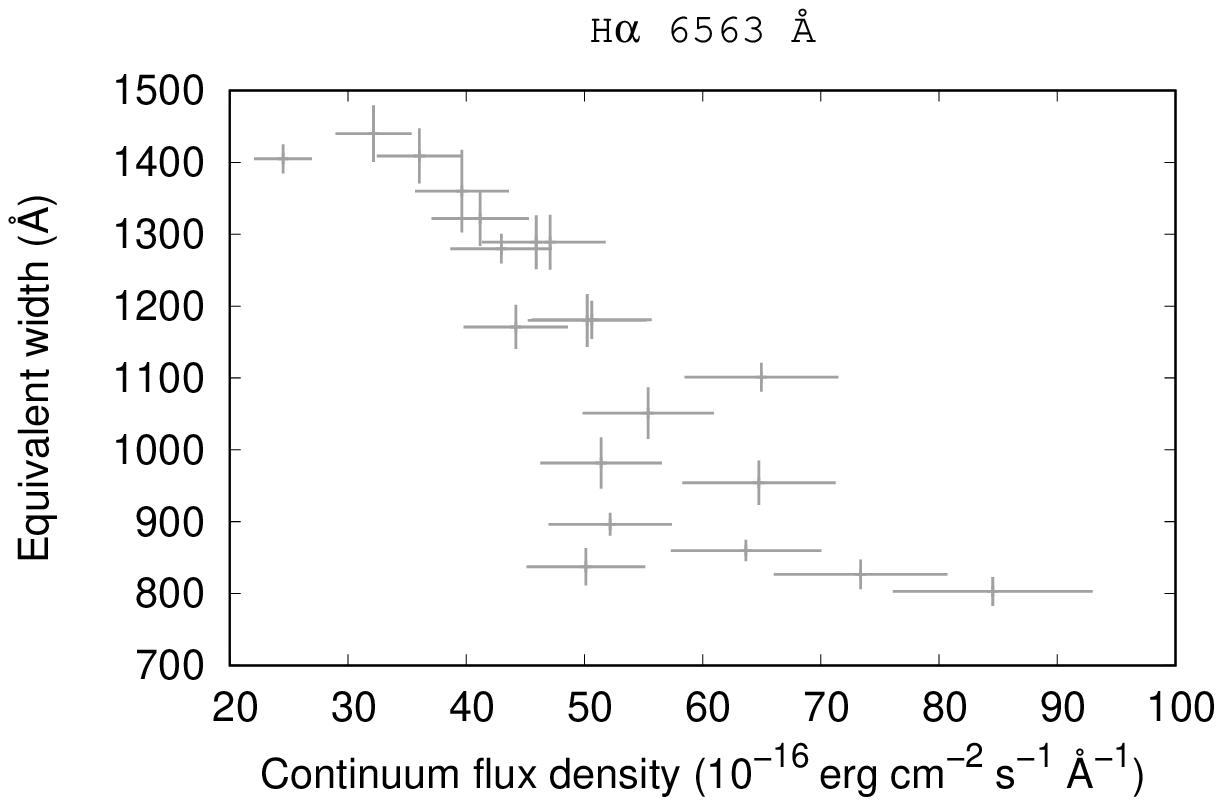}&\includegraphics[height=3.83cm,width=5.7cm]{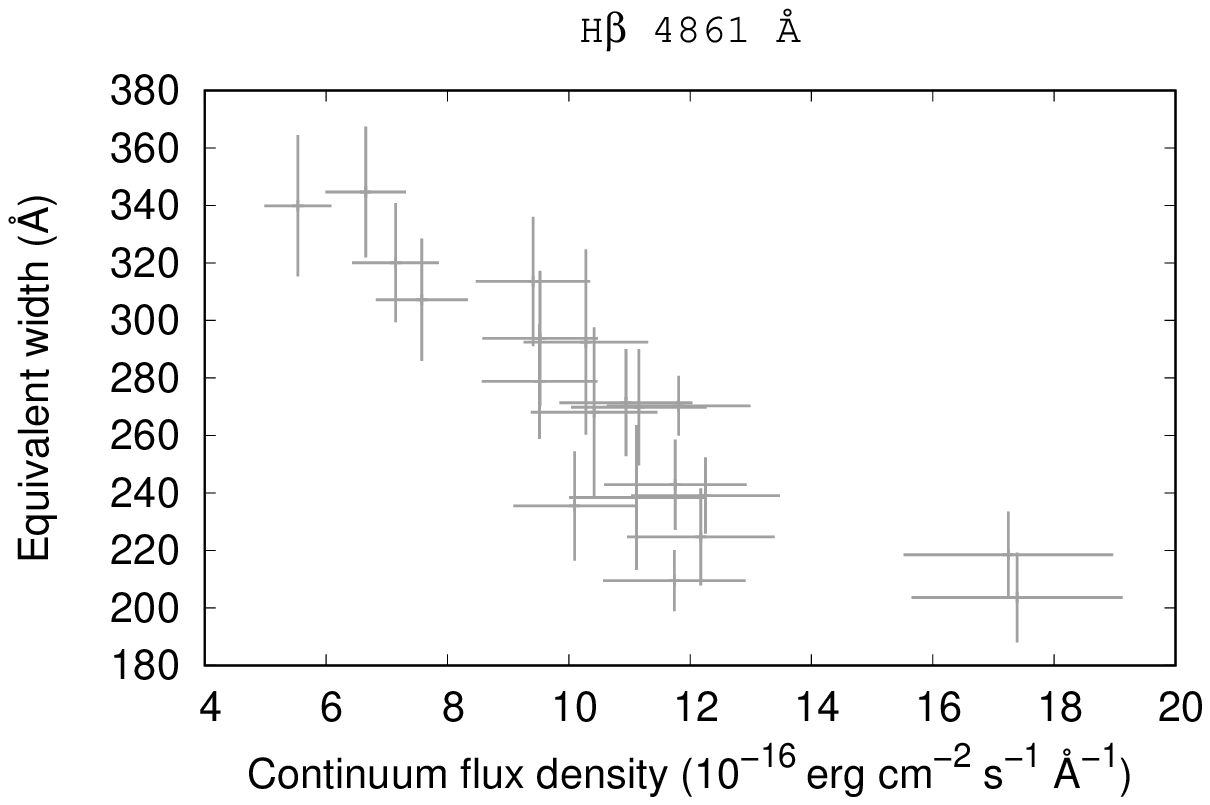}&\includegraphics[height=3.83cm,width=5.7cm]{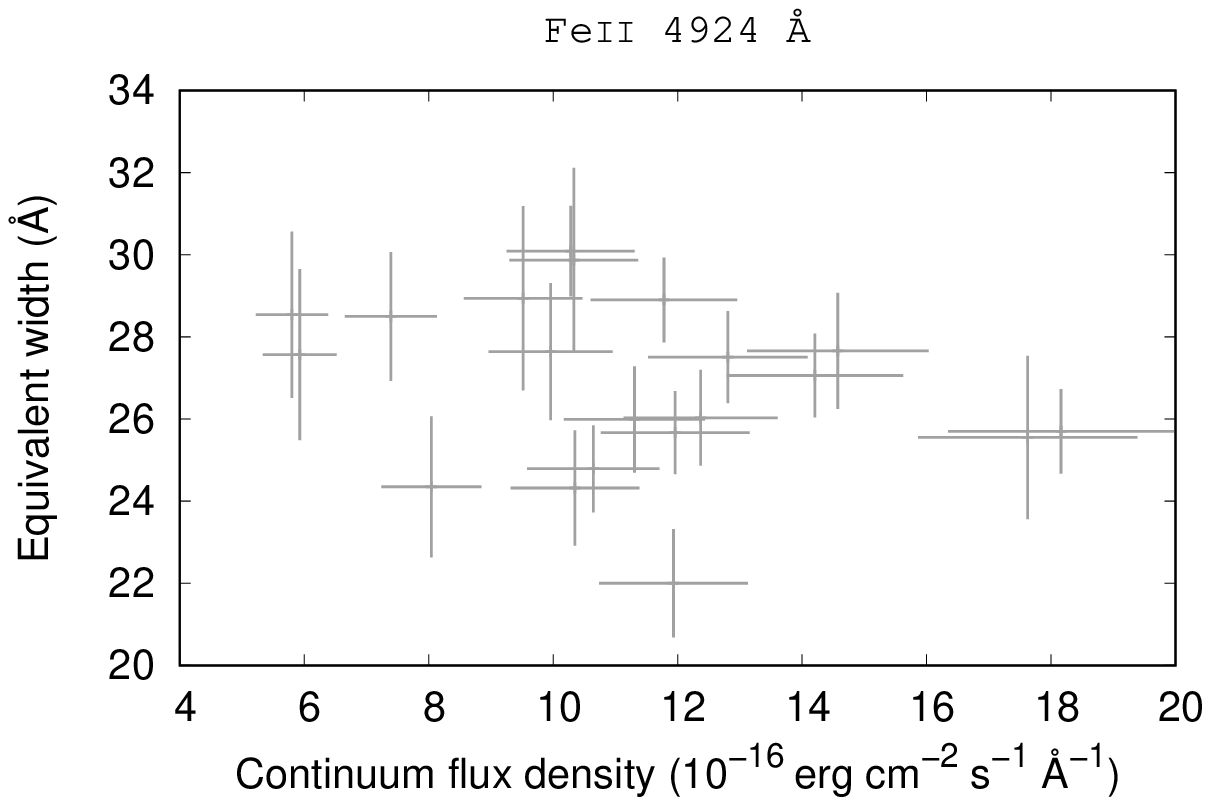}\\
\includegraphics[height=3.83cm,width=5.7cm]{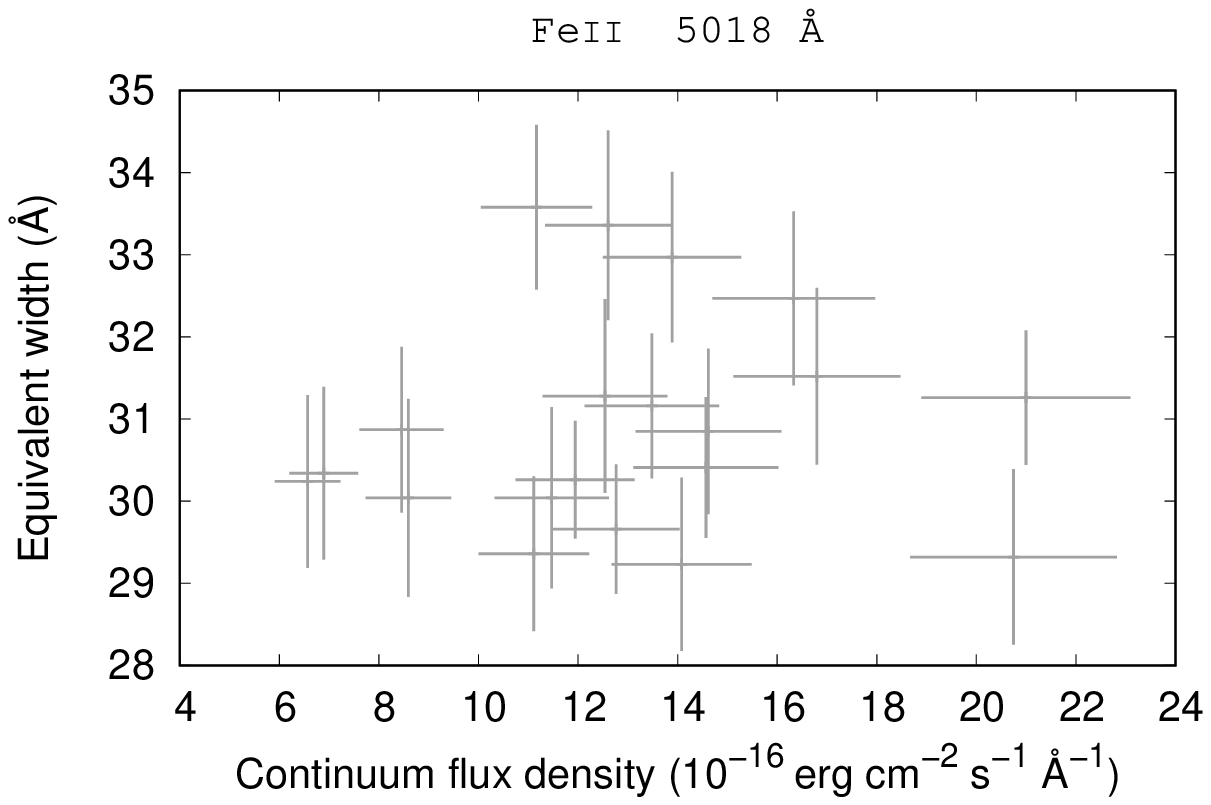}&\includegraphics[height=3.83cm,width=5.7cm]{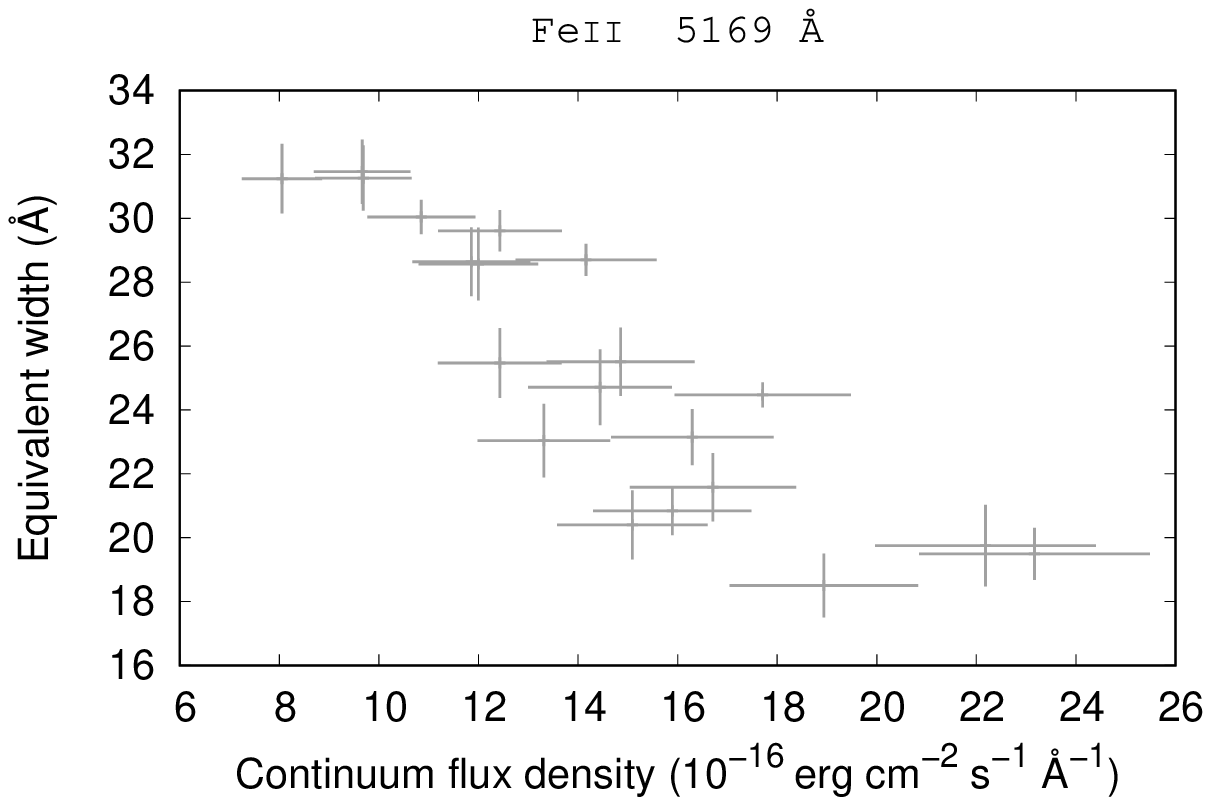}&\includegraphics[height=3.83cm,width=5.7cm]{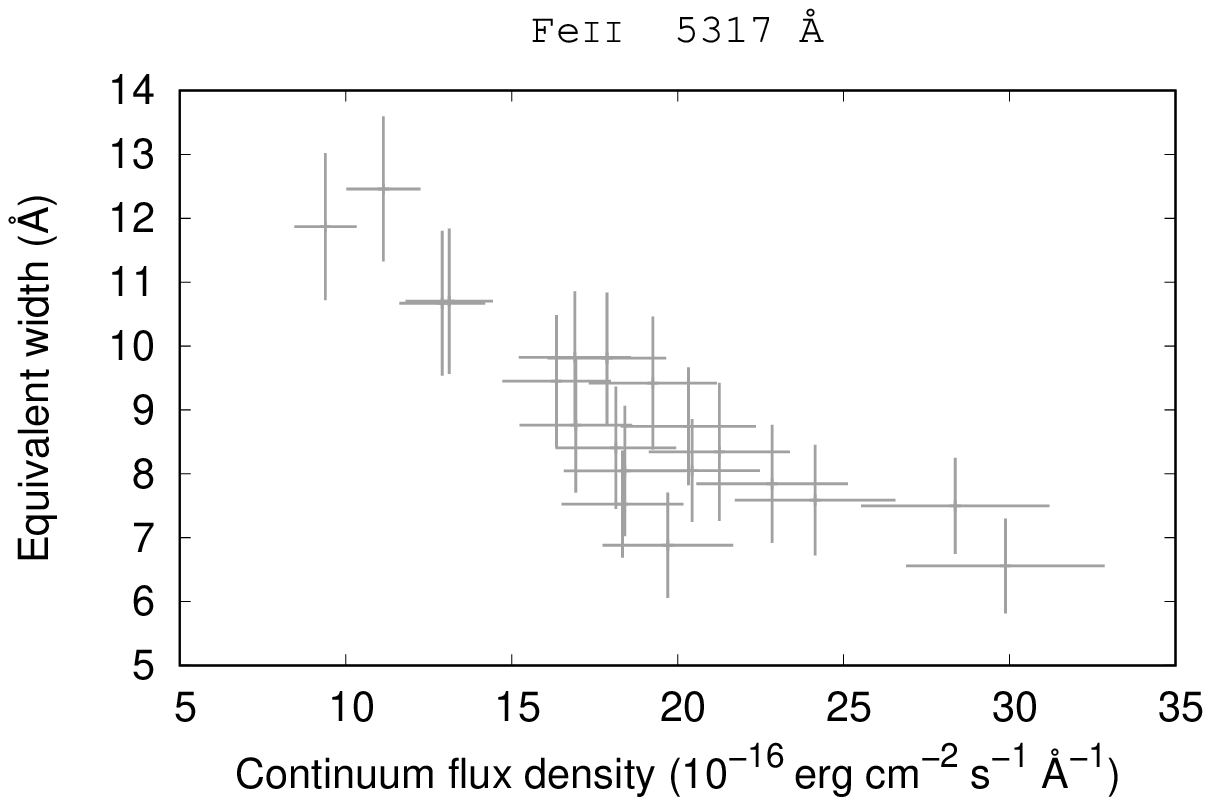}\\
\includegraphics[height=3.83cm,width=5.7cm]{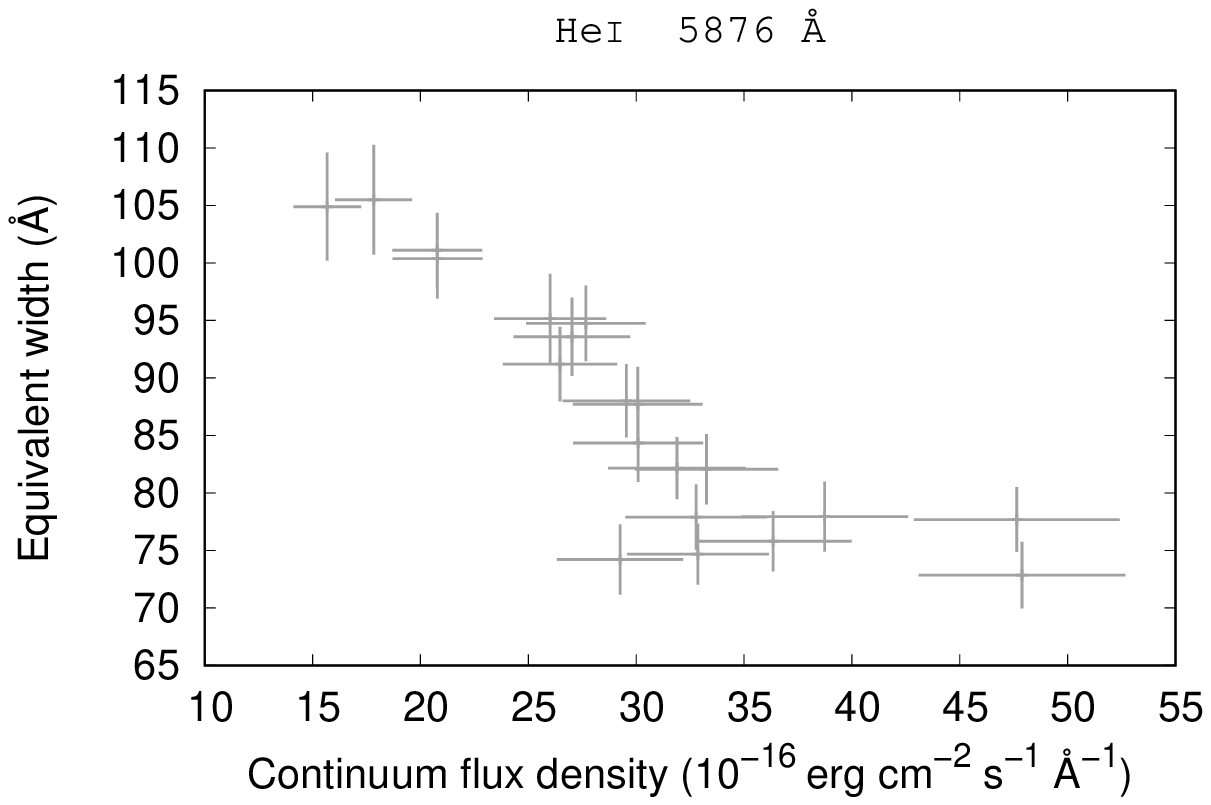}&\includegraphics[height=3.83cm,width=5.7cm]{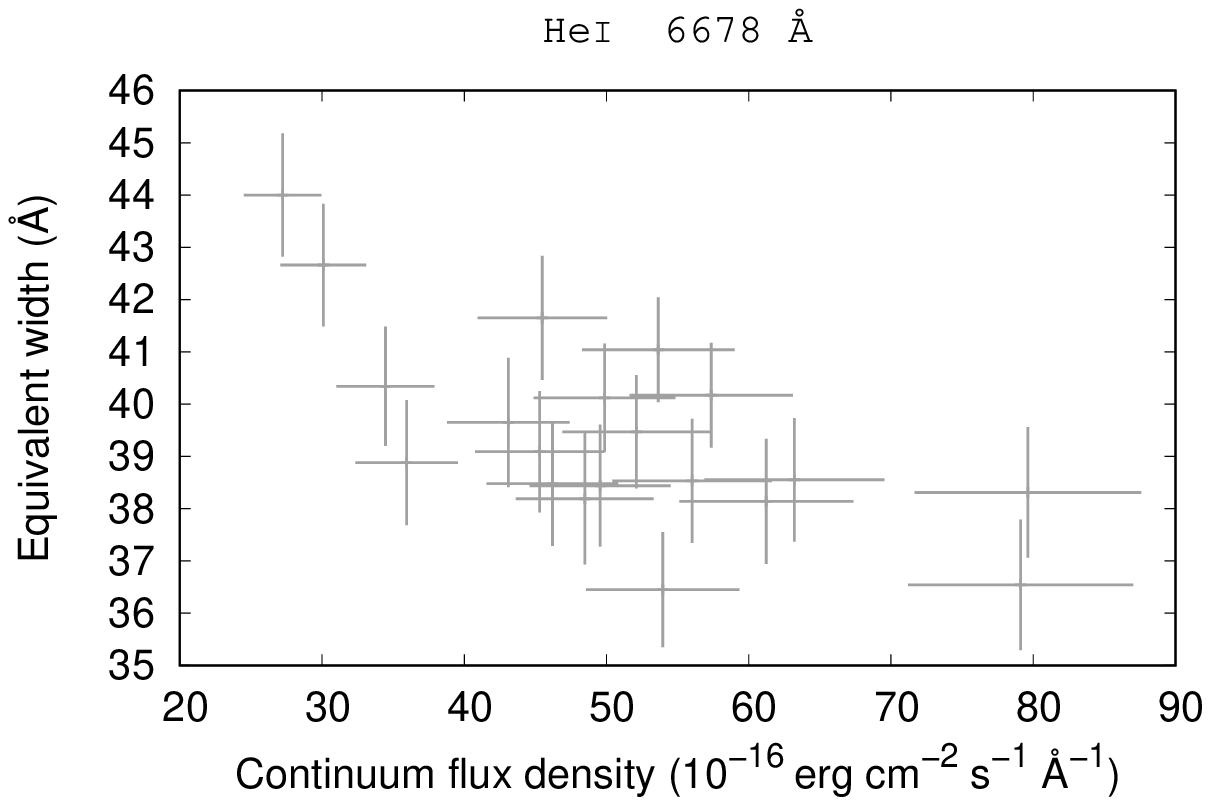}&\includegraphics[height=3.83cm,width=5.7cm]{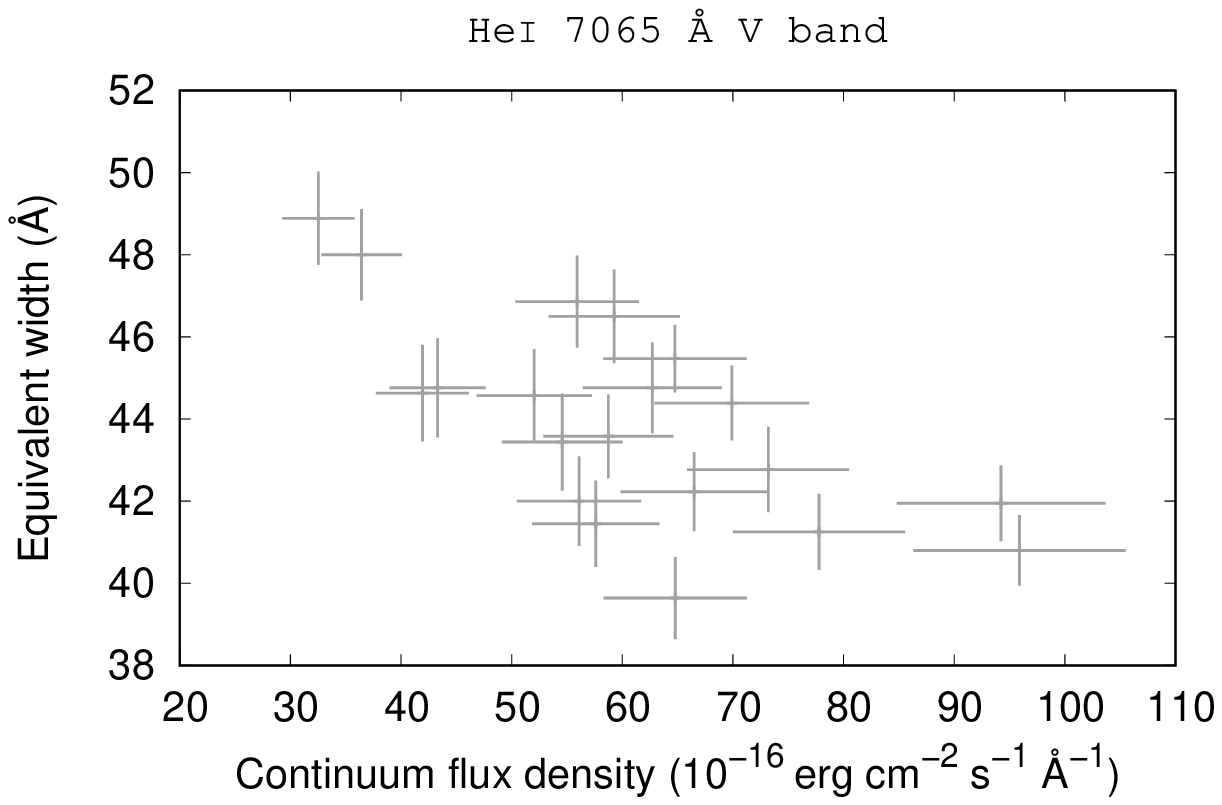}\\
\includegraphics[height=3.83cm,width=5.7cm]{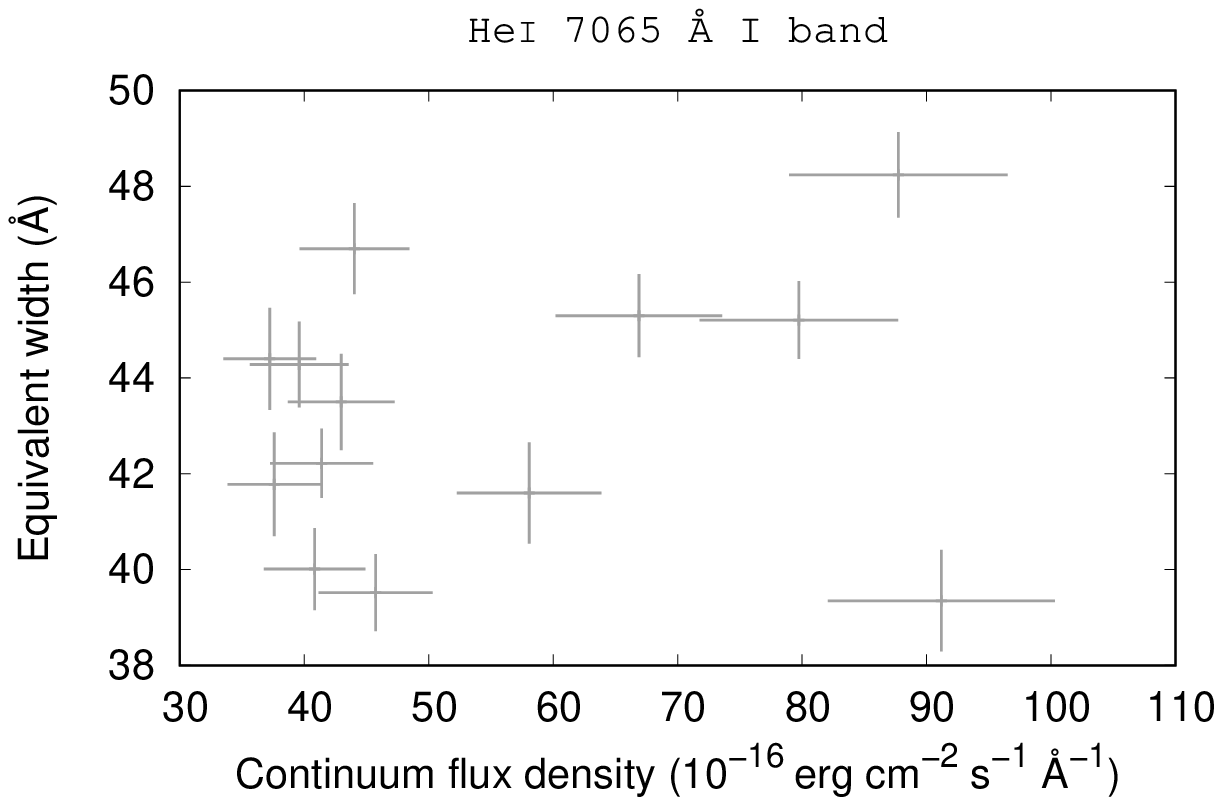}&\includegraphics[height=3.83cm,width=5.7cm]{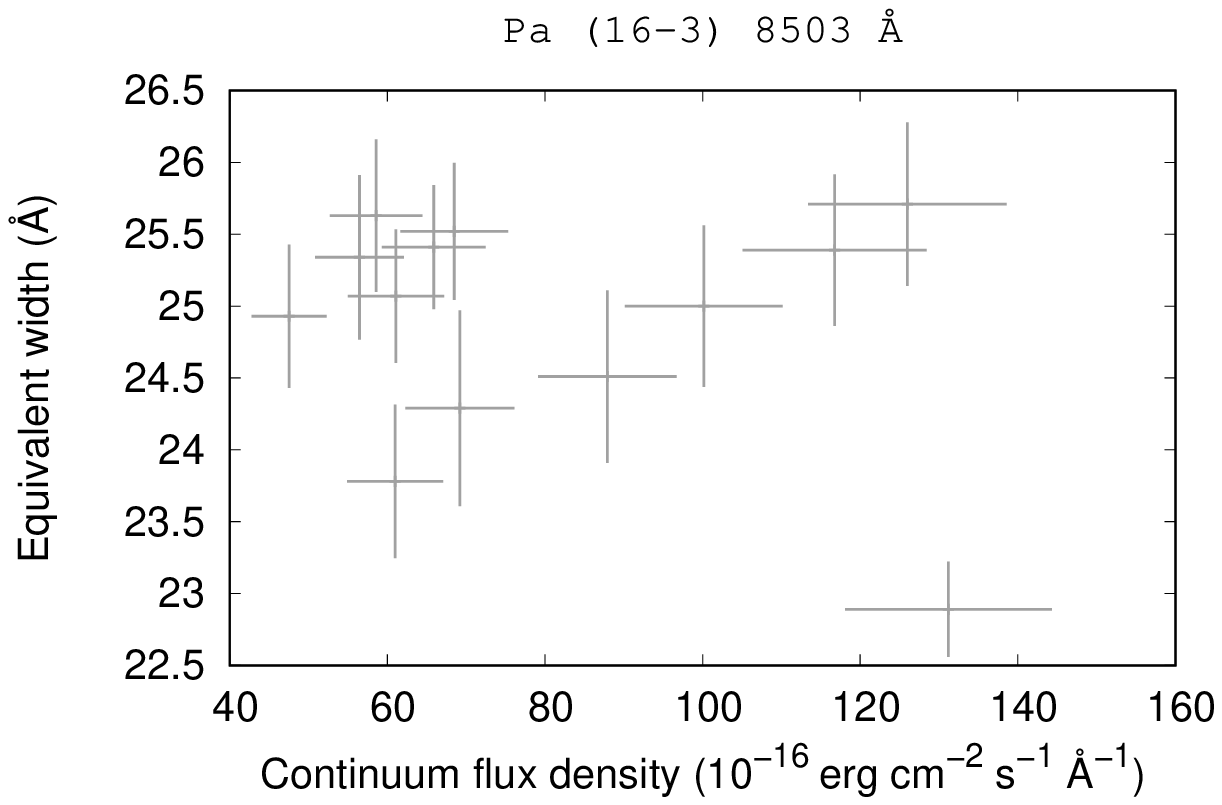}&\includegraphics[height=3.83cm,width=5.7cm]{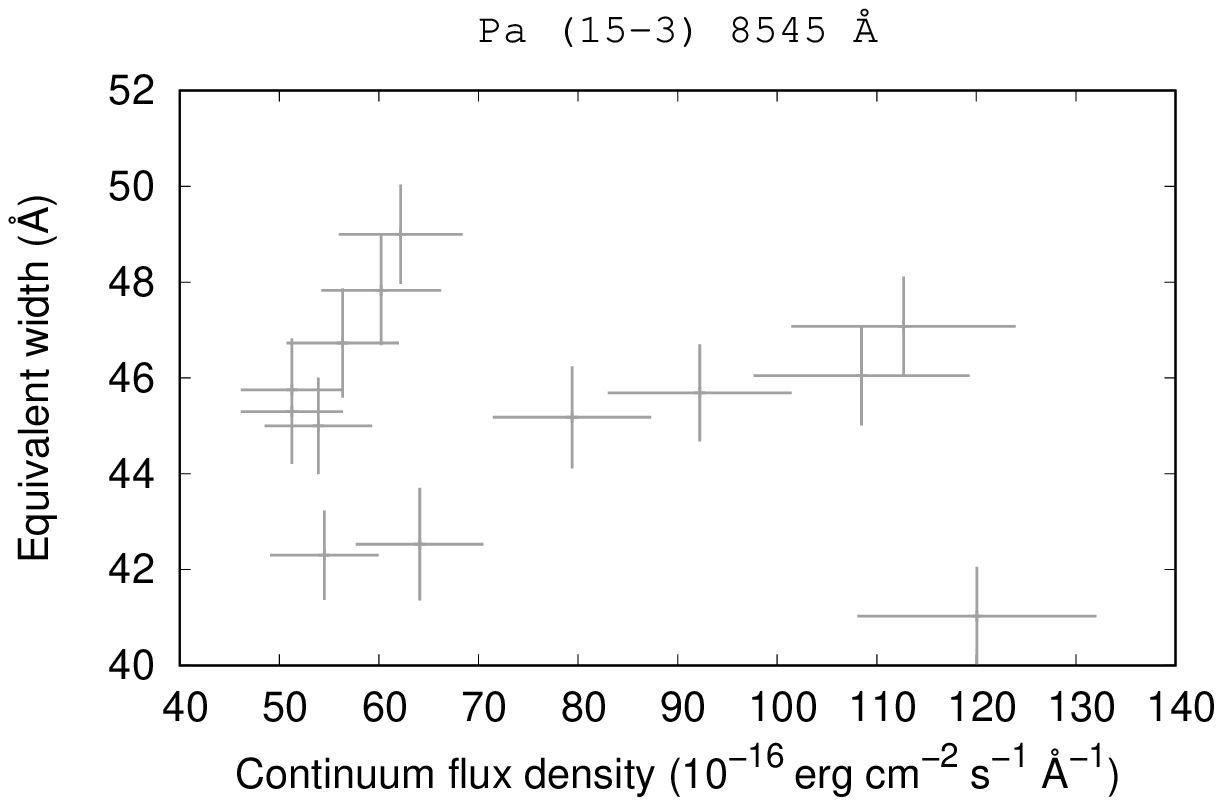}\\
\includegraphics[height=3.83cm,width=5.7cm]{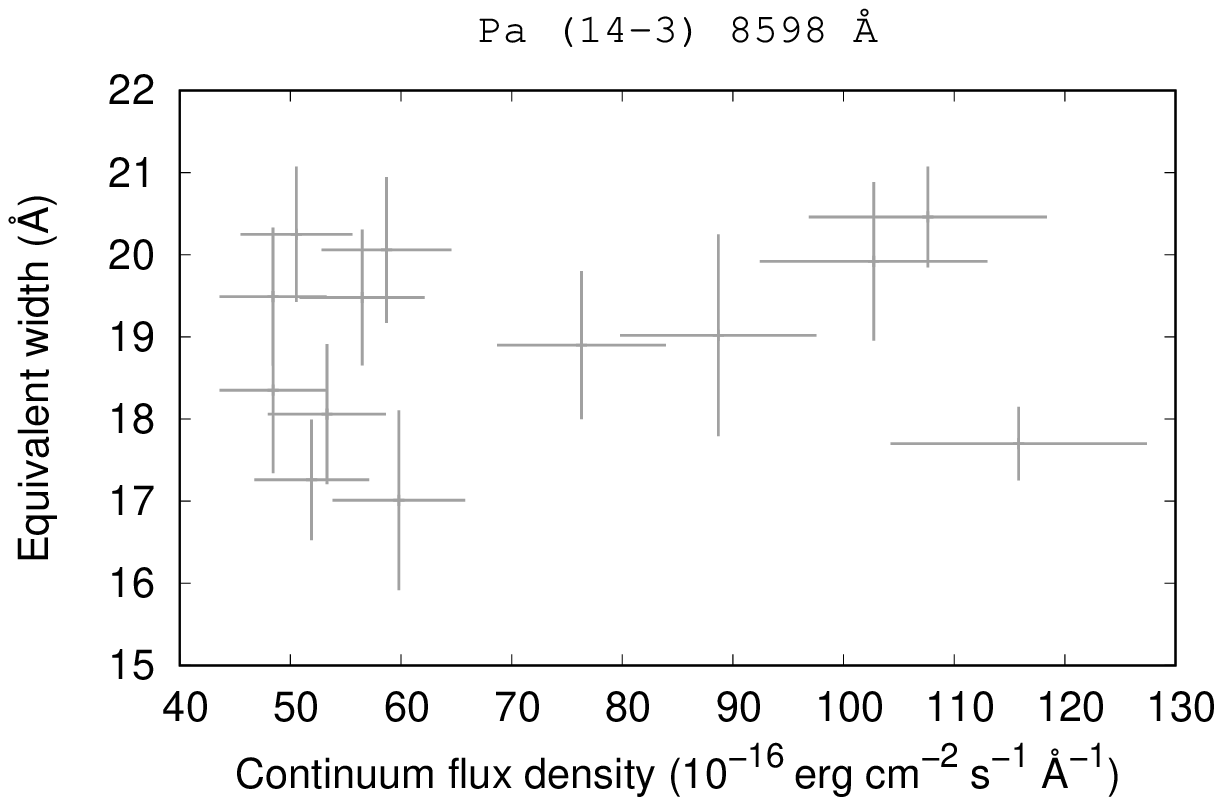}&\includegraphics[height=3.83cm,width=5.7cm]{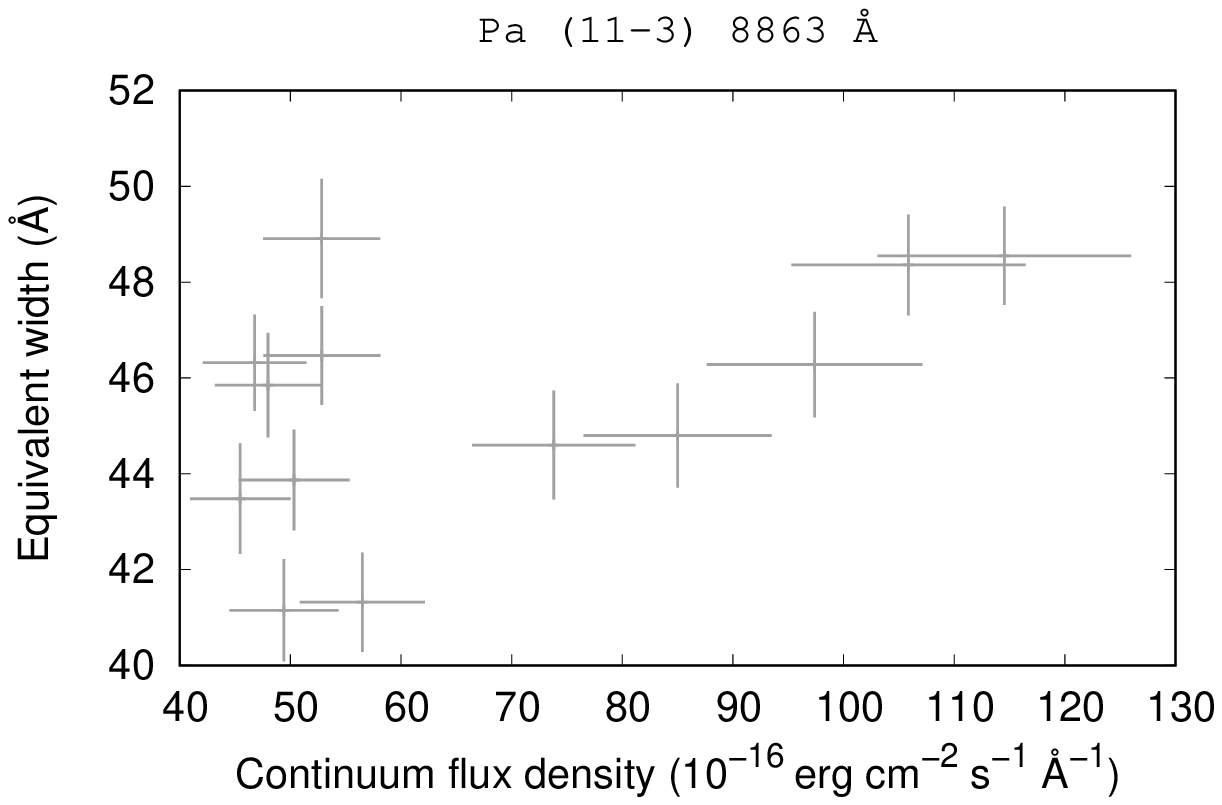}&\includegraphics[height=3.83cm,width=5.7cm]{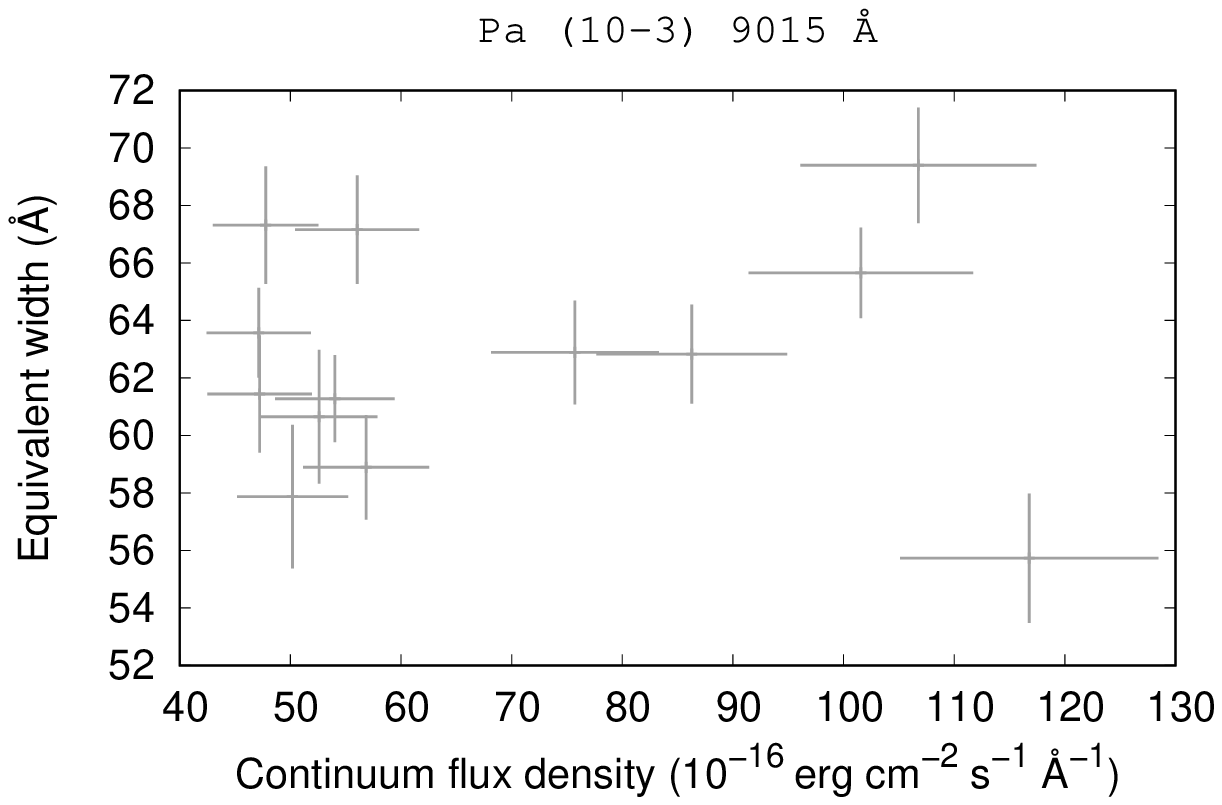}\\
\end{tabular}
\caption{Variations of the equivalent width of the emission lines
in our sample in function of their underlying continuum.} 
\label{var_line_eqw}
\end{center}
\end{figure*}

\begin{figure*}
\begin{center}
\begin{tabular}{ccc}
\includegraphics[height=3.83cm,width=5.7cm]{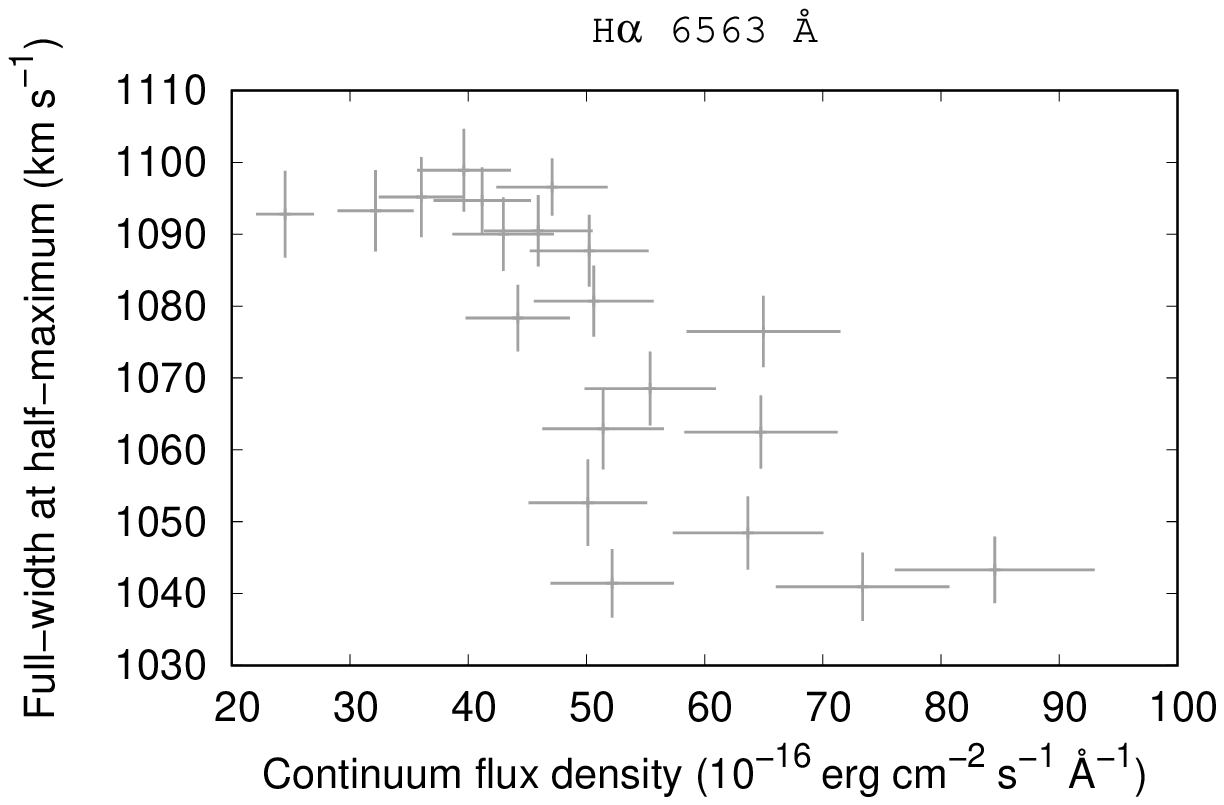}&\includegraphics[height=3.83cm,width=5.7cm]{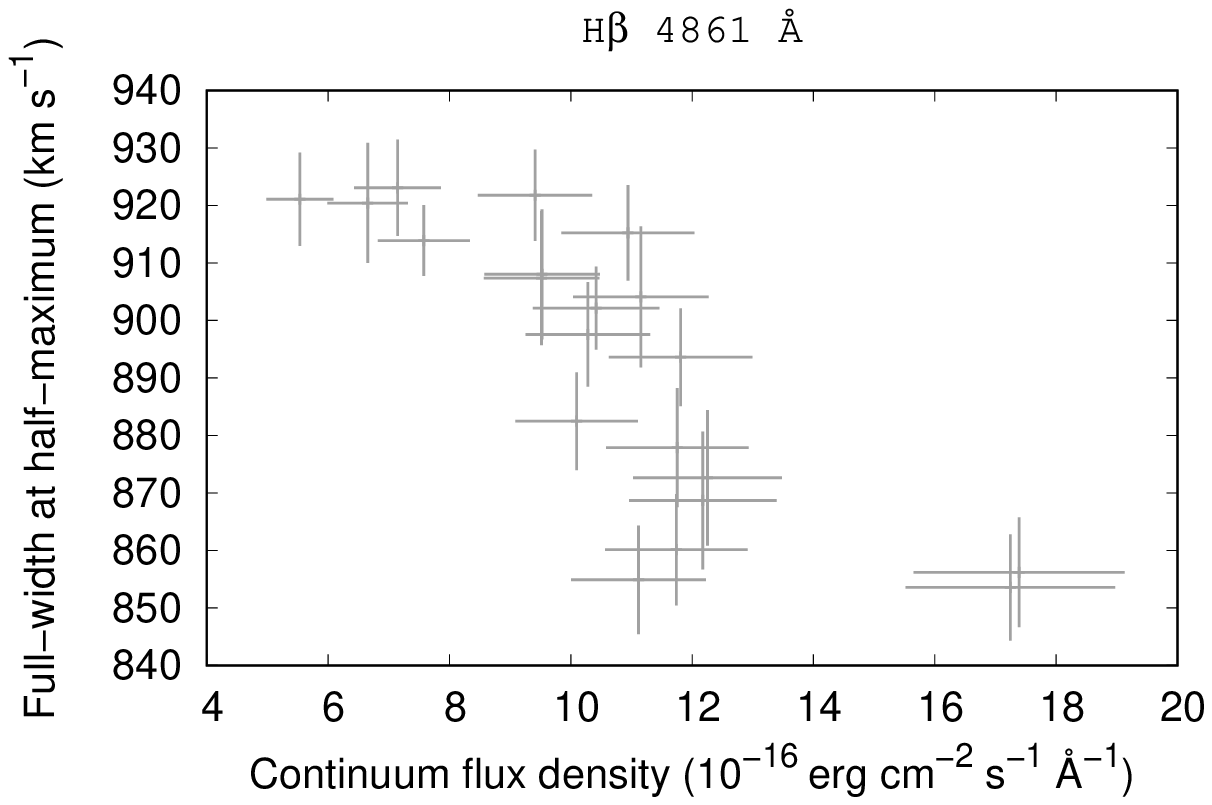}&\includegraphics[height=3.83cm,width=5.7cm]{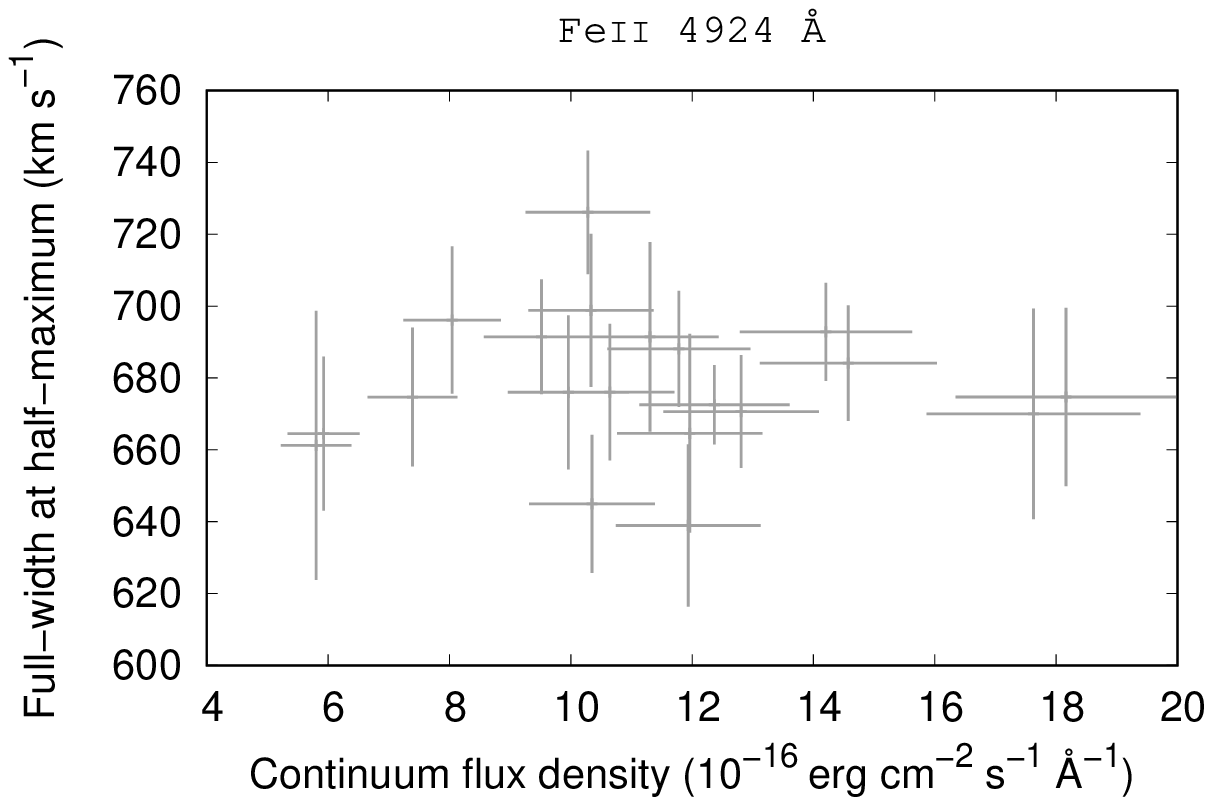}\\
\includegraphics[height=3.83cm,width=5.7cm]{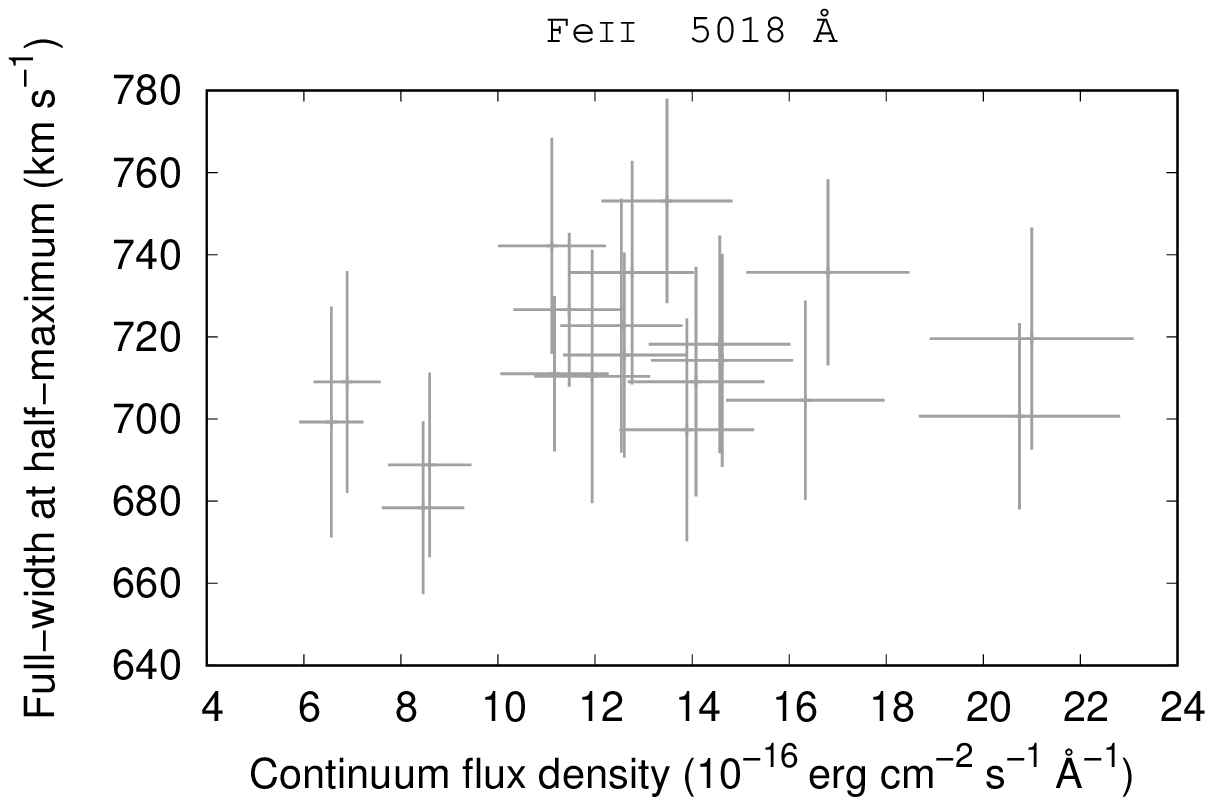}&\includegraphics[height=3.83cm,width=5.7cm]{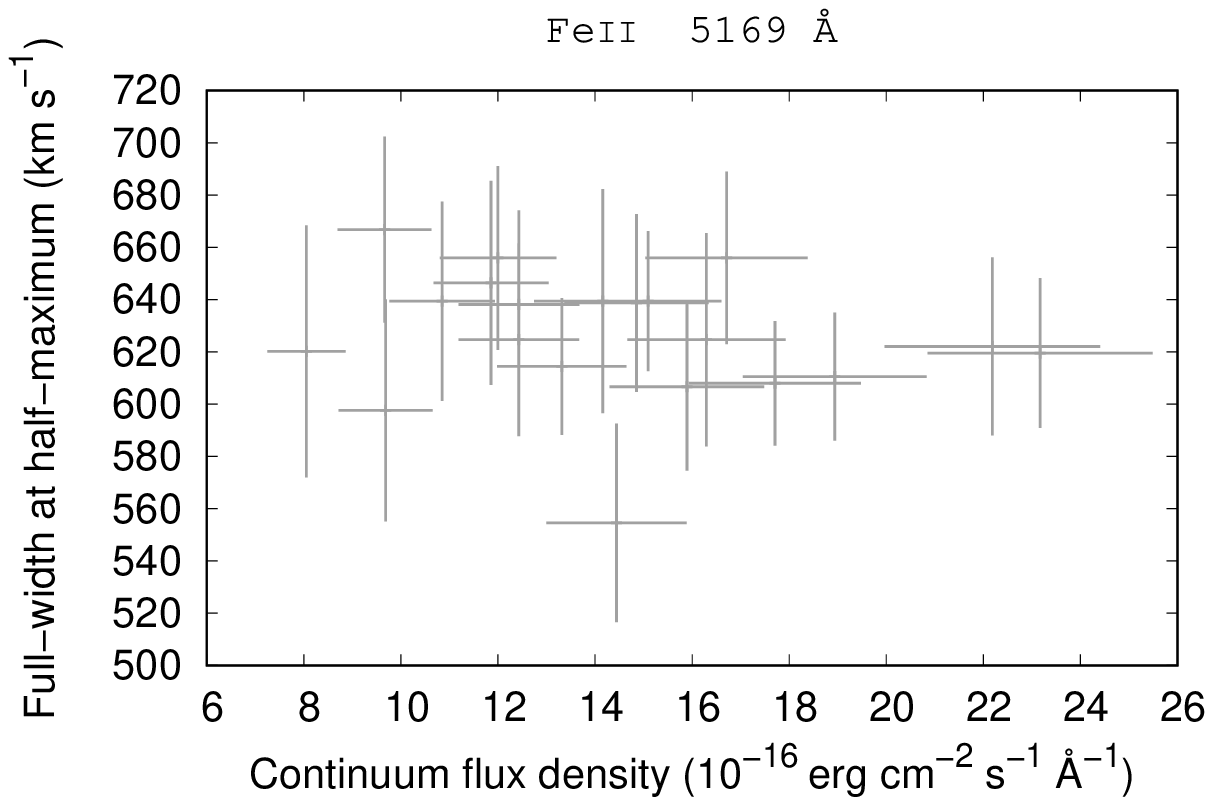}&\includegraphics[height=3.83cm,width=5.7cm]{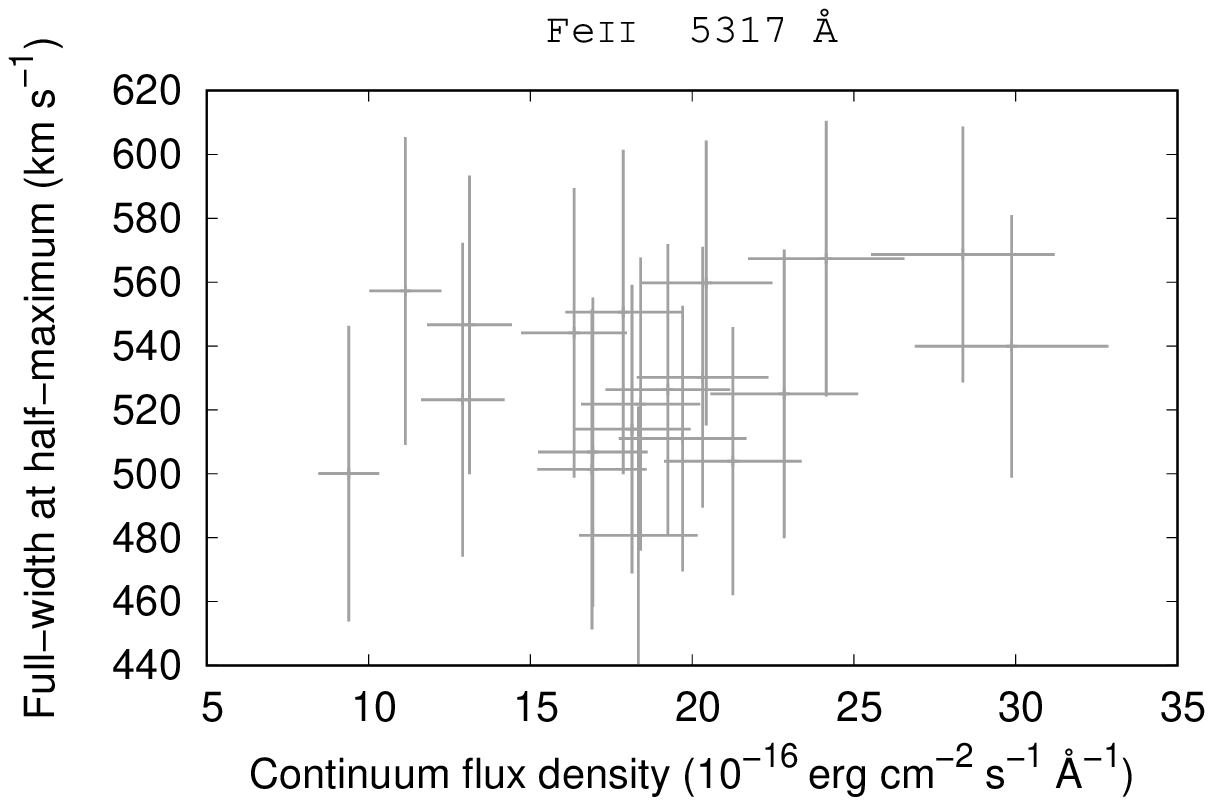}\\
\includegraphics[height=3.83cm,width=5.7cm]{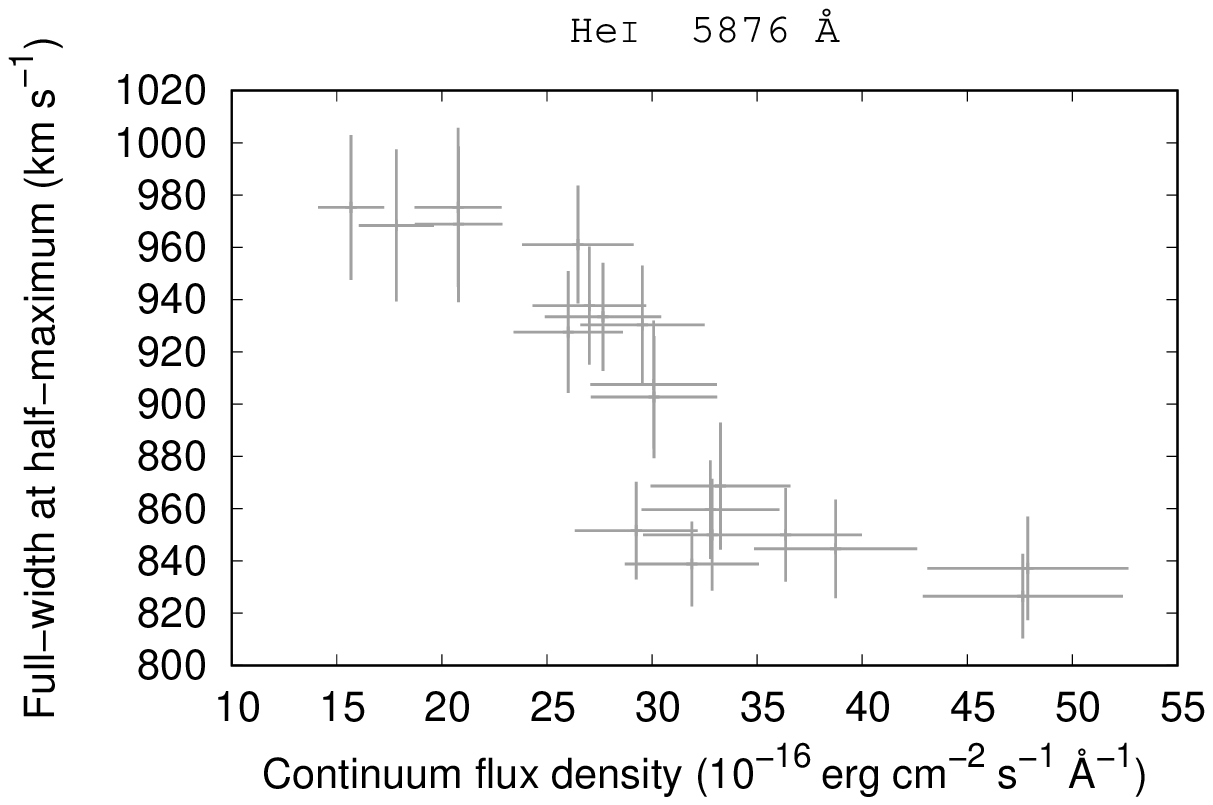}&\includegraphics[height=3.83cm,width=5.7cm]{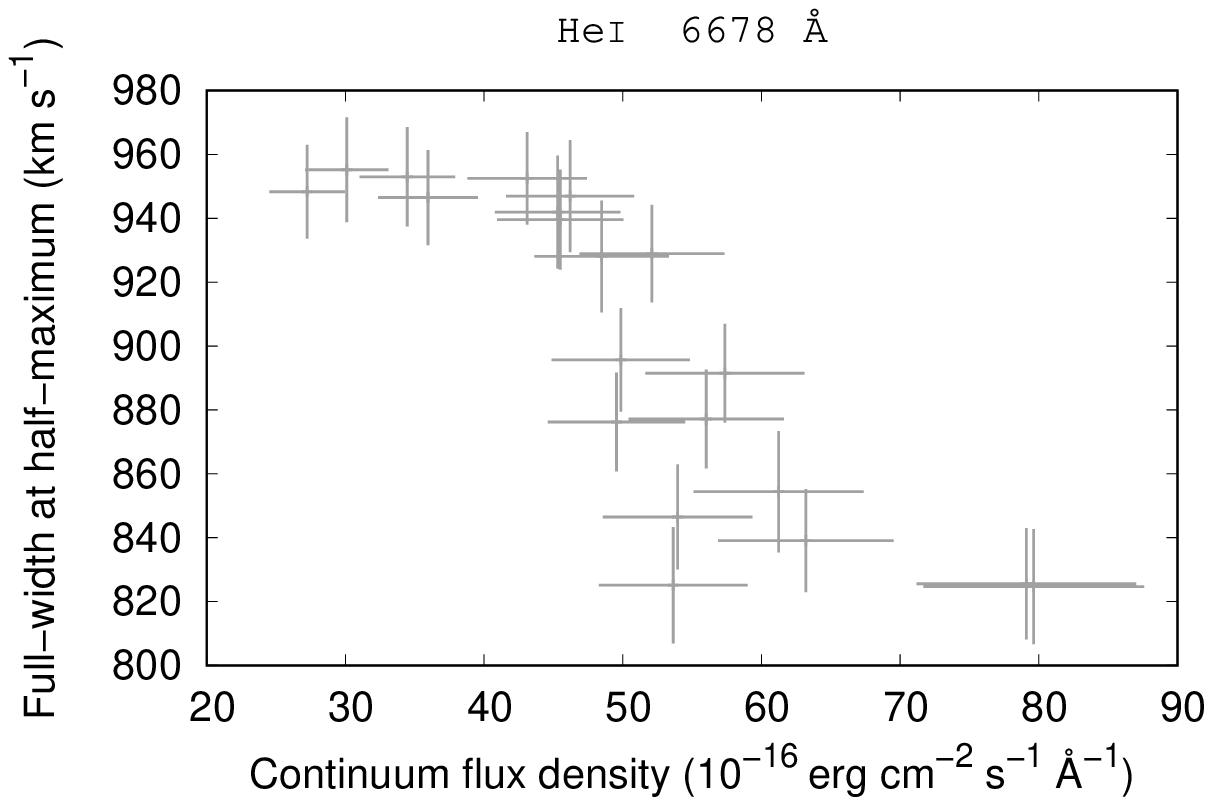}&\includegraphics[height=3.83cm,width=5.7cm]{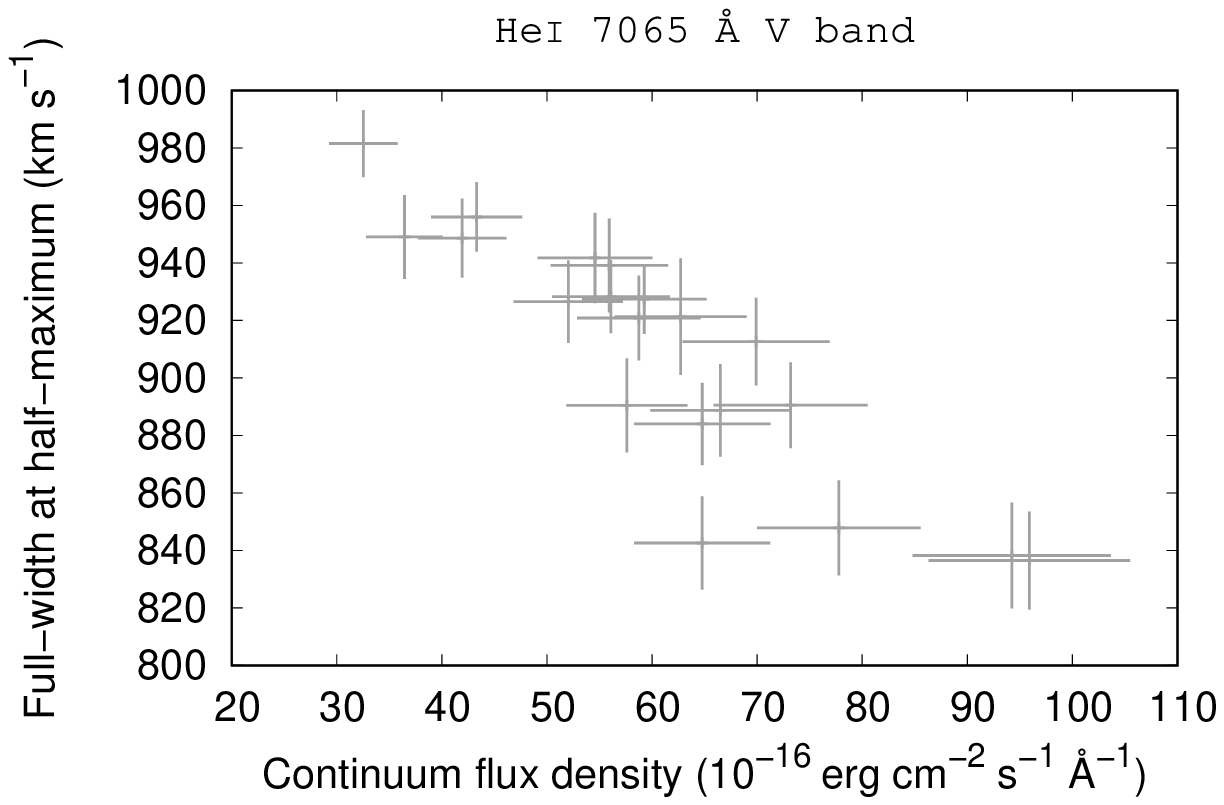}\\
\includegraphics[height=3.83cm,width=5.7cm]{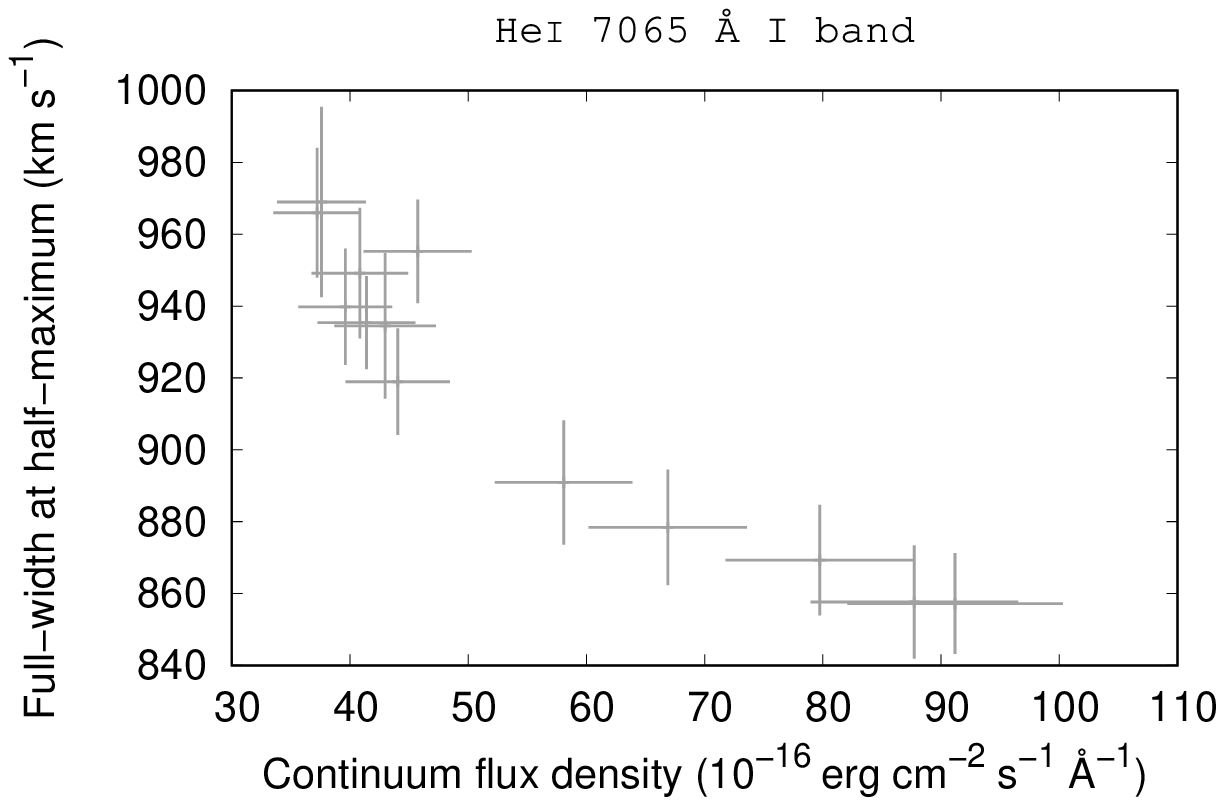}&\includegraphics[height=3.83cm,width=5.7cm]{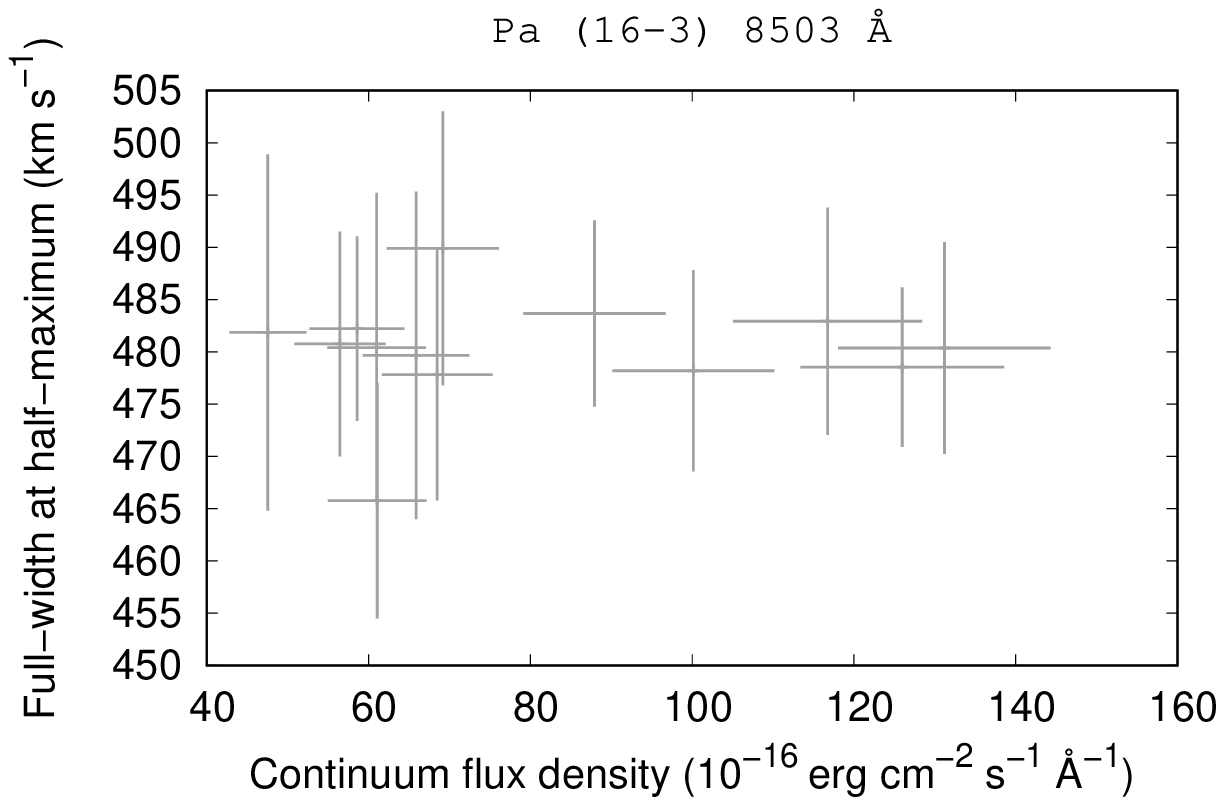}&\includegraphics[height=3.83cm,width=5.7cm]{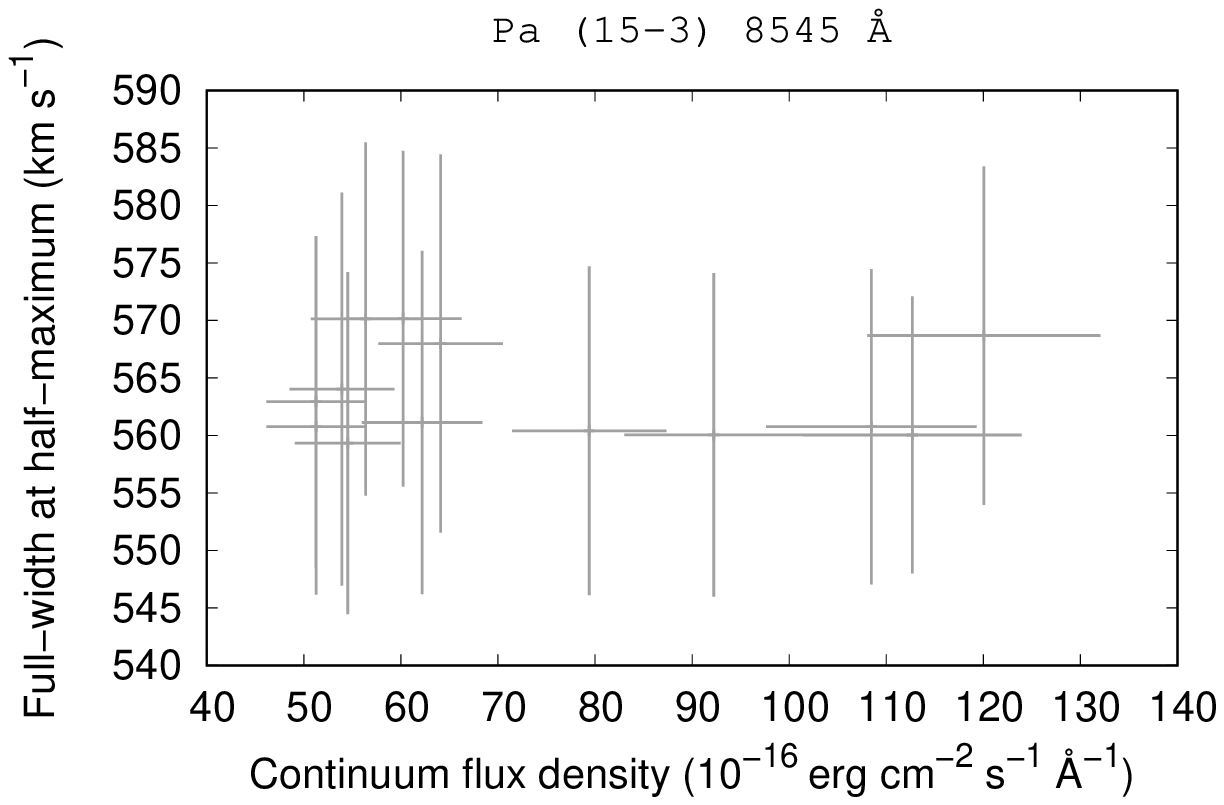}\\
\includegraphics[height=3.83cm,width=5.7cm]{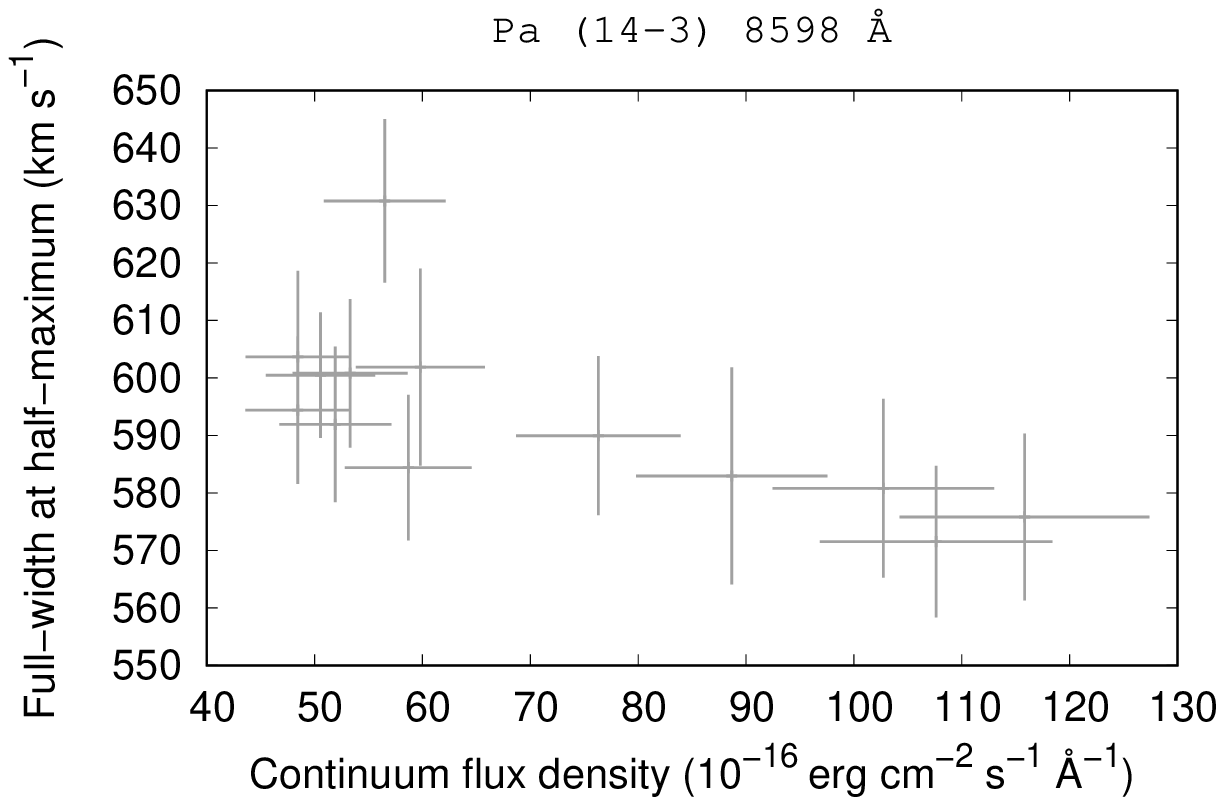}&\includegraphics[height=3.83cm,width=5.7cm]{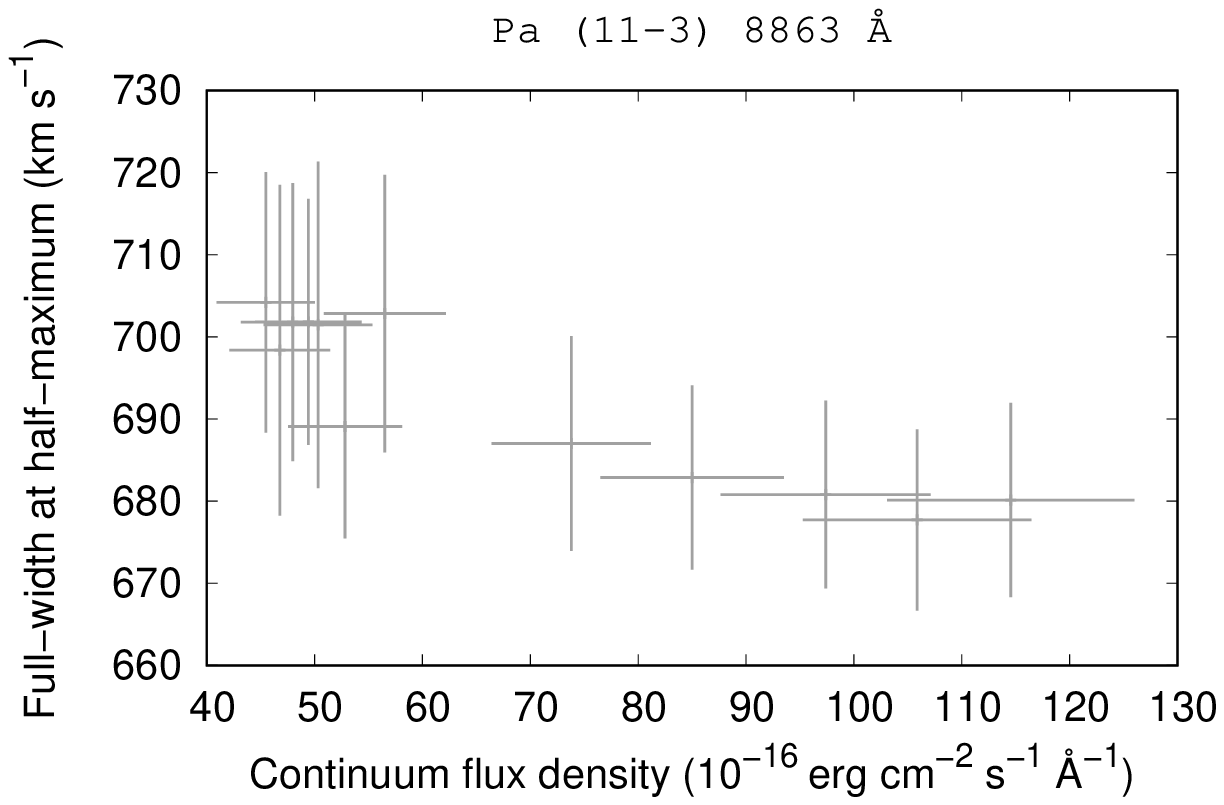}&\includegraphics[height=3.83cm,width=5.7cm]{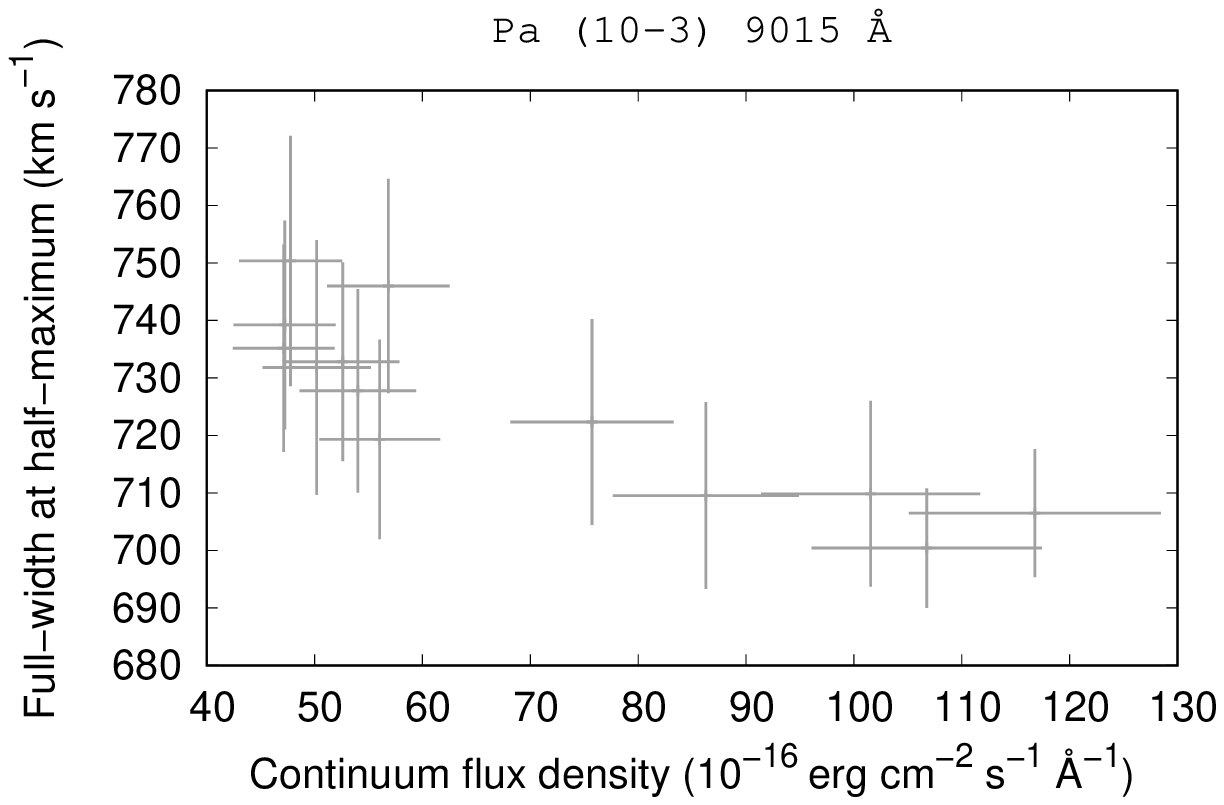}\\
\end{tabular}
\caption{Variations of the FWHM of the emission lines
in our sample in function of their underlying continuum.} 
\label{var_line_fwhm}
\end{center}
\end{figure*}

We thus selected the brightest \ion{H}{i} and \ion{He}{i} as well as four \ion{Fe}{ii} features and the way we measured the line parameters differs from that used for the average spectrum. Indeed, to ensure that the choice of continuum had the smallest effect as possible, a given line was fitted, in all the sub-spectra, within the exact same window with a single Gaussian and its base continuum was extrapolated with a first-order polynomial from the exact same wavelengths. Likewise, uncertainties were measured from the real errors as given after the extraction process. The first result is that none of the lines, including the red-shifted ones, exhibit detectable centroid variations. Second, there is an obvious correlation between the intrinsic fluxes and local continua (\autoref{var_line_flux}), which clearly hints at a common origin for the continuum and line variability. While this could also be interpreted as an evidence that all the features come from the same component, the evolution of their respective equivalent widths hints at two different behaviours (\autoref{var_line_eqw}). Indeed, those of \ha, \hb, \ion{He}{i}~$\lambda5876$, as well as \ion{Fe}{ii}~$\lambda5169$ and $\lambda5317$ are unambiguously anti-correlated with their underlying continua, whereas the others show little changes. Likewise, the FWHMs of the first three lines plus those of \ion{He}{i}~$\lambda6678$ and \ion{He}{i}~$\lambda7065$ (in both $V$ and $I$), show a similar anti-correlation (\autoref{var_line_fwhm}), whereas the four \ion{Fe}{ii} lines have constant FWHMs.

\begin{figure*}
\begin{center}
\begin{tabular}{ccc}
\includegraphics[width=6.3cm]{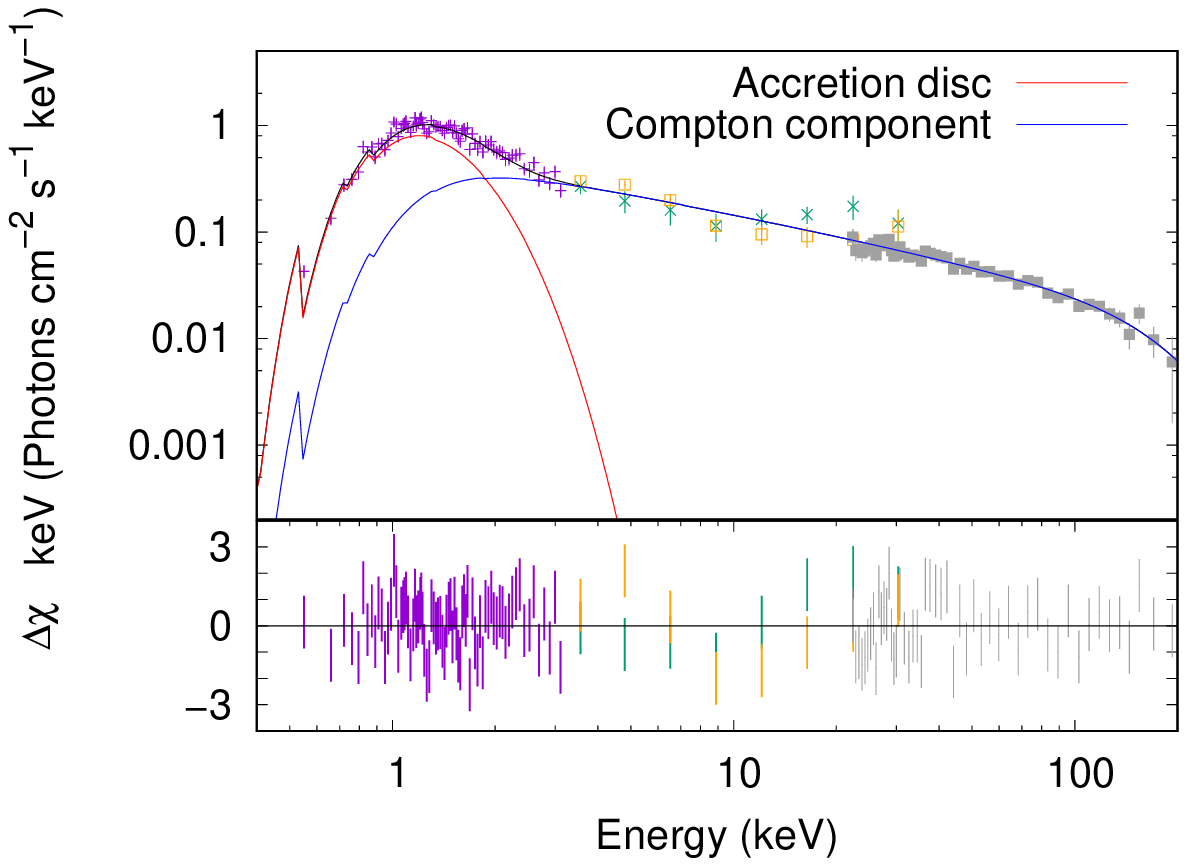}&\hspace{-1cm}\includegraphics[width=6.3cm]{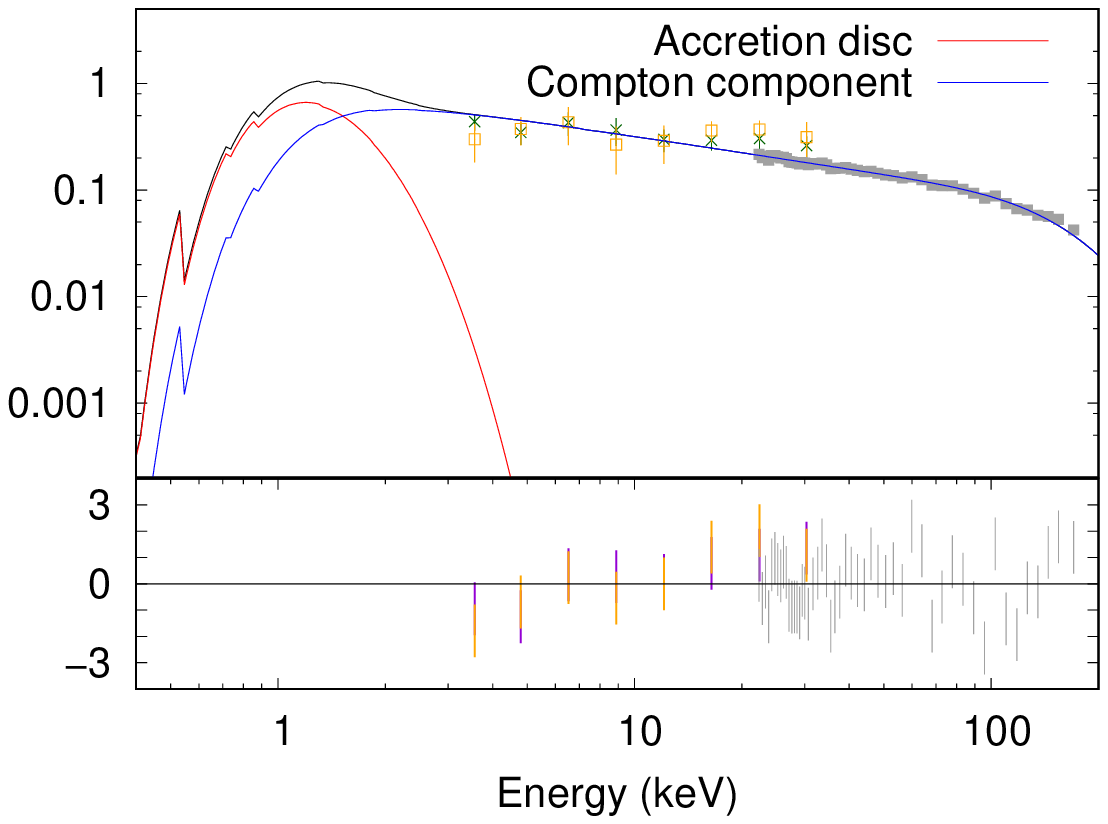}&\hspace{-1cm}\includegraphics[width=6.3cm]{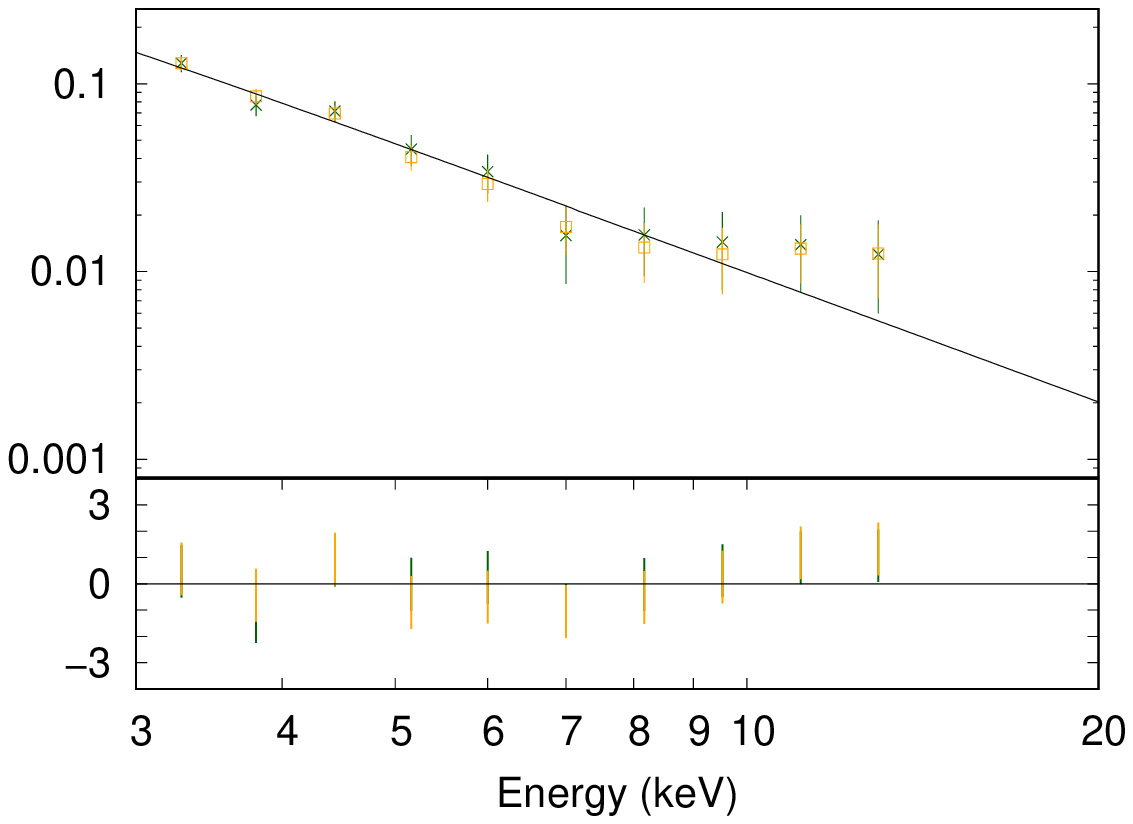}\\
\end{tabular}
\caption{Best fits to thenthe \v404\ X-ray spectra combining \sw/XRT (magenta) with \intl/JEMX-1 (green) and JEMX-2 (orange), and ISGRI (grey) data. Only the third broadband X-ray spectrum (right) is simultaneous to the FORS2 observations, the two others having been obtained about 2.2 hours (left) and 0.5 hour (middle) before.}
\label{plotbestfitx}
\end{center}
\end{figure*}

\subsection{The X-ray emission}

\begin{table}
\begin{center}
\caption{Best-fit parameters to the
   \v404\ \sw/XRT and \intl/JEMX+ISGRI combined spectra with {\sc tbabs$\times$(ezdiskbb+comptt}). Spec3 is simultaneous to the FORS2 observations, whereas Spec1 and Spec2 were obtained 2.2 hours and 0.5 hour before, respectively. The error bars are given at the 90\% confidence level. \label{parbestfitx}}
\begin{tabular}{lccc}
\hline\hline
&Spec1&Spec2&Spec3\\
\hline
\nh$^a$&$0.85_{-0.09}^{+0.10}$&0.85 (fixed)&0.85 (fixed)\\
$kT_{\rm max}^b$&$0.28\pm0.03$&--&--\\
$R_{\rm in}^c$&$108.1_{-28.5}^{+41.9}$&--&--\\
$kT_{\rm 0}^d$&tied to $kT_{\rm max}$&0.28 (fixed)&--\\
$kT_{\rm e}^e$&$38.5_{-6.7}^{+22.7}$&$36.4_{-2.1}^{+2.6}$&--\\
$\tau^f$&$3.46_{-1.19}^{+0.66}$&$4.41_{-0.21}^{+0.22}$&--\\
$\Gamma^{g}$&--&--&$3.31_{-0.24}^{+0.26}$\\
$C_{\rm JMX2}^{h}$&$1.21_{-0.18}^{+0.23}$&$0.80_{-0.18}^{+0.21}$&$1.11_{-0.13}^{+0.15}$\\
$C_{\rm ISGRI}^{i}$&$1.18_{-0.18}^{+0.22}$&$0.99_{-0.12}^{+0.16}$&--\\
$C_{\rm XRT}^{j}$&$1.01_{-0.18}^{+0.22}$&--&--\\
$F_{4-15}^{k}$&0.24&0.57&0.03\\
$F_{4-150}^{l}$&1.13&3.37&--\\
$\chi_{\rm r}^2$ (d.o.f)&1.02 (136)&1.12 (61)&0.76 (17)\\
\hline
\end{tabular}
\end{center}
\raggedright{\hspace{0.2cm}{{$^a$}{\scriptsize Column density in units of $10^{22}$~cm$^{-2}$\\}}
\hspace{0.2cm}{{$^b$}{\scriptsize  Accretion disc maximum temperature (keV) \\}}
\hspace{0.2cm}{{$^c$}{\scriptsize Inner radius in units of $f^2 D_{\rm BH}/ \sqrt{\cos{i}}$~km, with $i$ the inclination,\\ \hspace{0.2cm} $D_{\rm BH}$ the distance in unit of 10~kpc, and $f$ is the \\ \hspace{0.2cm} color-to-effective temperature ratio\\}}
\hspace{0.2cm}{{$^d$}{\scriptsize Input soft photon temperature (keV)\\}}
\hspace{0.2cm}{{$^e$}{\scriptsize Electron temperature (keV)\\}}
\hspace{0.2cm}{{$^f$}{\scriptsize Electron optical depth\\}}}
\hspace{0.2cm}{{$^g$}{\scriptsize Power law photon index\\}}
\hspace{0.2cm}{{$^h$}{\scriptsize JEMX-2 multiplicative constant\\}}
\hspace{0.2cm}{{$^i$}{\scriptsize ISGRI multiplicative constant\\}}
\hspace{0.2cm}{{$^j$}{\scriptsize XRT multiplicative constant\\}}
\hspace{0.2cm}{{$^k$}{\scriptsize 4-15 keV flux in units of $10^{-8}$~\ergcms \\}}
\hspace{0.2cm}{{$^l$}{\scriptsize 4-150 keV flux in units of $10^{-8}$~\ergcms \\}}
\end{table}

As shown in \autoref{logobs}, only the last \intl\ pointing is strictly simultaneous with the FORS2 observations. However, \v404\ is barely detected by \intl\ at this epoch and we can only make use of a 3.6-16~keV JEM-X spectrum. While this is enough to perform a phenomenological spectral fit, it cannot allow us to get a better view of the X-ray properties of the source. This is the reason why we also use the remaining X-ray spectra, which were obtained 0.5 and 2.2 hours before. Concerning the non-simultaneity of the X-ray spectra, we stress that we are not interested in performing a very detailed spectral analysis and that our aim here is only to obtain a rough estimate of the nature of the high-energy emission from \v404.

We first combined the \sw/XRT (\#00031403061) and \intl/JEM-X+ISGRI (\#155700300010) data sets to build a 0.5--150~keV spectrum (hereafter Spec1) and fit it, using {\tt Xspec}~v.~12.9.0, with a model consisting of a viscous disc \citep[{\sc ezdiskbb},][]{2005Zimmerman} and a spherical Comptonisation component \citep[{\sc comptt},][]{1994Titarchuk}. Both were modified by the ISM extinction, modelled with {\sc tbabs} for the abundances and cross-sections given in \citet{2000Wilms} and \citet{1996Verner}, respectively, and multiplicative constants were introduced to take calibration discrepancies into account. We restricted the XRT spectrum to the 0.5--3.2~keV energy range, as including data points up to 5~keV resulted in an abnormally low XRT multiplicative constant. During the fitting process, the seed photon temperature $kT_{\rm 0}$ of {\sc comptt} was moreover tied to the maximum accretion disc temperature $kT_{\rm max}$. The best-fit parameters are listed in the first column of \autoref{parbestfitx} and the best-fit model is displayed in the left panel of \autoref{plotbestfitx}. The fit is satisfactory, with a reduced \chis\ of 1.02, and points towards a relatively cold accretion disc ($kT_{\rm max}=0.28\pm0.03$~keV) dominating the soft X-rays below 3~keV whereas a warm and thick Compton component ($kT_{\rm e}=39_{-7}^{+23}$~keV and $\tau=3.46_{-1.19}^{+0.66}$) is responsible for the bulk of the emission beyond 4~keV. The inferred inner radius $R_{\rm in}$ for $f\sim1.7$ \citep{1995Shimura}, $i\sim67^\circ$, and $M_{\rm BH}\sim9$~\msun\ as given in \citet{2010Khargharia} ranges between 3 and 6 $R_{\rm S}$, i.e. we do not see evidence for significant truncation of the inner accretion disc, although we stress that a $f=1.7$ value may not be fully appropriate when the hard X-ray continuum dominates the emission \citep[see, e.g.][]{2013Reynolds}. We also note that the derived column density $\nhe=0.85_{-0.09}^{+0.10}\times 10^{22}$~\cm2, is consistent with an ISM extinction \Av\ between 3.4 and 4.3 once converted with the relationship given in \citet{2009Guver}, i.e. a similar value than that derived via optical spectroscopy. 

We used the same model to fit the second \intl/JEM-X+ISGRI 3.6--150~keV spectrum (\#155700320010, hereafter Spec2) but the lack of soft X-ray data prevented us from constraining the accretion disc parameters. We therefore kept {\sc comptt} only and fixed $kT_{\rm 0}$ to the value derived for Spec1, assuming that the accretion disc had not significantly cooled down or heated up. We derive similar parameters (see \autoref{parbestfitx}, second column, and \autoref{plotbestfitx}, middle panel, for the best-fit parameters and best-fit model, respectively), and the fit is good, with a reduced \chis\ of 1.12. The only difference with Spec1 is that the 4-150~keV flux is three times higher,  ($3.37\times10^{-8}$~\ergcms\ vs $1.13\times10^{-8}$~\ergcms), hinting at an X-ray variability driven by the Compton component. 

Our third 3.6-16~keV JEM-X spectrum, simultaneous with our FORS2 observations (\#155700340010, hereafter Spec3), was fit with an absorbed power law, the column density being fixed to that found for Spec1. The best-fit parameters are listed in the third column of \autoref{parbestfitx} and the best-fit model is displayed in the right panel of \autoref{plotbestfitx}. We find that Spec3 is quite soft, with a photon index $\Gamma=3.31_{-0.24}^{+0.26}$, and a lot fainter than Spec1 and Spec2, with a 4-15~keV flux of about $0.3\times10^{-9}$~\ergcms\ against $2.4\times10^{-9}$~\ergcms\ and $5.7\times10^{-9}$~\ergcms, respectively. For comparison, we also fit the 3.6-16~keV parts of Spec1 and Spec2 with absorbed power laws and find that both spectra are harder than Spec3, with 1.5--2 photon indices. This softening is likely the reason why \v404\ is not detected by IBIS, and assuming that Spec3 also stems from Comptonisation, a possible explanation for such a phenomenon is a change in the physical properties of the Compton component. In particular, a steepening of the power law may, for instance, result from a drastic cooling at constant optical thickness or alternatively a drop in opacity at constant temperature. This behaviour may indicate that the source has entered a hard to soft transition following a substantial flux decrease between Spec2 and Spec3. These types of transitions have been observed in some persistent microquasars which are predominantly found in the hard state, such as GRS~1758$-$258 \citep[see, e.g.,][]{2001Smith}.

\section{Discussion}

The wealth of \ion{H}{i}, \ion{He}{i}, and higher ionisation lines in \v404\ optical spectrum, the strength of the Balmer emission signatures, and the variability pattern at short time scale are exceptional and confirm the peculiarity of the \v404\ outburst with respect to more classical BH XRBs. In the following sections, we investigate the origin of the spectroscopic content and draw a possible scenario to explain this behaviour. 

\subsection{Where do the emission lines come from?}

In outbursting microquasars, optical emission lines are thought to stem from the irradiated chromosphere of the outer accretion disc. In this case and assuming that they are principally rotationally broadened, we have the following relationship between the Keplerian velocity $V_{\rm out}$ at the outer radius $R_{\rm out}$ and their FWHMs \citep[see, e.g.,][]{2016Casares}: 
\begin{equation}
V_{\rm out}\sin i=\sqrt{\frac{GM_{\rm BH}}{R_{\rm out}}}\sin i\,\leq\,\frac{FWHM}{2}\,\,\rm{km}\,\rm{s}^{-1}
\label{eqnfwhm}
\end{equation}
where $i$ is the system's inclination. Although \v404\'s outer radius is unknown, its derived orbital parameters are accurate enough to estimate the typical Keplerian velocities $V_{\rm L}$ and $V_{\rm C}$ at the Roche lobe and circularisation radii $R_{\rm L}$ and $R_{\rm C}$, respectively, the former being larger and the latter  smaller than $R_{\rm out}$. Using the expressions given in \citet{1983Eggleton} and \citet{2002Frank} as well as the Kepler's third law, we derive:
\begin{equation}
V_{\rm L}\sin\,i=4.07\,10^5 \left(\frac{2 \pi G f}{P}\right)^\frac{1}{3}\sqrt{(1+q)(0.7+q^\frac{2}{3}\ln(1+q^{-\frac{1}{3}})}
\end{equation}
and 
\begin{equation}
V_{\rm C}\sin\,i=2.85\,10^5 \left(\frac{2 \pi G f}{P}\right)^\frac{1}{3}\left(0.5-0.227\log q\right)^{-2}
\end{equation}
in \kms. Here $q$ is the secondary-to-primary mass ratio, $P$ is the orbital period in days, and $f$ is the orbital mass function of the BH in \msun. For \v404\ parameters as given in \citet{1994Casares} and using \autoref{eqnfwhm}, we can thus conservatively expect typical FWHMs between 400~\kms\ and 700~\kms\ for optical emission lines originating anywhere between the Roche lobe and circularisation radii (not necessarily the disc), and this is the case for most of the detected features listed in \autoref{speclines}. 
\begin{figure*}
\begin{center}
\begin{tabular}{cc}
\includegraphics[height=3.83cm,width=5.7cm]{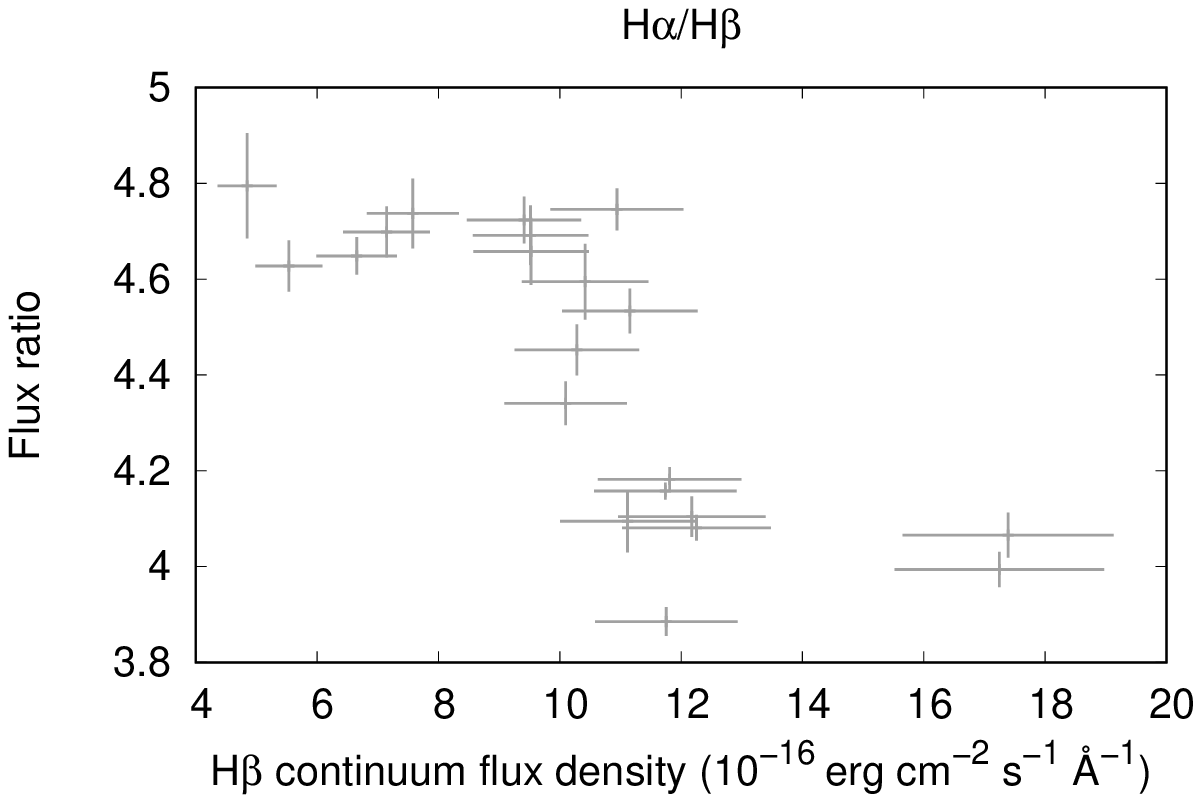}&\includegraphics[height=3.83cm,width=5.7cm]{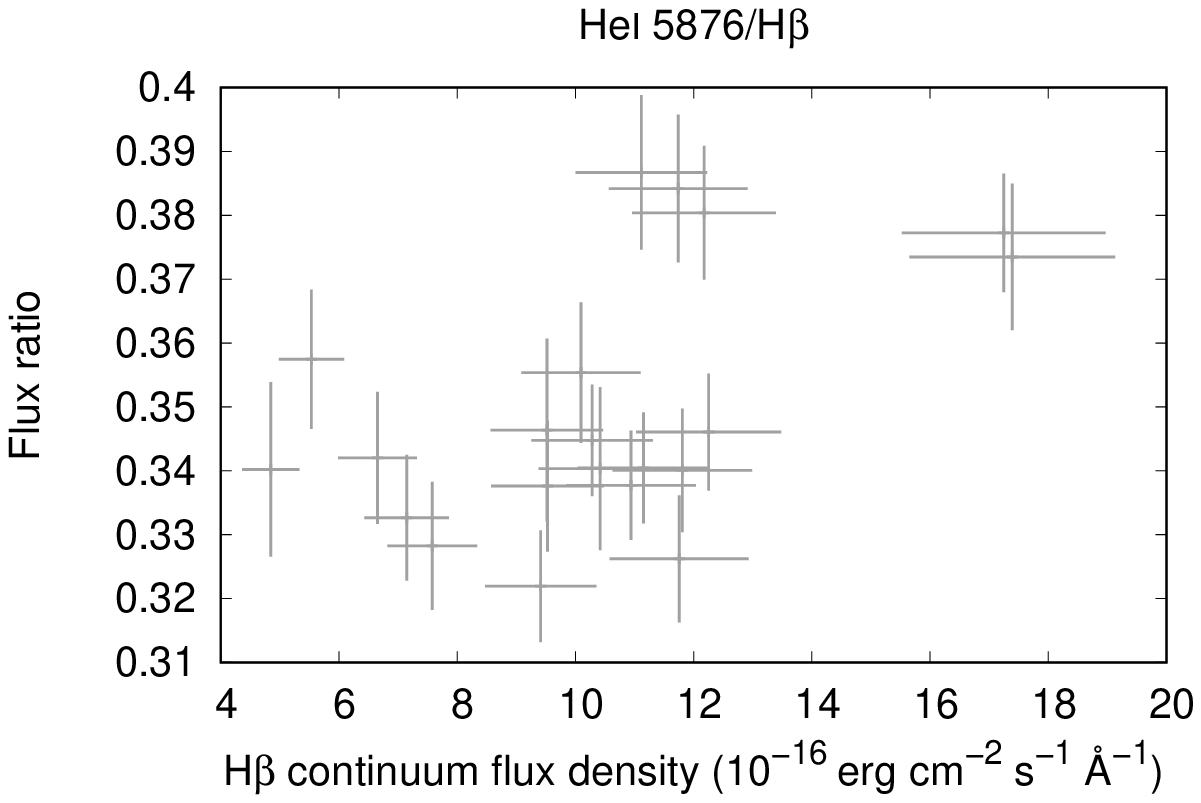}\\
\end{tabular}
\begin{tabular}{cc}
\includegraphics[height=3.83cm,width=5.7cm]{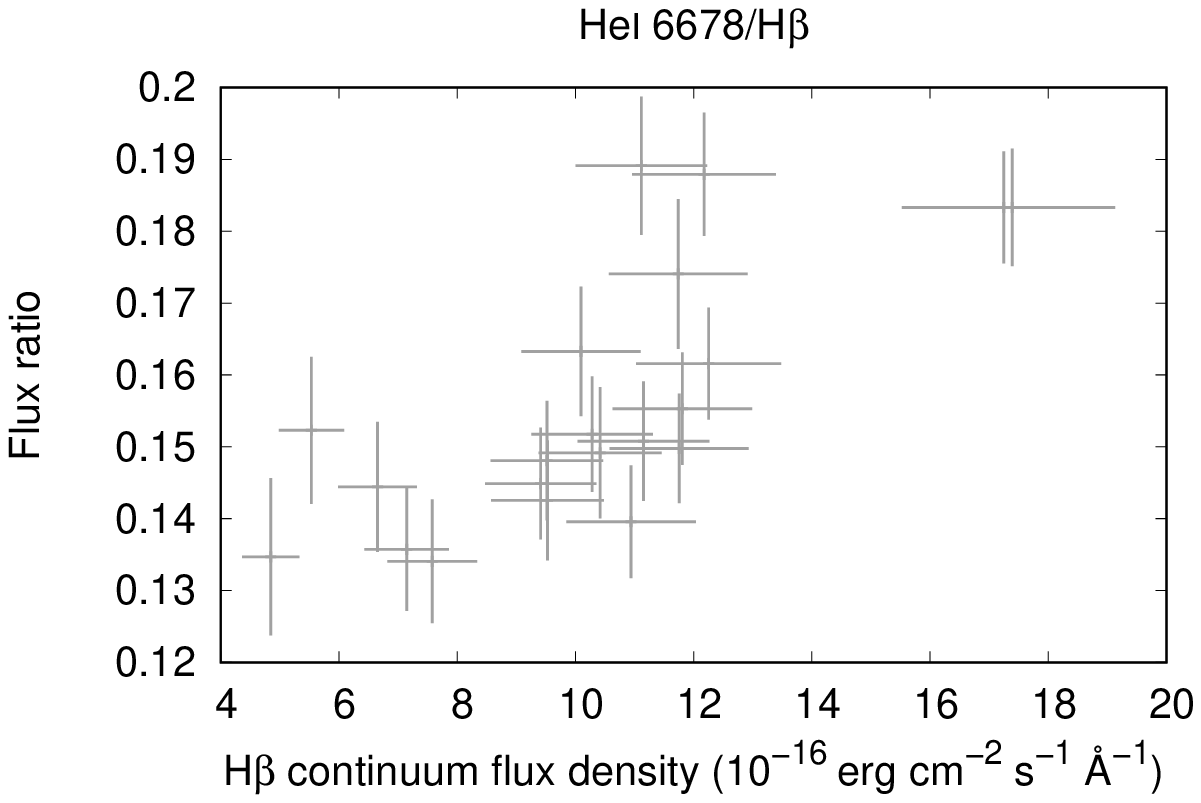}&\includegraphics[height=3.83cm,width=5.7cm]{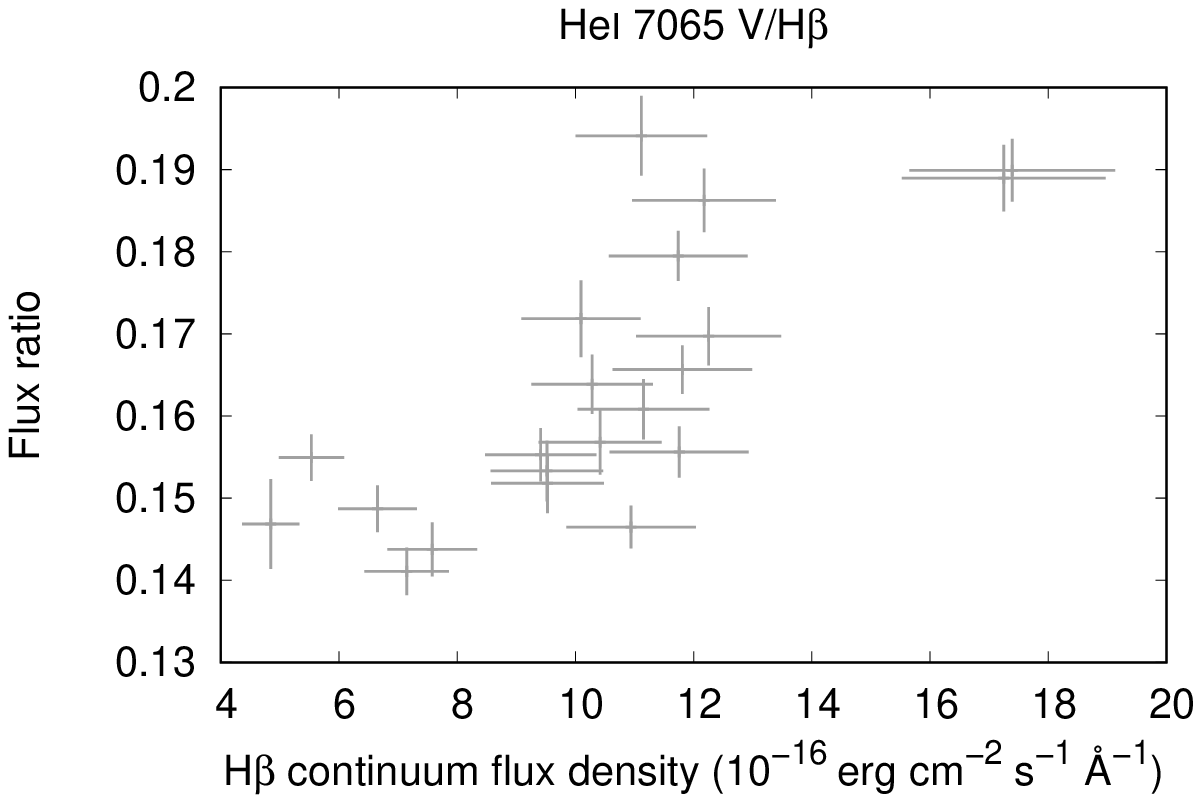}\\
\end{tabular}
\caption{\ha\ (top left), \ion{He}{i}~$\lambda5876$ (top right), \ion{He}{i}~$\lambda6678$ (bottom left), and \ion{He}{i}~$\lambda7065$ (bottom right) over \hb\ flux ratios variations in function of \hb\ underlying continuum.} 
\label{var_decr_v}
\end{center}
\end{figure*}
\begin{figure*}
\begin{center}
\begin{tabular}{ccc}
\includegraphics[width=5.7cm]{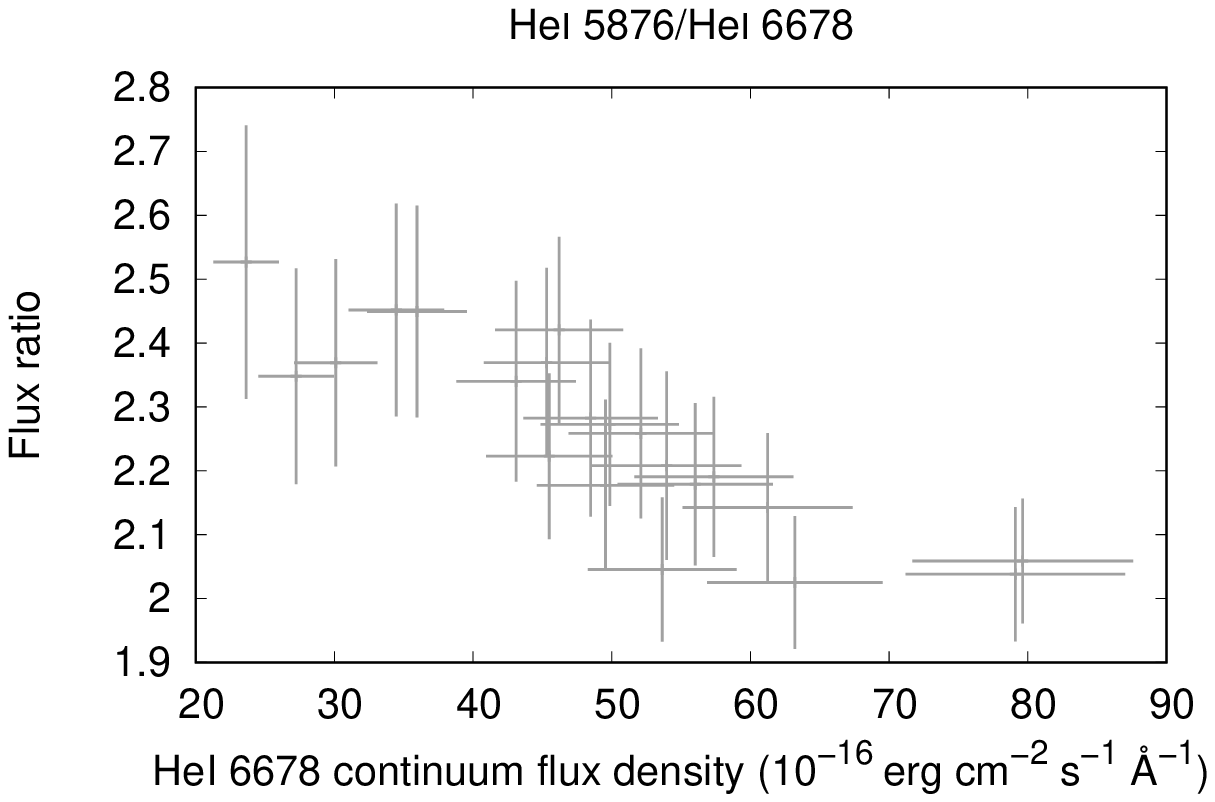}&\includegraphics[width=5.7cm]{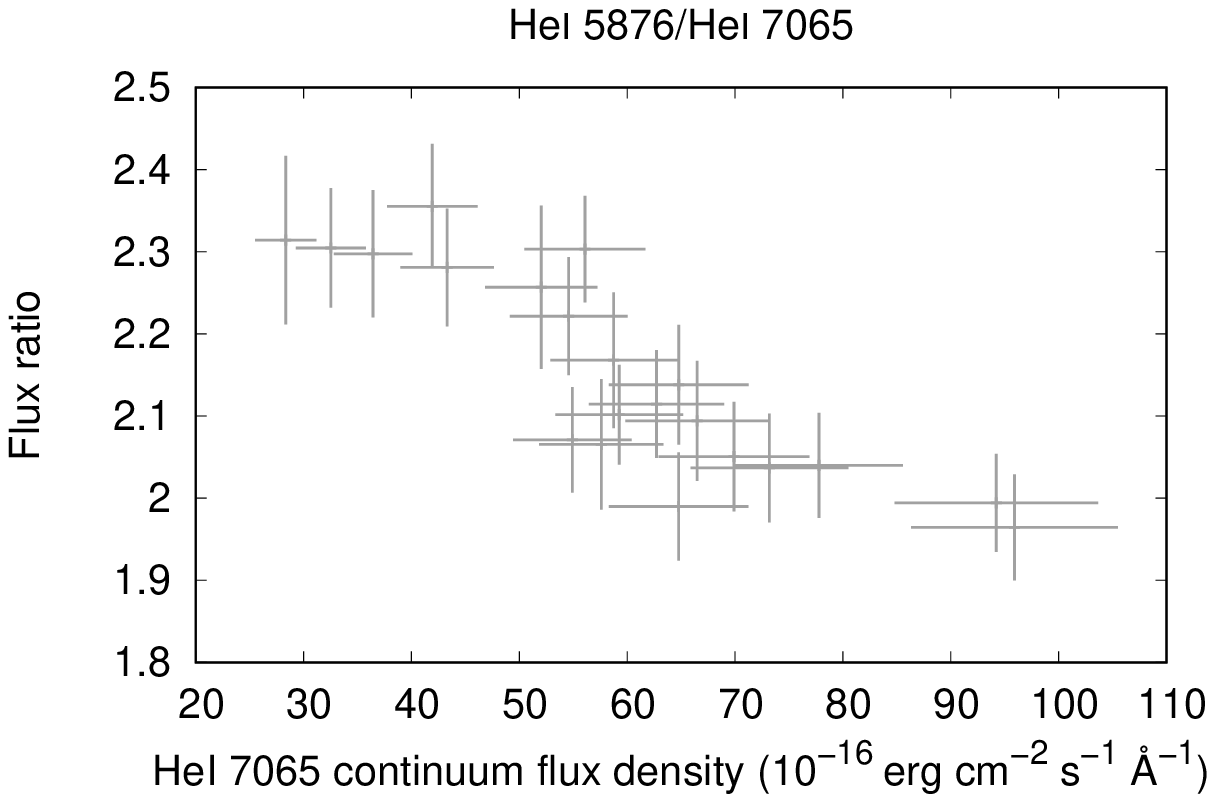}&\includegraphics[width=5.7cm]{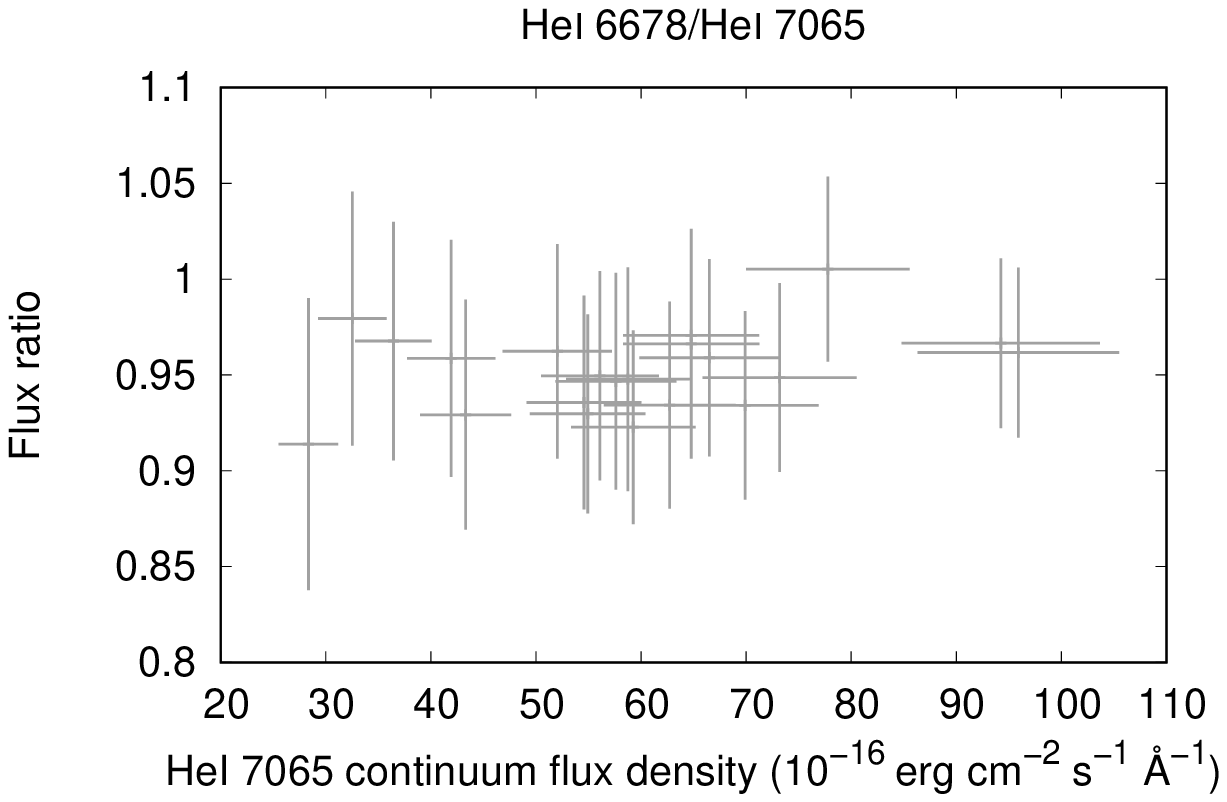}\\
\end{tabular}
\caption{\ion{He}{i}~$\lambda5876$/\ion{He}{i}~$\lambda6678$ (left), \ion{He}{i}~$\lambda5876$/\ion{He}{i}~$\lambda7065$ (middle), and \ion{He}{i}~$\lambda6678$/\ion{He}{i}~$\lambda7065$ (right) flux ratios variations in function of the denominator underlying continuum.}
\label{var_decr_he}
\end{center}
\end{figure*}

However, \ha, \hb, \ion{He}{i}~$\lambda$5876, \ion{He}{i}~$\lambda$6678, and \ion{He}{i}~$\lambda$7065 all are broader, with FWHMs$\ge900$~\kms, i.e. three to four times closer to the BH if we associate this broadening to a Keplerian rotation. Taking into account that these five \ion{H}{i} and \ion{He}{i} lines also are the only ones exhibiting significant radial velocity red-shifts and anti-correlations between their local continua and both their equivalent widths and FWHMs, we can speculate that they originate from a different emitting region than the outer system. For \ha\ and \hb\, this statement is further strengthened by the average Balmer decrement \ha/\hb$=4.61\pm0.62$, measured from the line fluxes corrected for the ISM extinction $\Ave=3.82\pm0.36$ as derived in Section~3.1. This value is larger than and inconsistent with the $\approx3.2$ value found just before the outburst initial rise \citep{2016Bernardini} and the canonical 2.86 for case B recombination of an ionised 10000~K nebula \citep{1938Baker}. It is also very different from unity, which is expected from optically thick irradiated accretion discs found in cataclysmic variables \citep{1980Williams, 1988Williams, 2016Tomsick} and some outbursting microquasars \citep[see e.g.][]{2014Rahouib}. As shown in \citet{1980Drake}, a possible explanation for such a large Balmer decrement is the combined effects of high optical depths in the Balmer lines and the relative importance of collisional excitation compared to radiative or recombination processes. This could also explain the anti-correlation between their equivalent widths and local optical continua. Indeed, if the lines are optically thick, any increase of the ionising radiation enhances their optical depths further, as shown in \citet{1979Ferland}, and the higher the ionising field, the flatter their flux evolution. Assuming an optically thin continuum emitting region, which is reasonable considering the broad \ha\ wings, this phenomenon effectively results in the aforementioned anti-correlation. Incidentally, if optical depth effects really are important, we expect the \ha/\hb\ decrement to be anti-correlated to the \hb\ local continuum, as \ha\ optical depth increases at a faster pace, and this is exactly the case (see top-left panel of \autoref{var_decr_v}). In contrast, the variations of the three \ion{He}{i}/\hb\ decrements in function of \hb\ continuum rather show a clear upwards evolution (see the remaining three panels of \autoref{var_decr_v}), which strengthens the hypothesis of a rapid optical depth increase in \hb. Likewise, \ion{He}{i}~$\lambda5876$/\ion{He}{i}~$\lambda6678$ and \ion{He}{i}~$\lambda5876$/\ion{He}{i}~$\lambda7065$ decrements point towards a similar enhancement for \ion{He}{i}~$\lambda5876$ optical depth, as as they evolve upwards too (see \autoref{var_decr_he}). Nonetheless, \ion{He}{i}~$\lambda6678$/\ion{He}{i}~$\lambda7065$ decrement is constant, hinting at the lack of significant optical depth variations, which may explain why the two lines do not exhibit the same unambiguous equivalent width/local continuum anti-correlation. 

\subsection{A classical nova-like nebula in \v404?}
\begin{figure*}
\begin{center}
\includegraphics[width=15cm]{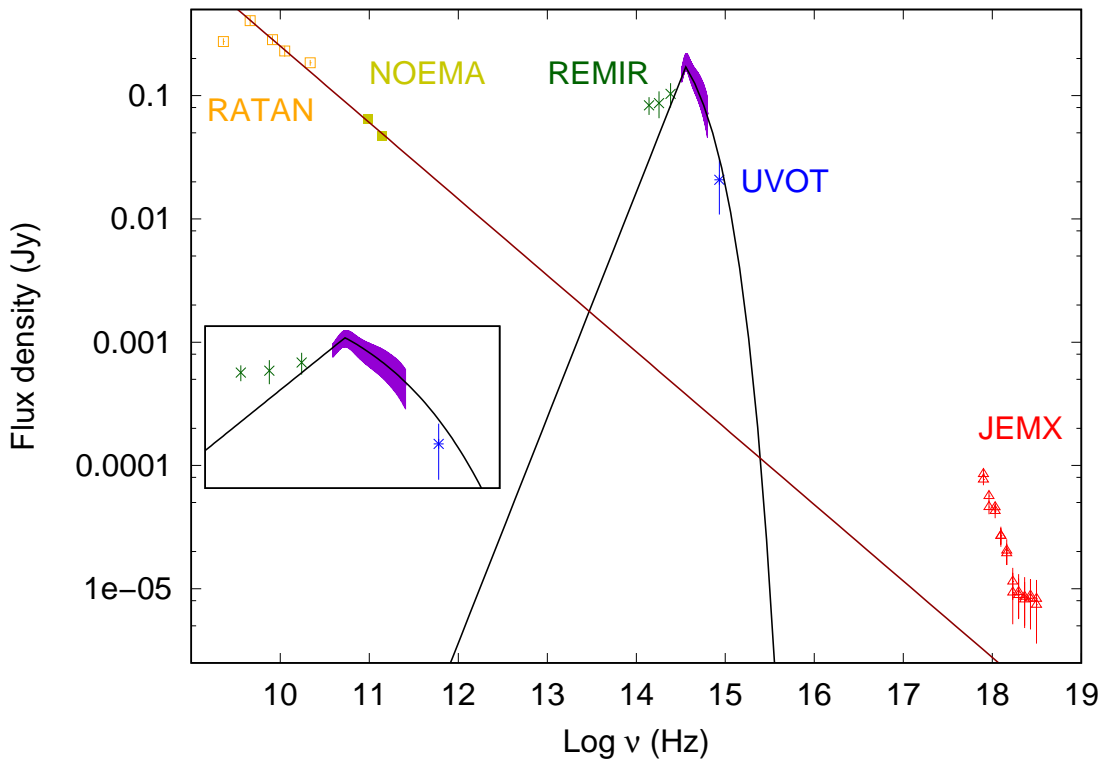}
\caption{\v404\ SED modelled with a self-absorbed optically thin bremsstrahlung in the optical and near-infrared domain. A $-0.62$ spectral index power law, consistent with optically thin synchrotron, is superimposed on the RATAN-600 AND NOEMA radio and sub-millimetre contemporaneous data. The inset displays a zoom-in centred on the REMIR, FORS2, and UVOT data.}
\label{specdered}
\end{center}
\end{figure*}

The same decrement analyses for all the narrow \ion{H}{i} and \ion{He}{i} lines (FWHM$\sim400-600$~\kms), which we argue likely stem from the outer regions, do not show any peculiarity and remain constant. Based on their equivalent widths, large radial velocity red-shifts and/or the evolution of their decrements, it is therefore tempting to associate \ha, \hb, \ion{He}{i}~$\lambda5876$, $\lambda6678$, and $\lambda7065$ to an expanding broad line region (BLR) in which collisional processes may be important, lines are optically thick, and the continuum optically thin. But what could this BLR be and how does this relate to the FWHM/local optical continuum anti-correlation?  A first explanation is that these FWHM variations trace the irradiation-induced expansion or contraction of the BLR, i.e. a decrease or an increase of its Keplerian velocity, similarly to what we would expect from the inflated base of a Compton-heated accretion disc wind \citep[ADW,][]{1983Begelman}. In this case, the maximum FWHM changes for the lines in $V$, i.e. when the optical continuum is increasing, imply an expansion of the emitting zones of about 20\% and 40\%, respectively. In contrast, the \ion{He}{i}~$\lambda7065$ FWHM increase in $I$, i.e. when the optical continuum is decreasing, gives a contraction of about 40\%. We note that \citet{2015King} reported the X-ray detection of blue-shifted emission lines and that P-Cygni profiles were observed in the optical domain \citep{2016Munoz}, which favours the ADW base hypothesis. It is however not clear to which extent we really could follow Keplerian velocity variations at such a short time scale. Moreover, without any better knowledge of the simultaneous X-ray properties of the source, it is difficult to say whether a 20-40\% radius increase in less than half an hour is realistic. Another possibility is that FWHM variations are thermally driven, i.e. the narrowing is due to cooling and the broadening to heating. Such an explanation may seem counter-intuitive as cooling and heating would be accompanied by an optical continuum increase and decrease, respectively, but this apparent paradox can be solved if we consider that the emission lines and the continuum arise from the same component and that the emission stems from optically thin bremsstrahlung or synchrotron. 

A third alternative, which would reconcile the two aforementioned scenarios, is the presence of a quasi-spherical optically thin nebula surrounding \v404, which was also proposed by \citet{2016Munoz}. Indeed, the optical spectrum strongly resembles that of \ion{Fe}{ii}-type classical novae in their post-outburst nebular phase \citep[see,][for a review]{2008Shore}. This phase is thought to occur once fast and optically thick ADWs stop being fuelled and massively ejected following a significant drop in mass-accretion rate and X-ray ionising emission. Their outer shell, previously formed, then starts growing and cooling down, becoming optically thin in the process and effectively turning into a nebula, while remnant inner optically thick ADWs keep expanding for some time.  Moreover, large Balmer decrement and equivalent width/continuum anti-correlations are also seen and explained in a similar way as we do for \v404, and a narrowing with time of the broad emission lines is also observed. This narrowing is mainly thought to stem from shocks occurring between the nebulae and much slower stellar winds. In the \v404\ case, we can expect slow winds from its cool sub-giant companion star, and the presence of many narrow \ion{Fe}{ii} emission lines, which behave differently from both the broad features and the other \ion{H}{i} and \ion{He}{i} signatures, hints at the existence of a stellar winds/nebula front, as suggested in \citet{2012Williams} for classical novae. However, it is important to remember that our data not only suggest a narrowing when the continuum is increasing, but also a broadening when it is decreasing. The spectroscopic content is also divided between narrow stationary features and broad red-shifted ones. Thus, another explanation that could account for both could be that the broad lines actually originate from deeper in the nebula, in a region, which could be the remnant ADW, moving faster than the outer optically thin shell and eventually colliding with it. The collision leads to the temporary narrowing of the thick emission lines, which re-broaden once the interaction is over because we see deeper in the nebula again. 

\subsection{The origin of optical continuum}

In the soft state, the optical continuum of outbursting low-mass XRBs is usually dominated by the viscous or reprocessed emission of the accretion disc, to which synchrotron radiation from the compact jets is added in the hard state. The companion star contribution may also be detectable but only when the source is close to quiescence. In the $\nu$ vs $F_{\nu}$ plan, this leads to a blue continuum from the disc, with a spectral index between roughly 0.3 and 2, and a potential red excess when jets are present \citep[see e.g.,][in the case of \gx\ and \swj]{2012Rahoui, 2015Rahoui}. 

\autoref{specdered} displays the extinction-corrected average continuum of \v404\ during our observations, corrected for the companion star emission, and it is clear that it does not correspond to the previous description. Indeed, it is characterised by a red increase and a blue exponential decay with rough spectral indices of $+2$ and $-1.5$, respectively, the turnover being located around $3.5\times 10^{14}$~Hz. We therefore performed a phenomenological fit of the \v404\ optical continuum with a function consisting of a power law of spectral index $\alpha$ up to frequency turnover $\nu_{\rm c}$ and an exponential decay of temperature $T$ beyond. To better constrain the fit, we also use the \sw/UVOT U and REM $J$, $H$, and $K_{\rm s}$ fluxes, all corrected for the companion star emission. The best-fit is displayed in \autoref{specdered}, to which we superimposed the contemporaneous radio and sub-millimetre fluxes obtained with RATAN-600 and NOEMA as well as the simultaneous JEMX spectrum to build the radio to X-ray SED. We infer $\alpha\approx1.86$, consistent with a Raleigh-Jeans tail, $\nu_{\rm c}\approx3.64\times10^{14}$~Hz, and $T\approx14000$~K, which is cool enough for Balmer and \ion{He}{i} emission line formation. Moreover, while the fit correctly accounts for the $J$-band flux density, a near-IR excess is present. Based on the radio and sub-millimetre flux levels and decay ($-0.62$ spectral index), this excess does not appear to stem from optically thin synchrotron from the compact jets, but the variability of the source is such that we believe it cannot be excluded. A dusty component might also account for this mismatch.

Our phenomenological fit confirms the peculiarity of the \v404\ optical spectrum with respect to other microquasars. But which physical processes does it describe? A first possibility is self-absorbed optically thin cyclo-synchrotron emission, similarly to that modelled for a magnetised corona above the accretion disc in \citet{1997Dimatteo}. However, for BH XRBs, such a mechanism is expected to peak in the UV domain, between $10^{15}-10^{16}$ Hz, and a peak around $3.6\times10^{14}$~Hz would yield very high magnetic field strengths $B\ge10^7$~G and/or temperatures $T\ge10^6$~K \citep{2000Wardzinski}. Similarly, optically thin synchrotron from an Advection Dominated Accretion Flow \citep[ADAF;][]{1995Narayan} was successfully used in quiescent sources but it is likely not applicable to an outbursting microquasar at the very beginning of the decay, especially considering that the mass-accretion rate we derive from our X-ray fit is $\dot{M}_{\rm acc}\approx 0.01\dot{M}_{\rm Edd}$ for \citet{2010Khargharia} parameters and $f=1.7$. In place of synchrotron-related phenomena, we rather believe that a more realistic physical process is self-absorbed optically thin bremsstrahlung, which gives rise to continuum radiation very similar to from observed from \v404. If true, it is tempting to associate this free-free emission to the geometry of the ADW cooling remnant we claim is present. Indeed, it is noteworthy that a consequence of shocks between slower optically thin and faster optically thick regions is the conversion of kinetic energy losses into a optically thin continuum emission, which would likely be bremsstrahlung. Using the expression of the free-free normalisation as given in \citet{1979Rybicki}, we infer an electron density $n_{\rm e}\approx10^{10}$~cm$^{-3}$ from our phenomenological fit, assuming that the nebula extends up to the accretion disc Roche lobe radius, which points towards a low-density. Incidentally, the shocks between faster and slower regions could also explain why the narrowing and broadening are anti-correlated to the local continua of the emission lines.

\section{Summary and conclusion}

We have presented a comprehensive study of the \v404\ optical spectrum at the very beginning of its 2015 outburst decay. The high S/N of each individual spectrum as well as the short exposure time has allowed us to perform a thorough analysis of the continuum and spectral variability pattern and to investigate the origin of the existing correlations. We find that the spectral features may originate in three different regions: (1) a fast-moving optically thick plasma, likely an ADW remnant, within a classical nova-like nebula for the broadest (FWHM$\sim900-1000$~\kms) and red-shifted ($\Delta V\sim70-120$~\kms) emission lines; (2) the slow-moving or stationary optically thin outer shell of the aforementioned nebula for the narrow high ionisation emission lines (FWHM$\sim400-600$~\kms); and (3) the outer accretion disc for the remaining narrow \ion{H}{i} and \ion{He}{i} features. This geometry assumes the existence of a massive, almost discrete optically thick ADW that was turned-off prior to the decay and was detected via X-ray emission lines and optical P-Cygni profiles. We further argue that the optical continuum is unlike any observed in other BH XRBs and likely arises from the aforementioned shocks, perhaps through optically thin bremsstrahlung cooling. 

Our results confirm the uniqueness of this \v404\ outburst in the microquasar family and point towards the dominant role played by massive ADW ejecta in the source properties. It is not clear if these ejecta can be held responsible for part of the extreme flaring activity the source exhibited during its outburst, or if they are only a consequence of the strong flares irradiating the accretion disc. In either case, the reasons behind this behaviour are unknown, but it is noteworthy that \v404, like the other extreme microquasar \grs, is characterised by a long orbital period and a very large accretion disc, which, according to \citet{2016Kimura}, could prevent sustained accretion in the inner regions and induce large disc instabilities. Whether or not this is the case, we note that \v404\ behaviour, including the various observed anti-correlations, is not only seen in the classical novae from which we draw a parallel, but also in some AGN and Seyfert galaxies \citep[see, e.g.][]{2013Lee} with the well-known Baldwin effect \citep{1977Baldwin}. This may point towards the importance of the universal presence of broad line regions in accreting systems spanning whole compact objects mass scale.

\section*{Acknowledgements}
We are very thankful to the referee for his/her very insightful comments and suggestions that helped improve this paper a lot. We also thank Mariko Kimura for kindly providing her $I$ band photometry. We acknowledge with thanks the variable star observations from the AAVSO International Database contributed by observers worldwide and used in this research. PC acknowledges support by a Marie Curie FP7-Reintegration-Grants under contract no. 2012-322259. This research has made use of data obtained from the High Energy Astrophysics Science Archive Research Center (HEASARC), provided by NASA's Goddard Space Flight Center. This research has made use of NASA's Astrophysics Data System, of the SIMBAD, and VizieR databases operated at CDS, Strasbourg, France.




\bibliographystyle{mnras}
\bibliography{../mybib_tot.bib} 





i


\bsp	
\label{lastpage}
\end{document}